\documentclass[journal=cgdefu,manuscript=article,layout=twocolumn]{achemso}
\usepackage{amssymb, amsmath, graphicx, verbatim, color, natmove, soul}
\usepackage[color=blue!20]{todonotes} \setcitestyle{super}
\usepackage{siunitx}
\usepackage{outlines}
\usepackage{booktabs}
\usepackage{caption}
\usepackage{rotating}
\usepackage{makecell}
\usepackage{etoolbox} 
\usepackage{url}
\patchcmd{\subsubsection}{\itshape}{\bfseries}{}{}
\patchcmd{\subsubsection}{\centering}{\flushleft}{}{}
\usepackage{float}
\usepackage[numbers,sort&compress,super]{natbib}
\usepackage{hyperref}
\hypersetup{
    colorlinks=true,
    linkcolor=blue,
    filecolor=magenta,      
    urlcolor=cyan,
}
\usepackage[inline]{enumitem}
\usepackage{threeparttable}
\usepackage{multirow}
\definecolor{lightyellow}{rgb}{1, 1, 0.65}
\sethlcolor{lightyellow}

\usepackage{cleveref}
\usepackage[version=4]{mhchem}

\newcommand{\eg}{\textit{e.g. }}
\newcommand{\ie}{\textit{i.e. }}
\newcommand{\etal}{\textit{et al. }}
\newcommand{\tu}[1]{\textup{#1}}

\newcommand{\beqn}{\begin{eqnarray}} 
\newcommand{\eeqn}{\end{eqnarray}}

\SectionNumbersOn

\author{Vikram Korede} 
\altaffiliation{denotes equal contribution}
\affiliation{Process \& Energy Department, Delft University of Technology, Leeghwaterstraat 39, 2628 CB Delft, The Netherlands}

\author{Nagaraj Nagalingam} 
\altaffiliation{denotes equal contribution}
\affiliation{Process \& Energy Department, Delft University of Technology, Leeghwaterstraat 39, 2628 CB Delft, The Netherlands}

\author{Frederico Marques}
\affiliation{Department of Chemical Engineering, KTH Royal Institute of Technology, Teknikringen 42, 114-28 Stockholm, Sweden}
\affiliation{Process \& Energy Department, Delft University of Technology, Leeghwaterstraat 39, 2628 CB Delft, The Netherlands}

\author{Noah van der Linden} 
\affiliation{Process \& Energy Department, Delft University of Technology, Leeghwaterstraat 39, 2628 CB Delft, The Netherlands}
 
\author{Johan T. Padding} 
\affiliation{Process \& Energy Department, Delft University of Technology, Leeghwaterstraat 39, 2628 CB Delft, The Netherlands}

\author{Remco Hartkamp}
\email{r.m.hartkamp@tudelft.nl} 
\affiliation{Process \& Energy Department, Delft University of Technology, Leeghwaterstraat 39, 2628 CB Delft, The Netherlands}

\author{Huseyin Burak Eral} 
\email{h.b.eral@tudelft.nl} 
\affiliation{Process \& Energy Department, Delft University of Technology, Leeghwaterstraat 39, 2628 CB Delft, The Netherlands}

\title{A review on laser-induced crystallization from solution}

\begin{document}

\begin{abstract} 
Crystallization is abound in nature and industrial practice. A plethora of indispensable products ranging from agrochemicals and pharmaceuticals to battery materials, are produced in crystalline form in industrial practice.  Yet, our control over the crystallization process across scales, from molecular to macroscopic, is far from complete. This bottleneck not only hinders our ability to engineer the properties of crystalline products essential for maintaining our quality of life but also hampers progress toward a sustainable circular economy in resource recovery. In recent years, approaches leveraging light fields have emerged as promising alternatives to manipulate crystallization. In this review article, we classify laser-induced crystallization approaches where light-material interactions are utilized to influence crystallization phenomena according to proposed underlying mechanisms and experimental setups. We discuss non-photochemical laser-induced nucleation, high-intensity laser-induced nucleation, laser trapping-induced crystallization, and indirect methods in detail. Throughout the review, we highlight connections amongst these separately evolving sub-fields to encourage interdisciplinary exchange of ideas.       
\end{abstract}
 
\section{Introduction}
Crystallization - how loosely correlated atoms in a solvent arrange themselves to flawless symmetric structures - has long captivated scientists and engineers alike. It is ubiquitous in nature and industrial practice, from the production of nanostructured materials, explosives, catalysts, organic electronics, pharmaceuticals to the formation of teeth and bones\cite{Myerson2002,Lewis2015,Kashchiev2000,10.1021/jacs.9b01883}. Despite its widespread use in industry as a separation and purification process, fundamental understanding of crystallization from solution and our ability to rationally dictate properties of emerging crystals is far from complete\cite{10.1073/pnas.1905929116,10.1038/nature25971,10.1126/sciadv.aao6283,10.1073/pnas.1700342114,10.1126/science.1243022}.
\par

Crystallization consists of two fundamental  phenomena: nucleation and growth. Nucleation, also referred to as primary nucleation, is the emergence of an ordered structure of solute molecules in solution. Classical nucleation theory (CNT) and two-step nucleation theory (TSN) are two models proposed to rationalize the nucleation process\cite{Kashchiev2000,10.1016/j.jcrysgro.2019.125300}. CNT, widely used due to its analytical simplicity, explains nucleation as a tug of war between the tendency to form a new phase and the energy cost associated with forming a new surface. In more formal terms, it describes nucleation as a one-step stochastic process dictated by the Gibbs free energy change for the phase transformation and the free energy change for the formation of a surface. Despite its simplicity, it remarkably, yet qualitatively, predicts experimentally observed trends\cite{doi:10.1021/ja4028814}. CNT is not free of shortcomings\cite{10.1103/PhysRevLett.98.145702}. Shortcomings of CNT in explaining observations, particularly in  protein crystallization experiments, led to the proposal of the two-step nucleation (TSN) model. In the TSN model, the formation of a sufficiently-sized amorphous pre-nucleation cluster is followed by its reorganization into an ordered structure\cite{10.1021/ar800217x,10.1021/cg1011633}. In industrial practice, nucleation and growth often occur in the presence of turbulent flows in well-stirred vessels. These two fundamental phenomena are always followed by secondary crystallization phenomena intimately related to coupled mass, momentum, and heat transfer - such as attrition, coalescence, and secondary nucleation unless the process is specially designed to suppress these secondary phenomena\cite{doi:10.1021/cm502834h}.  Nucleation, growth, and these secondary physical phenomena collectively dictate the crystal quality parameters: crystal size distribution, polymorphism, morphology, and purity, also referred to as the four pillars of industrial crystallization \cite{Myerson2002,Lewis2015,TerHorst2015,doi:10.1021/cm502834h,doi:10.1021/cg500250e}.\par   

In solution crystallization, nucleation plays a decisive role in determining crystal properties\cite{Kashchiev2000}. Light-material interaction experiments usually but not exclusively focus on nucleation. Approaches by-design targeting  growth and secondary nucleation phenomena exist in the literature, yet are less commonly encountered\cite{10.1038/nphoton.2016.202}. Due to the interconnected nature of nucleation and growth, it is hard to guarantee that a laser-induced method designed to steer nucleation does not influence growth. These light-material interaction experiments which we collectively refer to as laser-induced crystallization (LIC) have often, but not always, been conducted by exposing a solution carrying a  solute dissolved in a solvent to a light source of a given wavelength, intensity, exposure time, and pulse width. The exact experimental details including experimental geometry (e.g., container geometry, how the beam interacts with confining surfaces and solution), laser characteristics (intensity, wavelength, polarization, continuous or pulsed laser, pulse width), exposure time (ranging between femtoseconds to hours) and solution characteristics vary considerably across the literature. Moreover, distinct mechanisms based on molecular effects as well as continuum approaches have been proposed depending on these experimental details. Efforts to classify these constantly evolving LIC sub-fields in the literature may provide means to draw parallels among the proposed underlying mechanisms.\par

\begin{figure}[htbp!]
\centering	
\includegraphics[width=0.45\textwidth]{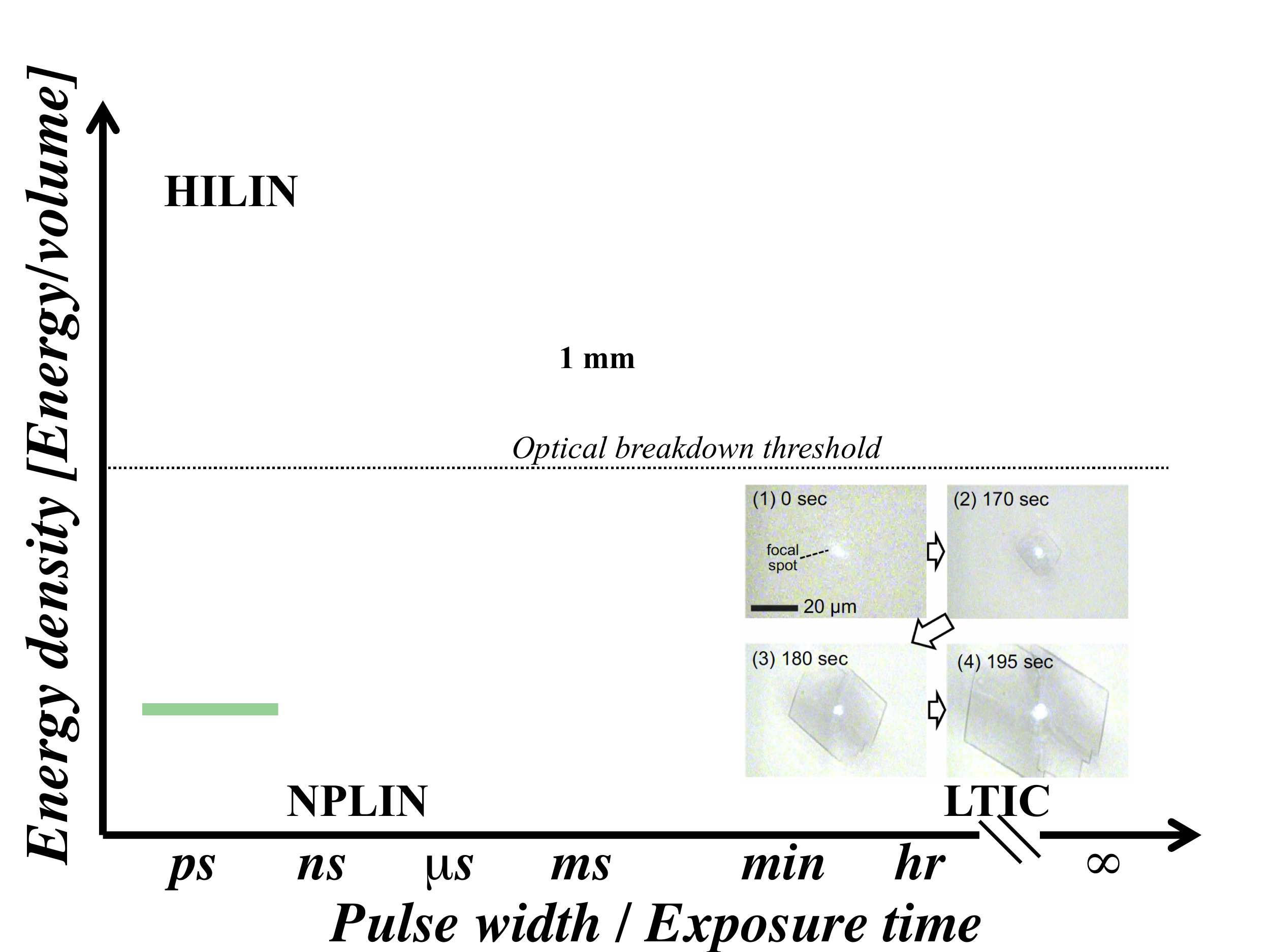}
\caption{\textbf{Laser induced crystallization phenomena.} Classification of non-photochemical laser-induced nucleation (NPLIN), high-intensity laser-induced nucleation (HILIN), and laser-trapping-induced crystallization with optical tweezers (LTIC) based on energy density and laser pulse width or exposure time in case of continuous lasers used in LTIC. The NPLIN time-lapse image, cavitation bubble produced by a high-intensity laser pulse, and LITC image are reproduced from \citet{10.1021/acs.cgd.7b01277}, \citet{doi.org/10.1007/s00348-016-2271-0} and  \citet{10.1007/s10043-015-0029-1} respectively. }
\label{fig1}
\end{figure}

In this review article, we summarize the existing experimental and computational literature, classify the reported experimental techniques, while discussing the proposed mechanisms. We will limit our classification to methods that are  non-photochemical in nature, at least not by intention. In photochemical approaches, the frequency of irradiation is intentionally tuned to trigger chemical reactions, whereas, in non-photochemical approaches, the frequency is chosen so that neither the solute nor the solvent ``significantly'' absorbs the irradiation. Figure~\ref{fig1} illustrates our early efforts to classify existing LIC literature based on the energy of the light irradiation and pulse width or exposure time when continuous lasers are concerned. In this rudimentary effort, we classify three different methods leveraging light-material interactions for LIC that have evolved as semi-independent  research fields, namely non-photochemical laser-induced nucleation (NPLIN), high-intensity laser-induced nucleation (HILIN) and laser-trapping-induced crystallization with optical tweezers (LTIC-OT) varying in energy density and pulse width.\par

We will first discuss NPLIN in \Cref{sec:NPLIN} and summarize the underlying mechanisms proposed in the literature.  Gareth, Myerson, and coworkers\cite{10.1107/S160057671401098X} serendipitously observed orders of magnitude faster nucleation kinetics upon light irradiation compared to unperturbed  supersaturated urea solutions at identical supersaturation. Since this report, the NPLIN effect has been reported in a considerable number of solute/solvent systems, yet the discussion on the underlying mechanism is not yet settled. A recent review by \citet{10.1063/1.5079328} provides a valuable source offering a summary of the proposed mechanism while  \citet{10.1107/S160057671401098X} provides a detailed account of experimental setups utilized. We extend both on the discussion of the proposed mechanism and the classification of experimental setups.  Following a detailed discussion on NPLIN, in \Cref{sec:HILIN}, we will focus on high-intensity laser-induced nucleation (HILIN) where the laser intensity of the pulse is orders of magnitude higher than for NPLIN. HILIN is characterized by energy densities above the optical breakdown threshold. At these high intensities, a plasma is formed and non-linear optical effects come into play. We have deliberately chosen the abbreviation HILIN as LIN will be used to refer to all laser-induced nucleation methods. This field of research has a history dating back to 1960s and early experiments with lasers connected to laser-induced boiling and energy research of pure substances\cite{PhysRevLett.78.1359}. Triggered by efforts to pinpoint the exact mechanism behind NPLIN, studies of HILIN on supersaturated solutions opened new alleys of investigation. In \Cref{sec:trapping}, we collect and summarize the efforts focused on laser-trapping-induced crystallization (LTIC), almost exclusively studied using optical tweezers operating with low-intensity continuous lasers\cite{10.1007/s10043-015-0029-1,10.1021/ar300161g,10.1039/c9sm01297d} as well as combinations of pulsed and continuous lasers\cite{10.1021/jacs.1c11154,10.1073/pnas.2122990119,10.1021/acs.cgd.9b01255} (as illustrated in Figure~\ref{fig1}). In \Cref{sec:indirect}, we will bring together indirect methods where auxiliary material-laser interactions are utilized to influence nucleation and growth mechanism. Next, we discuss molecular simulation efforts and opportunities to provide direct insight into molecular length and time scales that are often hard to access experimentally. In the summary section of each discussed LIC method, we attempt to highlight open scientific questions. Finally, \Cref{sec:concluding} offers a few concluding remarks. \par

\section{Non-photochemical laser-induced nucleation (NPLIN)} 
\label{sec:NPLIN}
\subsection{Phenomenology}
\label{NPLIN Phenomenology}

\begin{figure}[tbp]
	\centering
	\includegraphics[width=0.45\textwidth]{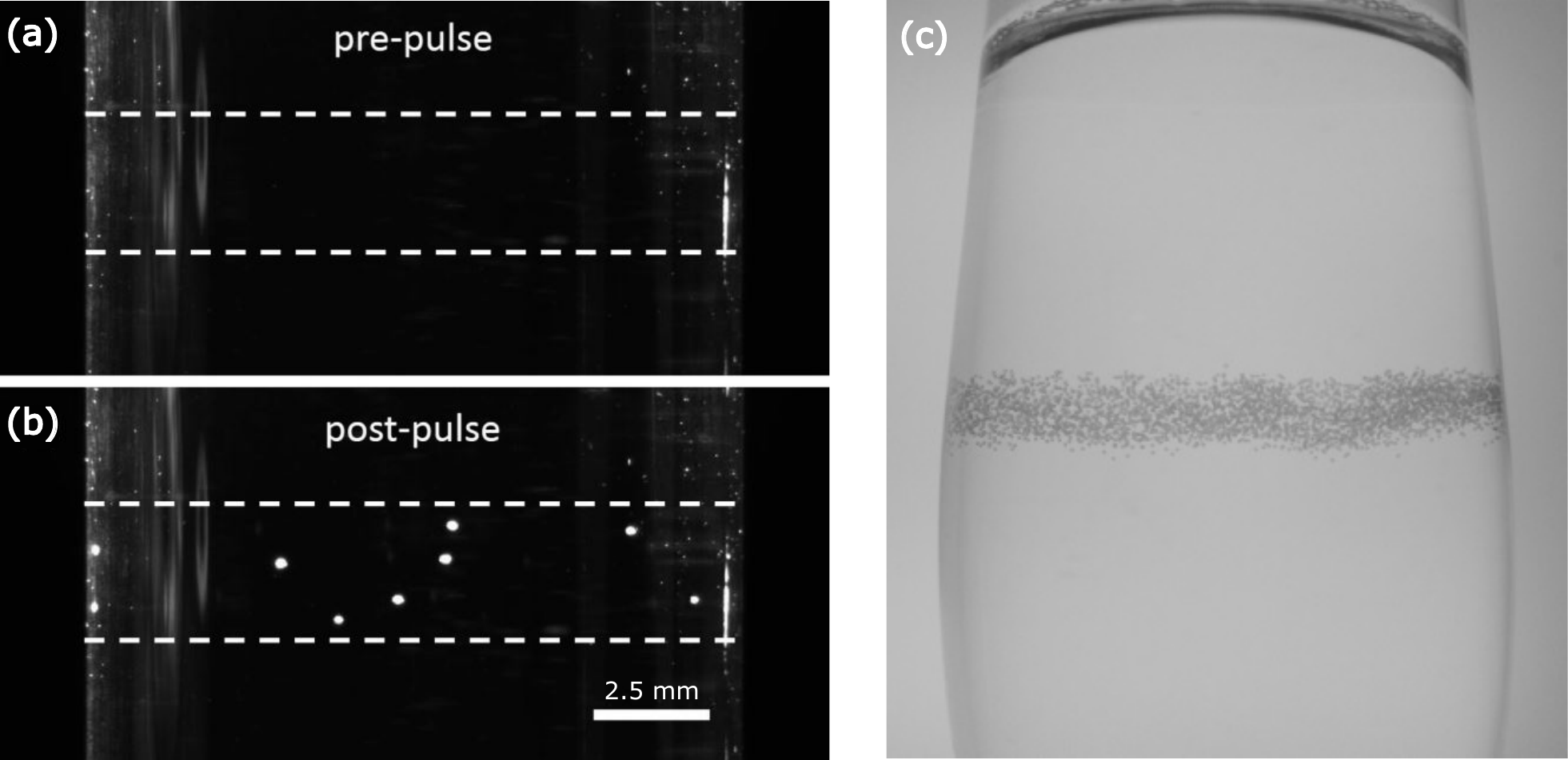}
\caption{\textbf{NPLIN in action} (\textbf{a,b}) Solution of NH$_4$Cl before and  approximately \SI{1.6}{s} after irradiating by a single laser pulse\cite{10.1021/acs.cgd.6b00882}. The path of the laser beam through the solution is indicated by the dashed white lines, with the start of nucleation visible as white dots between the lines. (\textbf{c}) Nucleation of carbon dioxide bubbles within carbonated solution caused by the passage of the laser pulse from left to right\cite{10.1063/1.4917022}.}
\label{NPLIN}
\end{figure}

In 1996, while attempting to observe second-harmonic generation in supersaturated aqueous urea solutions, Garetz \etal\cite{10.1103/PhysRevLett.77.3475}  noticed unexpected instantaneous crystallization upon light irradiation in solutions that would otherwise take several weeks to crystallize spontaneously. In this study, Garetz and co-workers exposed milliliter size vials as shown in Figure \ref{NPLIN}, using a series of linearly polarized, unfocused (50-\SI{250}{MW.cm^{-2}}), nanosecond light pulses with \SI{1064}{nm} wavelength.\par

Garetz \etal\cite{10.1103/PhysRevLett.77.3475} referred to this observed light-induced phenomenon as non-photochemical laser-induced nucleation, NPLIN. This phenomenon was considered non-photochemical as: (i) neither the solute nor the solvent has strong absorption bands at irradiated wavelengths and (ii) the applied laser intensity ($\sim \SI{}{MW.cm^{-2}}$) is considered too low to trigger photochemical reactions through non-linear optical effects. They reported the formation of needle-like urea crystals aligned with the polarization plane of the laser, suggesting an electric-field-induced origin of the underlying mechanism.  Over the following decades, various groups have reported enhanced nucleation probabilities upon laser irradiation, quantified by counting the fraction of vials nucleated after a given time, compared to spontaneous nucleation in a broad range of solute/solvent systems with comparable experimental parameters reported by Garetz \etal\cite{10.1103/PhysRevLett.77.3475} (one or more unfocused laser pulses of $\sim$\SI{}{ns} duration and \SI{532/1064}{nm} wavelength, see \Cref{table_NPLIN} for a detailed overview). Moreover, NPLIN has also been reported to offer control over the polymorphic form nucleating from solution\cite{10.1021/cg0055171,10.1021/cg800028v}.  Thus, the  observations of locally enhanced nucleation probability at the laser irradiation position, and the potential to control the polymorphic form, point out that NPLIN may be a promising primary nucleation control method (\Cref{NPLIN}). A solid understanding of the underlying NPLIN mechanism--a discussion yet to be settled in the literature, holds the key to fulfilling its potential as a broadly applied nucleation control method. In this section, we will summarize the common observations in NPLIN experiments reported in the literature, the proposed mechanisms, and classify the experimental setups and solutions studied. Particularly, we critically discuss to what extent the proposed mechanisms hold to explain the observations. Finally, we will highlight future directions in the summary section.\par

The experimental observations and observed trends in NPLIN experiments may shed light on the underlying mechanism. To this end, we present an extensive list of observations compiled from the literature. 
\begin{enumerate}
\item \label{abc}A broad range of compounds under NPLIN: NPLIN has been reported for a range of systems (predominantly in aqueous media), discussed in detail in section \cref{Reported solutions}, including small organics \cite{10.1021/cg050041c, 10.1021/cg050460+, 10.1021/cg800028v}, metal halides\cite{10.1021/cg300750c}, single component systems \cite{10.1103/PhysRevE.79.021701,10.1039/c1cp22774b}, dissolved gases\cite{10.1063/1.3582897, 10.1063/1.4917022}  and a macromolecule - lysozyme\cite{10.1021/cg800696u}.

\item \label{not_active}Not all solutions undergo NPLIN: Ward \etal\cite{10.1021/acs.cgd.6b00882}  reported that acetamide ($\textup{CH}_3\textup{CONH}_2$), an organic molecule with relatively high solubility and molecular structure similar to urea (CH$_4$N$_2$O), does not exhibit NPLIN. In the unpublished work of Barber\cite{EleanorRoseBarber}, aqueous sodium chlorate was also reported to not undergo NPLIN.

\item \label{Peak_intensity_dep}The NPLIN probability depends on laser peak intensity and supersaturation: the fraction of samples nucleated under NPLIN was reported to increase with both laser peak intensity and solution supersaturation. Along with others\cite{10.1103/PhysRevLett.77.3475}, Kacker \etal \cite{10.1021/acs.cgd.7b01277} reported that NPLIN nucleation probability is both supersaturation and laser peak intensity dependent for aqueous KCl solutions.

\item \label{pulsewidth} Laser pulse duration matters: for similar peak intensities ($j_\textup{peak}$ $\approx$ \SI{30}{MWcm^{-2}} per pulse), aqueous solutions of $\tu{CO}_2$, KCl, $\tu{NH}_4\tu{Cl}$ and CH$_4$N$_2$O exposed to unfocused femtosecond-laser pulses ( $\approx$ \SI{110}{fs}) did not nucleate while exposure to nanosecond ($\approx$ \SI{5}{ns}) pulses triggered nucleation\cite{10.1021/acs.cgd.6b00882}. Although both femtosecond and nanosecond pulses had the same peak intensity, the total energy per pulse (\SI{}{Jcm^{-2}}) is 5 orders of magnitude higher with the nanosecond pulse due to its longer pulse duration.

\item \label{wavelength_dep} Laser wavelength dependence: Kacker \etal \cite{10.1021/acs.cgd.7b01277} reported that nucleation probability of supersaturated aqueous KCl exposed to a single pulse of 355, 532 and \SI{1064} {nm} is not strongly dependent on the laser wavelengths. Yet, shorter wavelengths, namely 355/532 nm, led to slightly higher nucleation probability for KCl\cite{10.1021/acs.cgd.7b01277, 10.1021/cg300750c}, KBr\cite{10.1021/cg300750c} and urea\cite{10.1021/cg050041c}. \Cref{table_NPLIN} shows that almost exclusively the frequency multiples of the primary wavelength (\SI{1064}{nm}) of Q-switched Nd:YAG lasers are used in the NPLIN literature.

\item \label{dep_nc_peakI} Dependence of the number of crystals/bubbles on laser peak intensity: with an increase in the laser peak intensity, a linear increase in the number of crystals for KCl\cite{10.1021/ja905232m, 10.1021/acs.cgd.9b00362} and glycine\cite{10.1021/acs.cgd.0c00669}, and a quadratic increase in the number of $\tu{CO}_2$ bubbles\cite{10.1063/1.4917022} was observed.

\item \label{switching} Polarization switching: laser polarization is reported to influence  the polymorphic form of several simple organic molecules such as glycine\cite{10.1021/cg050460+}, L-histidine\cite{10.1021/cg800028v}, carbamazepine\cite{10.1021/cg500163c} and sulfathiazole\cite{,10.1021/acs.cgd.5b01526}. However, this observation could not be reproduced for glycine by Liu \etal\cite{10.1039/c7cp03146g}, nor later by Irimia \etal\cite{10.1021/acs.cgd.0c01415}, indicating a subtle effect in the experimental conditions is at play.

\item \label{threshold}Laser intensity threshold: several authors report a threshold laser intensity below which laser irradiation does not trigger nucleation\cite{10.1103/PhysRevLett.77.3475,10.1021/cg050041c,10.1039/c7cp03146g}. Moreover, this laser intensity threshold is observed to be dependent on the solute, the wavelength of the laser light, and the temperature\cite{10.1021/cg050041c, 10.1021/cg300750c}. Between solutes, small organics\cite{10.1021/cg050041c,10.1021/cg050460+} such as urea and glycine are observed to have a higher laser peak-intensity threshold ($>$\SI{50}{MW. cm^{-2}}) compared to metal halides\cite{10.1021/cg300750c} ($>$\SI{3}{MW. cm^{-2}}) such as KCl and KBr.

\item \label{aging} Dependence on solution aging: several authors\cite{10.1021/acs.cgd.0c00669, 10.1021/cg0055171} report that aging of glycine aqueous solutions improved the nucleation probability under NPLIN. However, the nucleation probability of metal halides was found to be invariant to aging\cite{10.1021/cg8007415}.

\item \label{filtration} Effect of filtration and nanoparticle doping: Ward \etal\cite{10.1021/acs.cgd.6b00882} studied how the filtration and intentional addition of impurities, namely $\tu{Fe}_3\tu{O}_4$ nanoparticles, alter NPLIN probability. Filtration decreased the NPLIN probability while the addition of nanoparticles increased the NPLIN probability reported at a fixed observation time.

\item \label{alignment} Product crystal alignment: in the experiments performed in aqueous urea by Garetz \etal\cite{10.1103/PhysRevLett.77.3475}, the direction of the needled-shaped crystals of urea were reported to be aligned with the polarization plane of the laser. However, Liu \etal\cite{10.1039/c6cp07997k}, in their experiments using aqueous urea, observed the angle between crystal alignment and laser polarization to be random.

\item \label{pathway} Irradiation pathway matters: when compared to laser intensity threshold values reported in the literature, Clair \etal \cite{10.1107/S160057671401098X} observed a lower value in experiments when passing the laser light via an air-liquid interface (from the vial top) through a supersaturated aqueous glycine solution. Unfortunately, in the same experiments, no trials were performed to pass the laser through the glass-liquid interface for comparison.

\item \label{contact} Direct solution-laser interaction matters:  Kacker \etal\cite{10.1021/acs.cgd.7b01277}, in their experiments with aqueous KCl, measured the pressure signal after a laser pulse at a fixed distance from the laser path within the vial. Even though the samples where laser light was masked with a black tape recorded higher radiation pressure compared to the samples which allowed the laser to pass through the solution, the former was not observed to undergo NPLIN.

\end{enumerate}

 


\subsection{Proposed mechanisms}
\label{Proposed mechanisms}

In this section, we critically discuss the proposed NPLIN mechanisms and to what extent these mechanisms explain the experimental observation listed above.   \Cref{Optical Kerr effect,Dielectric polarization} discuss molecular scale mechanisms based on how the field induces polarizability of metastable pre-nucleating clusters are considered to drastically reduce the induction time for nucleation\cite{10.1021/acs.cgd.6b00046,10.1021/acs.cgd.0c01415}. On the other hand, the proposed mechanism in \Cref{Nanoparticle heating} explains NPLIN as a result of laser heating of impurities present in the solution, a mechanism bridging the molecular and macroscopic scale. 



\begin{figure}[tbp]
	\centering
	\includegraphics[width=0.45\textwidth]{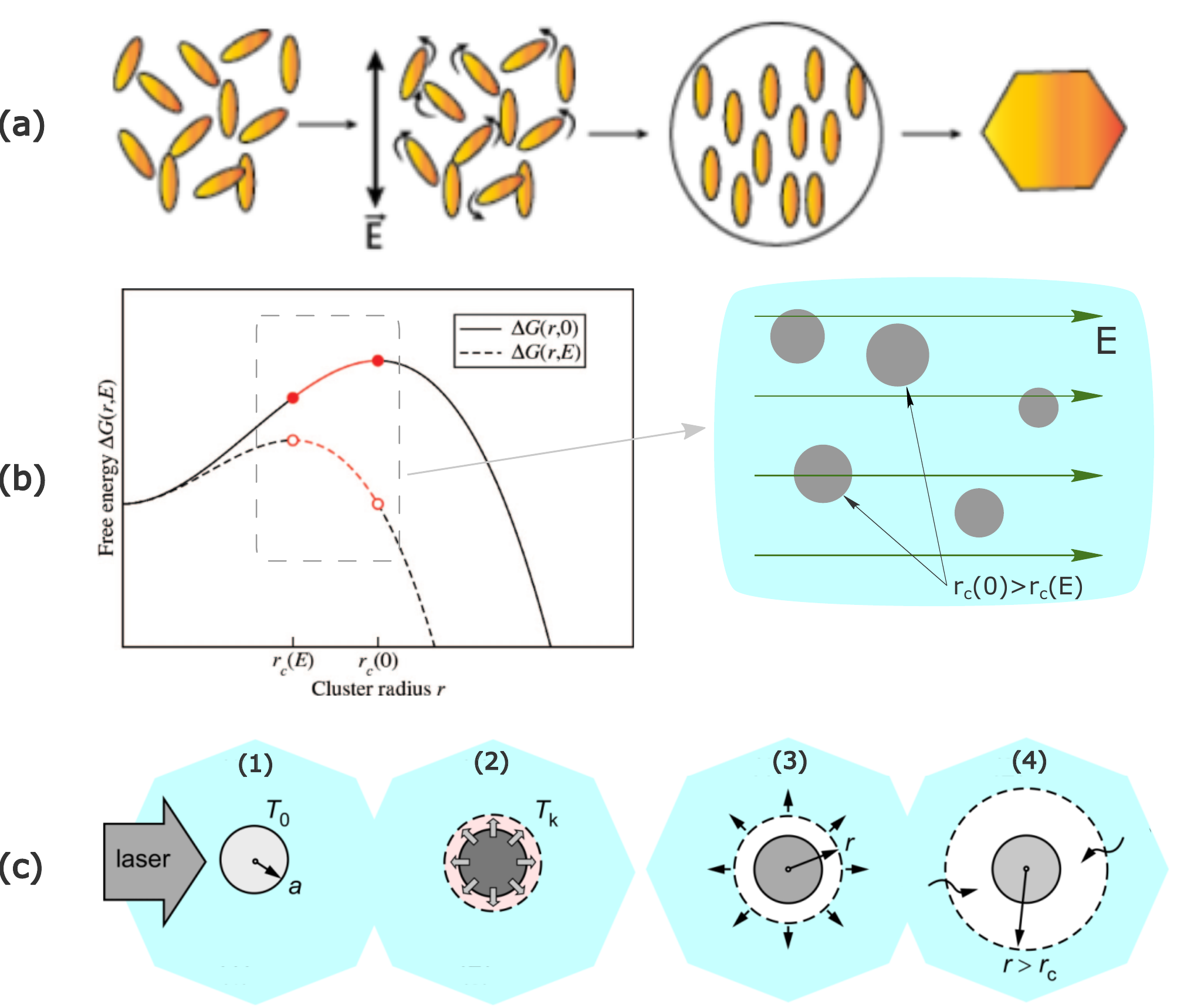}
\caption{\textbf{Plausible mechanisms for NPLIN.} (\textbf{a}) Field-induced alignment of molecules - optical Kerr effect.  (\textbf{b}) Stabilization of otherwise subcritical clusters under electric field\cite{10.1063/1.4917022} - dielectric polarization (where $r_c(0)$ and $r_c(E)$ are the critical cluster radius in the absence and presence of laser light, respectively).  (\textbf{c}) Evaporation of solvent surrounding a nanoparticle due to local heating.}
\label{Mechanisms_NPLIN}
\end{figure}

\subsubsection{Optical Kerr effect (OKE)}
\label{Optical Kerr effect}
The first hypothesized mechanism was based on the optical Kerr effect. This hypothesis states that the laser produces a weak torque that aligns all anisotropically polarizable molecules (or clusters of molecules) with their most polarizable axis parallel to the direction of polarization of the incident light (\Cref{Mechanisms_NPLIN}a). For instance, the observed alignment of urea crystals to the laser direction by Garetz \etal\cite{10.1103/PhysRevLett.77.3475} was argued based on the urea molecule's ability to align their $\textup{C}_2$ axes parallel to an applied laser's electric field. Consequently, it was proposed that the electric-field-induced alignment reduces the free energy barrier for nucleation. However, the permanent dipole moment of a molecule does not contribute to the Kerr effect since the rotational timescales of solute molecules\cite{10.1021/acs.jpcb.9b01904} are much larger than the duration of the change in the electric field of the laser ($\sim 10^{-14}\,$ s). \par

The interaction energy induced by the optical Kerr effect on a molecule is given by $-\Delta\alpha\ E^2/2$, where $\Delta\alpha$ is the polarizability anisotropy and $E$ the local electric field\cite{10.1103/PhysRevLett.89.175501}. The degree of alignment of the molecular axes along the $x-$, $y-$, and $z-$ direction can be quantified by the order parameters $K_x$, $K_y$ and $K_z$ respectively, which for relatively weak light-solute interaction is defined by\cite{10.1021/cg050460+}
{\scriptsize \begin{equation}
\label{order parameters}
\begin{gathered}
K_x = \frac{1}{3} + \frac{\Delta\alpha E^2}{90k_BT}\left(\frac{2-e^2}{1+e^2}\right),\hspace{0.1cm} K_y = \frac{1}{3} + \frac{\Delta\alpha E^2}{90k_BT}\left(\frac{2e^2-1}{1+e^2}\right),\\\hspace{0.1cm} K_z = \frac{1}{3} - \frac{\Delta\alpha E^2}{90kT},
\end{gathered} 
\end{equation}}

\noindent where $k_B$ is Boltzmann's constant and $e$ is the elliptical polarization of light, with the order parameters satisfying $K_x+K_y+K_z=1$. The order parameters are $1/3$ each for an isotropic distribution in the absence of an electric field. For linearly polarized light ($e=0$) along $x$, in the limit of zero temperature or infinite interaction energy, we obtain perfect alignment of rod-like building blocks in the $x$-direction, with $K_x=1$ and $K_y=K_z=0$. Similarly, we have another uniaxial case for circularly polarized light ($e=1$) with the electric field rotating in the $xy$-plane and the symmetry axis along the $z$-direction favoring disk-like arrangement with $K_x=K_y=1/2$.
\par

For supersaturated solutions of urea, Matic \etal\cite{10.1021/cg050041c} reported a lower laser peak-intensity threshold and a higher nucleation probability for linearly polarized light over circularly polarized light.
This recorded difference between laser light polarization is consistent with the optical Kerr mechanism since urea was observed to exhibit rod-like arrangement in the experiments conducted by Garetz \etal\cite{10.1103/PhysRevLett.77.3475}. The effect of the laser's polarization on the polymorph of the product crystal has also been observed for several other simple organic molecules such as glycine, L-histidine, carbamazepine, and sulfathiazole \cite{10.1021/cg050460+,10.1021/cg800028v,10.1021/cg500163c,10.1021/acs.cgd.5b01526} (\Cref{Small organic molecules}). This observed product polymorph dependence on laser polarization further suggests that the mechanism of NPLIN is not photochemical. \par

To have a quantitative estimate of the reported glycine dependence on laser polarization\cite{10.1021/cg050460+}, using the known values for glycine's polarizability, $\Delta\alpha = \SI{2e-40}{F.m}$, the interaction energy per molecule is calculated to be $\Delta\alpha E^2/kT = 4 \times 10^{-5}$. This value for interaction energy is much less than unity. Therefore, it cannot account for the large order parameters (from \cref{order parameters}) necessary for the observed nucleation rates. When compared to a single molecule, the cooperative effects among groups of glycine molecules within a pre-nucleating cluster could have enhanced polarizabilities in the stacking direction. However, Monte Carlo simulations using Potts lattice gas
model by Knott \etal\cite{10.1063/1.3574010} suggest that this mechanism, even if bound together by intermolecular forces in a very large cluster, seem too weak to explain the observed nucleation rate enhancements\cite{10.1063/1.5079328} (see \Cref{sec:MD}).\par

Although the Kerr effect hypothesis supports the observed correlation between the urea crystal and laser polarization by Garetz \etal, Liu \etal\cite{10.1039/c6cp07997k} reported the crystal orientation angle to be quite random under similar experimental conditions (observation \ref{alignment}). Sun \etal\cite{10.1021/cg050460+} in their experiments with glycine, observed a narrow window of temperature and supersaturation within which the circularly polarized light favored $\alpha$-glycine while linearly polarized light favored $\gamma$-glycine (observation \ref{switching}). This reported influence of laser polarization on glycine polymorphism by Sun \etal contradicts the results from Irimia \etal\cite{10.1021/acs.cgd.0c01415}. When using a single laser pulse in aqueous glycine solutions, Irimia \etal did not observe any effect of laser polarization on polymorphism. On comparing the results of Irimia \etal with others who employed hundreds of laser pulses, one might suggest that the interaction of laser light with  microscopic crystals after nucleation can trigger polymorphic transitions through polarization-dependent ablation and secondary nucleation. Interestingly, the experiments of Irimia \etal, when employing multiple pulses of \SI{1064}{nm}, showed an increase in the solution temperature. However, the effect of temperature rise on polymorph control is yet to be quantified. In addition, the Kerr effect hypothesis fails to explain the reported laser peak-intensity threshold and the weak wavelength dependence on the nucleation probability. The whole basis of the Kerr effect lies in the ability of laser light to polarize a solute molecule, yet the NPLIN of solutes without anisotropic polarizability, such as metal halides, lacks explanation. Thus below we present the dielectric polarization hypothesis that attempts to explain the observed NPLIN of potassium halides such as KCl\cite{10.1021/cg8007415} and KBr\cite{10.1021/cg300750c}.

\subsubsection{Dielectric polarization (DP)}
\label{Dielectric polarization}
The dielectric polarization mechanism suggests that isotropic polarization of pre-nucleating clusters by an electric field modifies the cluster's free energy by which it becomes stable. This means that a dielectrically homogeneous cluster larger than a critical size, $r_\mathrm{c}$, is stabilized by an electric field when its dielectric constant exceeds that of the surrounding medium (\Cref{Mechanisms_NPLIN}b). Unlike OKE which works on induced polarization of solutes under laser light, DP stems from differences in the dielectric permittivity of solutes compared to solvents. Including this effect in classical nucleation theory (CNT), the free energy of a cluster of radius $r$ in the presence of an electric field $E$ is given by\cite{10.1021/cg8007415}
{\footnotesize \begin{gather}
\label{CNTeqn}
\Delta G (r,E) = 4\pi r^2 \gamma - \frac{4}{3}\pi r^3 (A \ln{S}+aE^2),
\end{gather}}

\noindent where $\gamma$ is the solution-crystal interfacial tension, $A=\rho RT/M$, in which $\rho$ is the mass density, $R$ is the gas constant, $M$ is the molar mass of the solid and $S$ is the supersaturation ratio. The coefficient $a$ defines an effective dielectric constant
{\footnotesize \begin{gather}
a = \frac{3\epsilon_0\epsilon_s}{2}\left(\frac{\epsilon_p-\epsilon_s}{\epsilon_p+2\epsilon_s}\right).
\end{gather}}

For a particle with dielectric constant $\epsilon_p$ immersed in a medium of dielectric constant $\epsilon_s$, the free energy is lowered in the presence of an electric field provided that $\epsilon_p$ is greater than $\epsilon_s$ - a critical criterion for DP to work. Assuming a Poisson distribution, the probability of obtaining at least one nucleus is calculated using
{\footnotesize \begin{gather}
\label{linearprob}
p(n\geq1) = 1 - e^{-m\, j_{peak}}
\end{gather}}

\noindent where $m\,j_{peak}$ is equal to the mean number of nuclei produced by a given laser peak-intensity $j_{peak}$ and $m$ is the lability. For lower peak intensities, using a truncated Taylor series for the exponential term in the above equation, a linear relation between probability and $j_{peak}$ can be achieved\cite{10.1021/cg8007415}. From CNT, the probability of a cluster of size $r$ within the solution is expressed using the Boltzmann distribution, $e^{{-\Delta G (r,E)}/{(k_BT)}}$. In the presence of laser light, under the conditions where the change in a cluster's bulk energy due to the light's electric field is significantly small ($aE^2 \ll A \ln{S}$), we can analytically calculate the lability as\cite{10.1021/cg300750c}
{\footnotesize \begin{gather}
m = \frac{3N_\textup{molecules}\gamma a}{2\pi \rho^3(k_BT\ln S)^2} \times \frac{e^{-\Delta G(r_c,0)/(k_BT)}}{\int_0^{r_c(0)} r^3 e^{-\Delta G(r,0)/(k_BT)} dr}
\end{gather}} \par
\noindent where $r_c(0)$ is the critical radius in the absence of an electric field. $N_\textup{molecules}$ is the number of ion pairs within the volume illuminated by the laser, indicating an increase in the nucleation probability with an increase in the irradiated volume.\par
 
The dielectric polarization model successfully predicts the linear relation of the nucleation probability to low laser peak intensity for KCl\cite{10.1021/cg300750c} (\Cref{linearprob}). By doing so, it also hypothesizes a mechanism under which ionic solutes such as KCl and KBr\cite{10.1021/cg300750c}, that have no preferred orientation under laser, can nucleate under NPLIN. Yet, it cannot explain the experimentally observed intensity threshold, $j_0$ (observation \ref{threshold}). Therefore, phenomenological models use a corrected value for the number of nuclei produced, $m(j_{peak}-j_0)$, to replicate the observed zero probabilities below $j_0$. Together with the laser peak-intensity threshold, the dielectric polarization model fails to answer the observed probability dependence on laser pulse duration, wavelength, and the polymorph selectivity under different laser polarizations\cite{10.1103/PhysRevLett.89.175501,10.1021/cg050460+,10.1021/cg800028v} (observations \ref{pulsewidth}, \ref{wavelength_dep}, and \ref{switching}). Moreover, NPLIN of dissolved gases shown in \Cref{NPLIN}c in which the dissolved gas phase has a lower dielectric constant than water cannot be explained by the dielectric polarization hypothesis (observation \ref{dep_nc_peakI}). To explain the observed NPLIN of dissolved gases and the effect of impurities on NPLIN probabilities of NH$_4$Cl\cite{10.1021/acs.cgd.6b00882}, we present below the impurity heating hypothesis, which attempts to explain NPLIN as a function of inherent impurities rather than the solute-laser interaction. \par

\subsubsection{Impurity heating (IH)}
\label{Nanoparticle heating}
The impurity heating hypothesis suggests that the interaction of the laser irradiation with impurities plays a significant role in NPLIN. This hypothesis emerged from the inability of the OKE and DP mechanisms to explain certain observations common in NPLIN experiments, particularly the existence of a threshold below which no nucleation is observed and the pronounced effect of filtration on NPLIN (observation \ref{filtration}). In a nutshell, this hypothesis assumes a scenario where insoluble impurities such as nanoparticles absorb laser energy at the wavelength of irradiation and rapidly heat and evaporate the surrounding solution. This phenomenon is expected to trigger nucleation by locally enhancing the supersaturation. In order to test this, Ward \etal\cite{10.1021/acs.cgd.6b00882} studied how the intentional addition of impurities, namely $\tu{Fe}_3\tu{O}_4$ nanoparticles and a surfactant (polyethylene glycol, $M_\mathrm{r} = 8000$), alter the NPLIN probability and number of crystals nucleating in supersaturated aqueous $\tu{NH}_4\tu{Cl}$ solutions. First, they compared filtered and non-filtered aqueous $\tu{NH}_4\tu{Cl}$  solutions. The filtration was carried out using a \SI{0.2}{\micro\metre} pore-size membrane with freshly prepared samples at high temperatures to justify that only the impurities were filtered out as opposed to the solute clusters. A stark difference in nucleation probability and the number of crystals was observed between filtered and unfiltered samples. The filtered samples showed a lower nucleation probability and a lower number of crystals. In the same work, a similar effect due to filtration was observed in other systems, such as in aqueous urea and glycine. Furthermore, supporting the role of impurities, the addition of both nanoparticles and surfactant showed an increase in the NPLIN probability. While the impurities due to nanoparticles and surfactant would serve as active sites for the local increase in supersaturation, the surfactant was also expected to stabilize the dispersion of impurities - promoting more viable nucleation sites. \par

Javid \etal\cite{10.1021/acs.cgd.6b00046} performed NPLIN experiments with glycine for both filtered and unfiltered samples. Irrespective of whether the solutions were irradiated or not, filtration of glycine solutions across all supersaturations resulted only in an $\alpha$-polymorph under the influence of the laser. The unfiltered samples at higher supersaturations (1.5 and 1.6) showed a significant presence of a $\gamma$-polymorph (40\%) when irradiated, while non-irradiated solutions nucleated almost exclusively the $\alpha$-polymorph at all supersaturations. \par

Ward \etal\cite{10.1021/acs.cgd.6b00882} reported that for systems with $\tu{CO}_2$, $\tu{KCl}$, $\tu{NH}_4\tu{Cl}$ and CH$_4$N$_2$O, NPLIN was not observed using unfocused femtosecond laser pulses (\SI{\sim 110}{fs}, $j_{peak}$ = \SI{30}{MW.cm^{-2}}), while nanosecond pulses (\SI{\sim 5}{ns}, $j_{peak}$ = \SI{12}{MW.cm^{-2}}) induced nucleation. The total energy per pulse, $\sim j_{peak} \times \textup{pulse duration}$, limits the energy available for a nanoparticle to absorb. This absorbed energy is hypothesized to evaporate the solvent surrounding the nanoparticle and form a local vapor-filled cavity (\Cref{Mechanisms_NPLIN}c). Consequently, a region of high solute concentration at the vapor-liquid interface is expected to emerge, due to the solvent that evaporated. This increased solute concentration at the vapor-liquid interface is expected to contribute to a higher local supersaturation and therefore trigger nucleation. The observed differences in nucleation probabilities between the aforementioned pulse durations were argued based on the energy available for the local cavity formation surrounding the nanoparticle (observation \ref{pulsewidth}). This supports the hypothesis that heating solid nanoparticle impurities, which are intrinsically present within a solution, act as a precursor to nucleation. \par

The nanoparticle heating hypothesis explains the observed laser intensity threshold because enough energy must be supplied to induce cavitation around a nanoparticle. The nanoparticle heating hypothesis however fails to explain two sets of reported experimental results, namely the alignment of urea crystals\cite{10.1103/PhysRevLett.77.3475} and the influence of laser polarization on polymorphic form\cite{10.1021/cg050460+} (observation \ref{switching}). Interestingly, Liu \etal\cite{10.1039/c6cp07997k} failed to reproduce this alignment effect in aqueous urea upon exposure to linearly polarized nanosecond pulses. One possible explanation for this observed alignment of crystals might be due to hydrodynamic interactions between the crystal and the surrounding fluid. The Marangoni flow induced by local heating of the solution could apply torque and align the crystals. At the current time, this explanation is merely speculation without quantification of the flow fields and the local sample heating under studied experimental conditions. Irimia \etal\cite{10.1021/acs.cgd.0c01415} quantified the temperature increase of aqueous glycine solutions when subjected to one or more pulses of 532 and 1064 \SI{}{nm}. Both Alexander \etal and Irimia \etal observed that laser polarization does not influence the polymorph formed. A possible explanation for the difference in the observations made could rely on the nature of the impurity rather than the solute. Moreover, the ability of nanoparticles to have a difference in absorption, based on the ellipticity of laser polarization (circular or linear), is left unexplored. This difference in laser absorption could dictate the magnitude of the local supersaturation and thus the polymorph formed.\par

Alternatively, a new mechanism inspired by dielectric polarization can be formulated by considering that the crystal grows on the surface of an inherent nanoparticle - which acts as a favorable site over which solute clustering is favorable. If we consider CNT (\Cref{CNTeqn}), the nanoparticle could essentially increase the magnitude of the electrostatic term ($a$) and modify the free energy profile via the interfacial term ($\gamma$). However, the composition of these nanoimpurities and their effect on NPLIN probability are yet to be determined.\par



\subsection{Experimental setups}
\label{Experimental setups - NPLIN}
\begin{figure*}[tbp]
	\centering
	\includegraphics[width=1\textwidth]{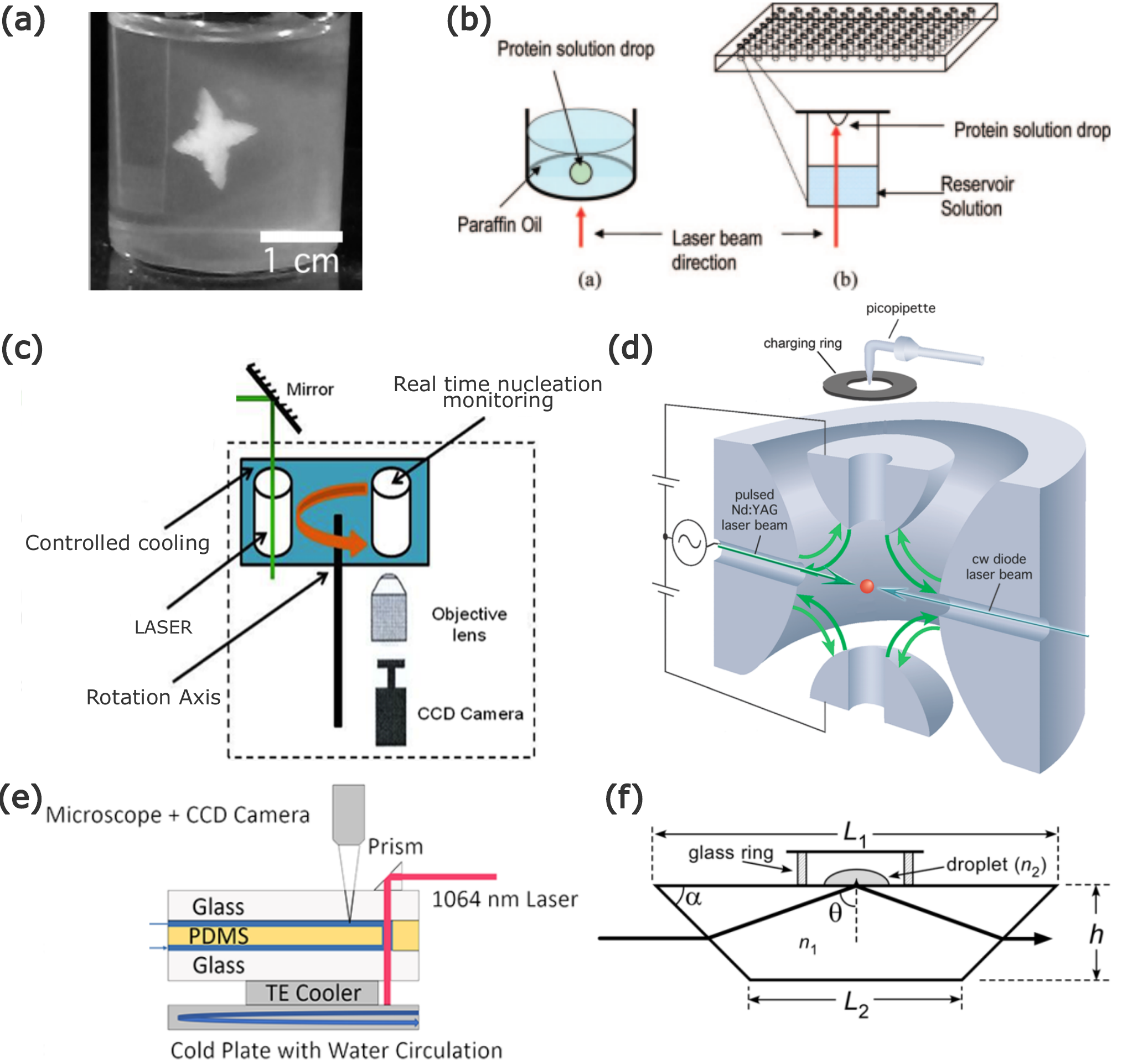}
	\caption{\textbf{Experimental setups for NPLIN.} (\textbf{a}) Gel medium: dendritic growth of glycine using supersaturated aqueous solutions in agarose gel\cite{10.1021/acs.cgd.8b00688}. (\textbf{b}) Micro-batch/hanging droplet: crystallization of hen egg-white lysozyme and Bovine pancreatic trypsin using microbatch and hanging droplet methods, respectively\cite{10.1021/cg800696u}. (\textbf{c}) Carousel glass vials: high-throughput controlled crystallization using a carousel\cite{10.1107/S160057671401098X}. (\textbf{d}) Levitated microdroplet: crystallization in a levitating microdroplet supersaturated with KCl\cite{10.1021/cg5004319}. (\textbf{e}) Continuous microfluidic setup: continuous production of KCl crystals using microfluidic setup\cite{10.1021/acs.cgd.9b00362}. (\textbf{f}) Using evanescent wave: crystallization of KCl in a microdroplet using evanescent wave\cite{10.1021/acs.cgd.5b00854}.}
	\label{ExpSetups_NPLIN}
\end{figure*}

Classical NPLIN experiments were performed using milliliter-sized cylindrical glass vials by manually replacing them in the path of the laser beam (\Cref{table_NPLIN}). The laser light always passes through the side walls of the container. These vials are then transferred to a temperature-controlled bath (if required) after exposure. Below are some innovative ways adopted by different authors to answer specific questions relating to proposed NPLIN mechanisms and to address the shortcomings of previously proposed experimental techniques.\par

\noindent \textbf{Gel medium}: In an experimental setup leveraging a gel medium, a canonical NPLIN setup with cylindrical glass vials filled with a mixture of agarose gel and supersaturated aqueous solutions was used (\Cref{ExpSetups_NPLIN}a). Gels are known to prevent the convection-enhanced rehomogenization of solutions. As a result, the formed crystals stay anchored to their point of origin. For KCl, Duffus \etal\cite{10.1021/ja905232m} used two low-intensity beams with a total energy equivalent to the threshold laser intensity. Within the gel medium, nucleation was only induced where the two beams crossed, thus opening the route to three-dimensional control of nucleation. Tasni \etal\cite{10.1021/acs.cgd.8b00688} reported that a laser, when passed through the air-gel interface, resulted in a much higher nucleation probability than through the glass-gel interface for glycine solutions. Also, the air-to-gel laser path always gave rise to tree-branch dendrites composed of pure $\alpha$-glycine. In contrast, when the laser was passed through the glass-gel interface, a stellar dendrite was formed in the solution bulk with $\gamma$-glycine concentrated in the core of the dendrite and $\alpha$-glycine dominating the exterior.\par

\noindent \textbf{Micro-batch/Hanging droplet}: For nucleation studies with difficult-to-synthesize or costly solutions, the use of smaller amounts is crucial. For such solutions, a conventional microbatch or hanging drop system is adapted for NPLIN studies (\Cref{ExpSetups_NPLIN}b). The droplet containing the supersaturated solution is irradiated using the laser to monitor crystallization\cite{10.1021/cg800696u}. The ease of preparing and handling larger sample numbers also allows one to have better statistics on nucleation probability and crystal quality.\par

\noindent \textbf{Carousel glass vials}: A carousel setup is an automated setup using the canonical NPLIN vials with additional control over container temperature and laser intensity (\Cref{ExpSetups_NPLIN}c). Developed by Clair and Bir\'e \etal\cite{10.1107/S160057671401098X}, it has a high throughput with in-situ monitoring of the vials. The continuous monitoring of the laser intensity and vials, together with the thermostated water bath for vials, reduce the uncertainty regarding the induction of nucleation. Just like the levitated microdroplet setup, the laser beam passes through the air-solution interface. Using this setup, Clair and Bir\'e \etal\cite{10.1107/S160057671401098X} reported that  glycine nucleated at the meniscus interface before falling down to the bottom of the container.\par

\noindent \textbf{Levitated microdroplet}: To study the volume dependence in the absence of container walls in NPLIN, Fang \etal\cite{10.1021/cg5004319} performed experiments using levitated micrometer-sized droplets containing supersaturated aqueous KCl (\Cref{ExpSetups_NPLIN}d). For the same peak-laser intensities as employed by Alexander \etal\cite{10.1021/cg8007415} in bulk solutions, NPLIN within droplets was reported only at dramatically higher supersaturation. This observation was attributed to the possible low number of nucleation events owing to the relatively low exposed volume.\par

\noindent \textbf{Continuous microfluidic setup}: Hua \etal\cite{10.1021/acs.cgd.9b00362} developed a microfluidic device operating under laminar flow conditions to have more statistics on crystal size, shape, growth, and polydispersity (\Cref{ExpSetups_NPLIN}e). For KCl, the mean crystal size and polydispersity were observed to be independent of laser intensity, while higher supersaturation resulted in a higher mean crystal size. Crystals formed were cubic or cuboid in shape, with a unimodal distribution in crystal size - evidencing the absence of secondary nucleation. For glycine\cite{10.1021/acs.cgd.0c00669}, the morphology of the crystals was found to switch from prism-like to plate-like with increasing supersaturation. NPLIN was observed only for glycine solutions which were aged for 24 hours, supporting the existence of pre-nucleating clusters.\par 


\noindent \textbf{Using evanescent waves}: A supersaturated liquid droplet placed over a glass enclosing a laser beam was used in this setup (\Cref{ExpSetups_NPLIN}f). For a laser beam undergoing total internal reflection within the glass, a medium present near the interface can absorb the electromagnetic field and direct it perpendicular to the interface. This directed wave is evanescent, meaning that within the medium its amplitude decays exponentially with distance from the interface. Using the short penetration depth characteristic of an evanescent wave, Ward \etal\cite{10.1021/acs.cgd.5b00854} performed NPLIN only in regions close to the glass-solution interface. Hydrophobization of the interface was found to suppress nucleation at the glass surface. This technique thus allows for localization of the nuclei production in two dimensions parallel to the surface. The threshold laser intensity determined for NPLIN was three factors higher than that of samples in glass vials. However, the increase in the nucleation probability with laser intensity by an evanescent wave was similar to previous works in bulk solution above a laser intensity threshold.

\subsection{Reported solutions}\label{Reported solutions}

In an attempt to understand the mechanism(s) behind NPLIN, several aqueous solutions involving different solute types: \begin{enumerate*}[label={\roman*)}] \item organic/inorganic, and \item solid/gas \end{enumerate*}, have been reported in the literature. Studies of NPLIN thus far have been carried out almost exclusively in aqueous solutions, limiting the understanding of the role of the solvent in NPLIN.

\subsubsection{Small organic molecules} \label{Small organic molecules}
So far, NPLIN has been observed for several small organic molecular solutes such as  urea, glycine, L-histidine, carbamazepine, and sulfathiazole. \par

\begin{figure}[tbp]
	\centering
	\includegraphics[width=0.45\textwidth]{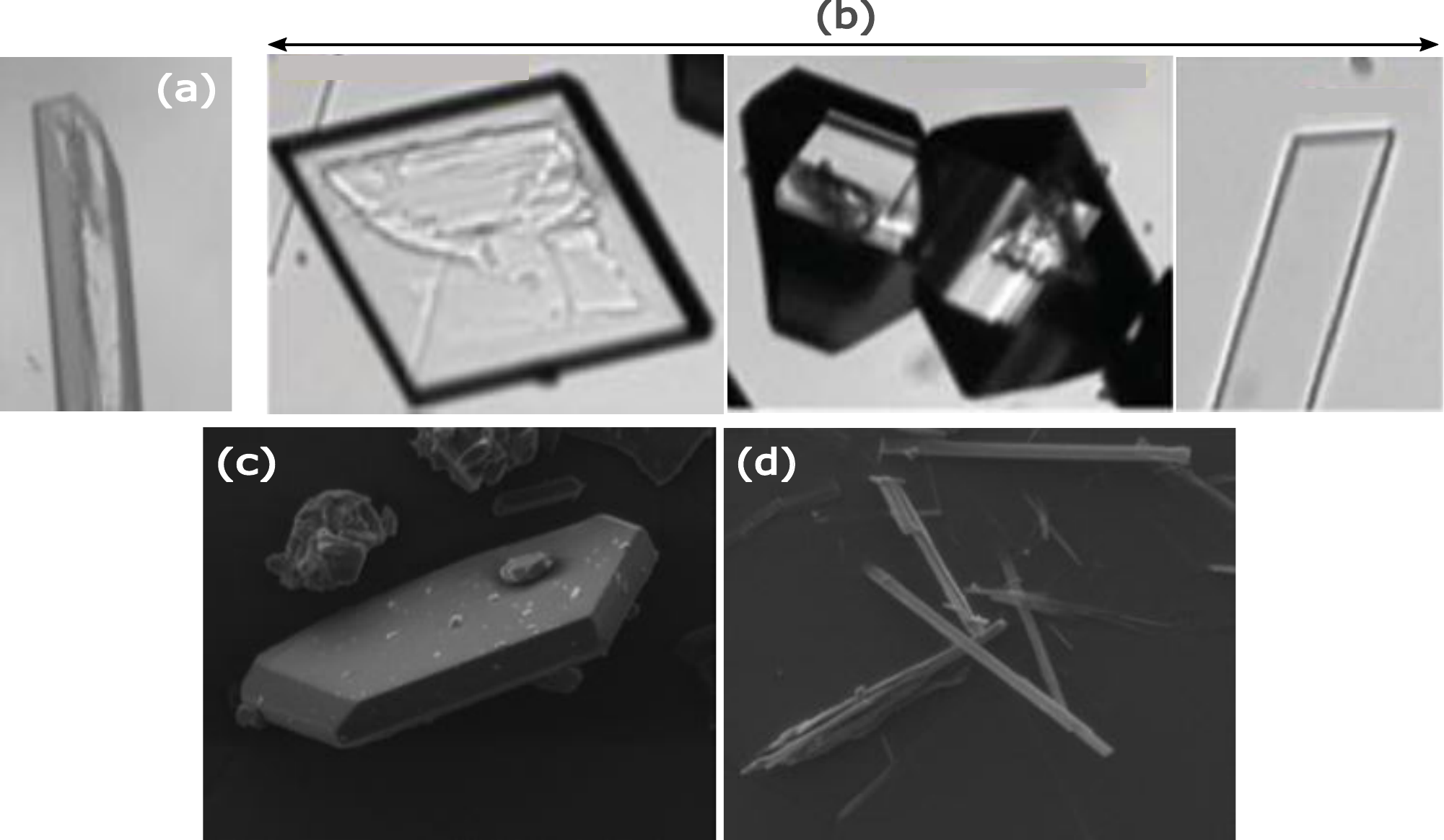}
\caption{\textbf{Morphology and polymorph control using NPLIN.} Morphologies of $\alpha$-glycine under\cite{10.1107/S160057671401098X}: (\textbf{a}) spontaneous and (\textbf{b}) NPLIN crystallization. Micrographs of carbamazepine crystals produced by NPLIN in acetonitrile\cite{10.1021/cg500163c}: (\textbf{c}) phase III and (\textbf{d}) phase I.} 
\label{polyandmorph}
\end{figure}

In the experiments performed using glycine by Zaccaro \etal \cite{10.1021/cg0055171}, it was observed that linearly polarized (LP) and circularly polarized (CP) laser light induces nucleation of different polymorphs. The effect was termed \textit{polarization switching} and was explained using the non-linear anisotropic polarizability of the pre-nucleating clusters - \ie the Kerr effect. Sun \etal\cite{10.1021/cg050460+} reported a supersaturation window for polarization switching under NPLIN of aqueous glycine. In their experiment, for light intensities of \SI{0.46}{GW.cm^{-2}} and a supersaturation window of 1.45-1.55, a small change in elliptical polarization of the incident laser light was found to induce an abrupt change in the polymorph formed. Consequently, LP and CP light was found to favor the formation of rod-like ($\gamma$-glycine) and disk-like ($\alpha$-glycine) glycine polymorphs, respectively\cite{10.1103/PhysRevLett.89.175501}. The reported supersaturation window range was observed to be considerably smaller at lower laser intensities (\SI{0.24}{GW. cm^{-2}}). In addition to favoring polymorphism, Clair \etal\cite{10.1107/S160057671401098X} found the laser to have an impact on the glycine morphology - where three distinct morphologies were obtained as opposed to the rod-like morphology obtained by spontaneous nucleation (\Cref{polyandmorph}). Results showed the influence of the CP light, with an increase in $\gamma$ nucleation above a supersaturation of $1.56$. However, this observation is counterintuitive with respect to the Kerr effect, where one might expect the opposite. In addition, a lower laser intensity threshold was recorded when compared to previously reported values from the literature. The observed differences in both the polymorph formation and laser intensity threshold could be because these experiments were performed with light passing through the air-liquid interface as depicted in the carousel setup in \Cref{Experimental setups - NPLIN}. This differs from the other conventional setups which have a laser beam enter and exit through the glass walls. \par

For L-histidine\cite{10.1021/cg800028v}, polarization switching was observed in the supersaturation range of $1.40-1.60$, where CP laser light and low supersaturation were found to favor the formation of a pure orthorhombic polymorph. On the other hand, LP laser light and high supersaturation were observed to result in a mixture of orthorhombic and monoclinic polymorphs. This emphasizes the field-induced re-arrangement of a pre-nucleating cluster as a likely mechanism for NPLIN. \par 

Ikni \etal\cite{10.1021/cg500163c} demonstrated the tendency of carbamazepine in acetonitrile to form polymorph (needle-like and prism-like) mixtures and only prism-like crystals under LP and CP light, respectively (\Cref{polyandmorph}). However, in the same experiments for carbamazepine in methanol, nucleation was reported to be polarization independent, which resulted in the formation of only the prism-like polymorph. In the experiments performed by Li \etal\cite{10.1021/acs.cgd.5b01526} on sulfathiazole in water/ethanol mixtures, LP light was observed to favor FIV polymorph formation, while CP light favored the FIII polymorph. Moreover, the number of sulfathiazole crystals was found to increase with increasing laser exposure time or supersaturation, while the mean crystal size decreased. 
\par

For experiments performed with aqueous urea\cite{10.1021/cg0055171,10.1021/cg050041c}, supporting the role of pre-nucleating clusters, the exposure of unaged solutions (solutions readily exposed to the laser after preparation) to amplified laser pulses (\SI{0.7}{GW.cm^{-2}}) did not induce any nucleation. For aged solutions, a higher nucleation probability and lower laser intensity threshold were observed for LP light when compared to CP light at both wavelengths (\SI{532/1064}{nm}). This observation is also consistent with the optical Kerr mechanism since the rod-like arrangement of urea molecules is expected to be favored by LP light over CP light. \par

Based on the observations made so far\cite{10.1021/cg050460+,10.1039/c7cp03146g}, it is evident that, for glycine, NPLIN can still operate even with a mismatch between the symmetry of glycine's polarizability and the polarization of the laser light (always resulting in $\alpha$ and $\gamma$ at low and high supersaturations, respectively). Irimia \etal\cite{10.1021/acs.cgd.0c01415}, in their recent experiments with glycine, also demonstrated the independence of the crystal polymorph formed to the polarization of the laser used when subjected to both single and multiple laser pulses (600 pulses). Interestingly, a higher percentage of $\gamma$ polymorph was observed within laser-irradiated samples when compared to crashed-cooled samples.\par

Although the NPLIN of small organic molecules could be attempted to be explained by the Kerr effect, simulations by Knott \etal \cite{10.1063/1.3574010} suggest that although the Kerr effect mechanism can lower the barrier to nucleation, the solute–solute interactions are simply too strong to allow significant alignment at the electric field strengths employed in NPLIN experiments. In addition, when experiments were repeated for urea\cite{10.1039/c6cp07997k}, the correlation between the direction of the laser and crystal formed appeared to be quite random at both laser wavelengths (\SI{532/1064}{nm}). In an attempt to study the directional influence of laser light together with mechanical shock and sonocrystallization on crystal formation, Liu \etal\cite{10.1039/c7cp03146g} also observed an increasing propensity to form $\gamma$-glycine at higher supersaturations. However, the results did not reproduce the binary polarization switching of glycine under NPLIN as reported by Sun \etal\cite{10.1021/cg050460+}. At the very least, the switch in preference from $\alpha$-glycine to $\gamma$-glycine was observed to happen over a smaller range of supersaturation for NPLIN compared to the mechanical shock and sonocrystallization techniques.

\subsubsection{Metal halides} \label{Metal halides}
In contrast to small organic molecules, in NPLIN of halide salts, there is no preferred polarization axis, as needed for the Kerr effect. To further support the claim on polarization independence, in experiments performed using KCl, no effect of laser polarization over the nucleation probability was observed\cite{10.1021/cg8007415}. In experiments using KCl, Alexander \etal\cite{10.1021/cg8007415} reported a significantly lower laser intensity threshold than was observed by Garetz \etal\cite{10.1103/PhysRevLett.77.3475,10.1021/cg050041c} for urea. In the same experiment, the solution aging was found irrelevant and it was possible to nucleate a single crystal of KCl with a single laser pulse. Therefore, for metal halides, the crystallization pathway was hypothesized to be based on the dielectric polarization mechanism.
Complying with the proposed mechanism, the observed intensity dependence for KCl was linear at low laser intensities\cite{10.1021/cg8007415} ($<$\,\SI{40}{MWcm^{-2}}). \par

The measure of how susceptible a solution is to NPLIN is called lability. Ward \etal\cite{10.1021/cg300750c} reported that a single pulse of \SI{532}{nm} laser light had a higher nucleation probability than \SI{1064}{nm} light. In addition, KBr samples yielded more crystals than KCl samples, with \SI{532}{nm} exhibiting relatively lower intensity thresholds for both salts. The observed reduction in the efficiency for \SI{1064}{nm} light was attributed to the higher absorption of near-infrared light by water. The heat generated within the sample due to light absorption is expected to increase the solubility, thus reducing the local supersaturation. In addition, for experiments performed using the same supersaturation, samples maintained at \SI{33}{\celsius} were significantly more labile than those at \SI{23}{\celsius} owing to a higher concentration. Experiments by Hua \etal\cite{10.1021/acs.cgd.9b00362} with KCl using \SI{1064}{nm} laser light, showed the nucleation probability to be independent of the number of laser pulses.
\par

Liu \etal \cite{10.1021/acsomega.0c04902} recently studied NPLIN of aqueous CsCl in presence of an acidic polymer (polyepoxysuccinic acid - PESA). It was observed that the added acidic polymer highly decreased the number of nucleation sites, leading to a fewer number of crystals compared to just aqueous CsCl under laser irradiation. Moreover, with PESA, the morphology of the CsCl crystals was found to be flower-like under NPLIN, while spontaneous nucleation resulted in cuboidal-shaped crystals. The observed effect of PESA on the number of nucleation sites and morphology was argued to be based on its potential to control crystallization from nucleation to crystal growth. While PESA is expected to increase the energy barrier to nucleation due to increased interfacial tension (from CNT, \cref{CNTeqn}), thus reducing the number of nucleation sites, the presence of PESA molecules surrounding a nucleus could also lower the crystal growth rate, affecting the morphology. 

\subsubsection{Macromolecules}
Organic molecules, due to their larger meta-stable zone width, exhibit spontaneous nucleation only at relatively high supersaturated conditions. The potential of laser irradiation to avoid poly-crystals and produce high crystalline order at low supersaturation is therefore intriguing. Yennawar \etal\cite{10.1107/S1744309110023857} observed improvement in crystal size, growth speed, quality, and resolution of diffraction for various proteins due to NPLIN, with no indication of a change in crystal packing compared to non-irradiated controlled samples. 
\par

Lee \etal\cite{10.1021/cg800696u} performed NPLIN on small droplets of supersaturated hen egg-white lysozyme (HEWL) solutions using picosecond (\SI{532}{nm}) and nanosecond (\SI{532/1064}{nm}) lasers. The nucleation efficiency was reported to be higher with \SI{532}{nm} and at higher peak intensities (0.223-\SI{0.257}{GW.cm^{-2}}) and shorter pulse duration (\SI{100}{ps}). While higher laser peak intensity would maximize the electric field strength, a \SI{532}{nm} laser with a shorter pulse duration is expected to minimize the energy absorbed by a droplet. Owing to the long molecular rotational timescales - larger than picoseconds\cite{10.1038/196057a0}, at first, the nucleation is unlikely to be attributed to electric-field-induced reorganization. However, a slight change in the degree of anisotropic interaction between protein molecules is reported to have a tremendous effect on the nucleation rates\cite{10.1063/1.1514221}. 
In conjunction, laser light might restrict any other alternative conformations of side chains, thus accelerating the arrangement of protein molecules into a crystalline structure. However, in the same experiments by Lee \etal\cite{10.1021/cg800696u}, the crystal number and size were observed to reduce upon aging, probably because the system became more homogeneous through diffusion. Thus, the clustering of globular proteins 
was expected to result in increased nucleation probability.

\subsubsection{Other systems}
NPLIN of single-component systems has been tested by a few authors. The absence of solvent reduces the complexity in the phase transition, making it more attractive for experimental and theoretical study. Sun \etal\cite{10.1103/PhysRevE.79.021701} studied the non-photochemical laser-induced phase transition in supercooled 4'-$n$-pentyl-4-cyanobiphenyl (5CB) liquid crystal using linearly polarized picosecond laser pulses. Slightly below the nematic-isotropic temperature (\SI{308.4}{K}), only those liquid domains whose directors were along the polarization of the laser light and whose size were greater than a critical value were observed to nucleate. In glacial acetic acid, for low laser intensities (\SI{< 100}{MW.cm^{-2}}), Ward \etal\cite{10.1039/c1cp22774b} reported a linear relation between the fraction of samples nucleated and the employed laser intensity. \par

Knott \etal\cite{10.1063/1.3582897} performed nucleation of $\textup{CO}_2$ bubbles in carbonated water, where the threshold laser intensity was observed to decrease with increasing supersaturation and was not a strong function of solution purity or laser wavelength. Nucleation theory of solids from solutions is not appropriate for gases since the pre-nucleating cluster is surely less dense and cooperative effects between solute molecules are ruled out. In addition, Knott \etal observed that for water co-supersaturated with argon and glycine, the bubbles escaping the water induced crystal nucleation even without a laser. It should be noted that NPLIN experiments are distinct from experiments where cavitation is induced deliberately to cause crystallization, e.g., by focusing a beam of light. Ward \etal\cite{10.1063/1.4917022} reported that the number of bubbles nucleated increases quadratically with laser intensity, with a generally lower laser intensity threshold for unfiltered samples and a clear trend of decreasing lability with better filtering and cleaning. In addition, femtosecond pulses (\SI{\sim 110}{fs}) at \SI{28} {MW.cm^{-2}} or \SI{11}{GW.cm^{-2}} did not produce any bubbles. These results support the claim that the mechanism for NPLIN of $\textup{CO}_2$ is non-photochemical since the high intensity of the femtosecond pulses would be expected to favor non-linear or multiphoton ionization processes. \par

\begin{figure}[tbp]
	\centering
	\includegraphics[width=0.35\textwidth]{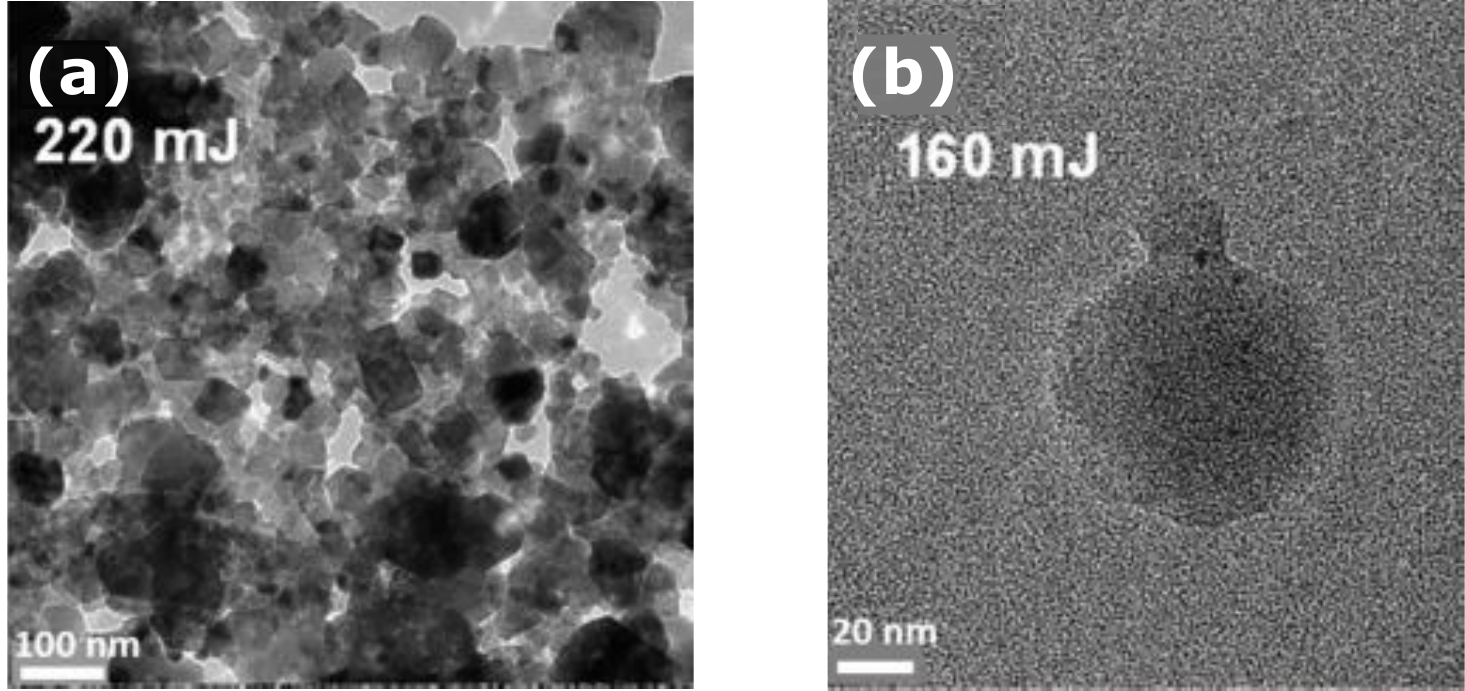}
\caption{\textbf{Nucleation of hematite nanocrystals using a single nanosecond laser pulse\cite{10.1039/c8nr03069c}.} (\textbf{a}) Hematite crystals  under \SI{220}{mJ} laser pulse. (\textbf{b}) Amorphous particle under \SI{160}{mJ} laser pulse}
\label{hematitecluster}
\end{figure}

In an effort to understand the nucleation pathway, Liu \etal\cite{10.1039/c8nr03069c} studied the formation of hematite nanocrystals from electrolyte - a photo-thermal process where crystallization was thermally activated only for a short period of time by a single laser pulse (\Cref{hematitecluster}). A mechanism based on the strength of the inter-molecular forces was used in a comprehensive nucleation theory built on two-step nucleation (TSN): first, the formation of liquid-like clusters of solute molecules, followed by the rate-limiting organization of this cluster into a protocrystal. Based on the difference between the diffusion energy barrier and the nucleation energy barrier, this pathway was reported to offer an effective route to synthesize ultrafine and coarse nanocrystals by using multiple low-intensity pulses and a single high-intensity pulse, respectively. \par

Ward \etal\cite{10.1063/1.3637946} observed that under laser-induced nucleation of molten sodium chlorate, the sample crystallized into the enantiomorph of the original sample prior to melting. Under similar experimental conditions, spontaneous nucleation did not exhibit such retention of enantiomorphism. Moreover, the molten samples made from fresh solid crystals, rather than being powdered first, were reported to be more resistant to NPLIN. During the melting process, the possibility of the formation of small particles of NaCl was attributed to helping the original sodium chlorate in evading melting, thus providing an active site to seed nucleation under NPLIN. It was also reported that no polarization effect was observed and not all samples could be nucleated by the laser at higher intensities (\SI{0.24}{GW.cm^{-2}}).


\subsection{Summary}


 Owing to the work of a large number of research groups, a considerable amount of observations has been accumulated in the literature.  \Cref{NPLIN Observation} provides a visual summary of observation and the extent to which each proposed mechanism can explain a given observation.  Despite the ever-growing list of systems exhibiting NPLIN, some basic questions remain unanswered. Can we \textit{a priori} predict whether a solution is NPLIN active or not, based on the physicochemical properties of the solute and solvent? For a given solution, what are the critical laser parameters for triggering NPLIN? As one can immediately see in \Cref{NPLIN Observation}, none of the proposed mechanisms explain all observations. In other words, more effort is required to extend our current understanding of NPLIN from both the theoretical and the experimental side. 
  
 Various solutes, including small organics (urea, glycine, L-histidine, carbamazepine, and sulfathiazole), metal halides (KCl, KBr), and gases ($\tu{CO}_2$) have been reported to undergo NPLIN (observation \ref{abc}). There is so far limited information on the list of solutions that do \emph{not} undergo NPLIN. To the best of our knowledge, only acetamide and sodium chlorate are reported to not undergo NPLIN (observation \ref{not_active}). Any contribution going beyond the proposed mechanism has to both explain observations on NPLIN active systems as well as rationalize why some systems do \emph{not} undergo NPLIN. Our ability to exploit potential advantages of NPLIN in industrial settings requires figuring out how laser-matter interaction(s), such as laser-solute interaction, laser-impurity interaction, or both, dictate the NPLIN phenomena. The answer to this question holds the key to explaining the observations mentioned in \Cref{NPLIN Phenomenology}. Below, we have made an effort to rationalize the observations so far, by classifying them into laser and solution parameters from a pragmatic application point of view.\par

\begin{table*}[tbp]
  \centering
  \caption{Observation and mechanism summary for NPLIN (inspired by Barber\cite{EleanorRoseBarber}). OKE = optical Kerr effect, DP = dielectric polarization, IH = impurity heating. Meaning of symbols: \checkmark - can be explained; x - cannot be explained; ? - needs more insight/analysis.}
    \begin{tabular}{|c|c|c|c|c|c|} \hline
    &\multicolumn{2}{|c|}{\textbf{Observation}} & \textbf{OKE} & \textbf{DP} & \textbf{IH} \\ \hline
    1 &\multicolumn{1}{|p{10.5em}|}{\multirow{3}{*}{\parbox{3.5cm}{NPLIN is reported for a broad range of systems}}} & Small organics & \checkmark     &   x    & \checkmark \\
    \cline{3-6}    &\multicolumn{1}{|c|}{} & Metal halides & x      & \checkmark     & \checkmark \\
    \cline{3-6}    &\multicolumn{1}{|c|}{} & Gases &   x    &   x    & \checkmark \\ \hline
    2 &\multicolumn{2}{|p{16em}|}{Not all solutions undergo NPLIN} &   x   &   x   & ? \\ \hline
    3 &\multicolumn{2}{|p{18em}|}{Dependence of NPLIN probability on laser peak intensity and supersaturation} & \checkmark    & \checkmark     & \checkmark \\ \hline
    4 &\multicolumn{2}{|p{16em}|}{Laser pulse duration matters} & \checkmark     &   x    & \checkmark \\ \hline
    5 &\multicolumn{2}{|p{16em}|}{Laser wavelength dependence} &    ?   &   x    &  ?\\ \hline
    6 &\multicolumn{1}{|p{10.5em}|}{\multirow{2}{*}{\parbox{3.5cm}{Product count vs laser peak intensity}}} & Crystals & \checkmark     &   \checkmark    & \checkmark \\
    \cline{3-6}    &\multicolumn{1}{|c|}{} & Bubbles & x      & x     & \checkmark \\\hline
    7 &\multicolumn{2}{|p{16em}|}{Polarization switching} & \checkmark     &   x    & ? \\ \hline
    8 &\multicolumn{2}{|p{16em}|}{Laser intensity threshold} &   x    &   x    & \checkmark \\ \hline
    9 &\multicolumn{2}{|p{16em}|}{Dependence on solution aging} & \checkmark     & \checkmark     & \checkmark \\ \hline
    10 &\multicolumn{2}{|p{16em}|}{Effect of filtration and nanoparticle doping} &  x   & ?   & \checkmark \\ \hline
    11 &\multicolumn{2}{|p{16em}|}{Crystal alignment} & \checkmark     &   x    & x \\ \hline
    12 &\multicolumn{2}{|p{16em}|}{Irradiation pathway matters} & \checkmark   &  \checkmark   & \checkmark \\ \hline
    13 &\multicolumn{2}{|p{16em}|}{Direct solution-laser interaction matters} & \checkmark   & \checkmark  & \checkmark \\ \hline
    \end{tabular}%
  \label{NPLIN Observation}%
\end{table*}%

\subsubsection{Laser parameters}
 For the reported aqueous solutions, irrespective of the solute type, increasing the laser peak intensity and  pulse duration have been observed to enhance NPLIN probability (observations \ref{Peak_intensity_dep} and \ref{pulsewidth}, respectively). For both conditions, it is evident that the amount of laser-matter interaction increases. The reported low influence of the laser wavelength on the nucleation probability (observation \ref{wavelength_dep}), and its role in laser-matter interaction is less conclusive. The absorption of near-infrared laser light by water does explain the reported lower NPLIN probabilities for \SI{1064}{nm} wavelength in aqueous solutions. But the relatively high probability for \SI{355}{nm} wavelengths compared to \SI{532}{nm} still lacks explanation.\par

The observed increase in the number of crystals with increasing laser peak intensity (observation \ref{dep_nc_peakI}) can be rationalized by all mechanisms as seen in \Cref{NPLIN Observation}. However, for the DP mechanism, when applied to glycine, the values of the experimental parameters such as laser peak intensity and supersaturation were not comparable to the theoretically calculated ones\cite{10.1021/acs.cgd.0c00669}. Similarly, the OKE mechanism fails to explain the observations for KCl since dissolved KCl lacks directional polarization under laser light. Moreover, for small organic solutes, the effect of laser polarization on the polymorph formed (observation \ref{switching}) is still under debate.\par

The lack of understanding of laser-matter interactions in NPLIN leaves us with an unanswered question: why is there a laser intensity threshold to NPLIN below which no nucleation occurs (observation \ref{threshold})? As the existence of this threshold cannot be explained by the OKE and DP mechanisms, this observation may hold the key to solving the puzzle of laser-matter interaction in NPLIN. As the nucleation probability increases with laser peak intensity above this threshold, one may pragmatically ask if rather any supersaturated solution, regardless of its chemical identity, may be forced to nucleate at high enough laser intensities.   

\subsubsection{Solution parameters}
For a fixed laser peak intensity above a threshold, the observed increase in NPLIN probability with solution supersaturation (observation \ref{Peak_intensity_dep}) can be explained using classical nucleation theory (CNT), because the average size of the pre-nucleating clusters will increase with supersaturation - favoring stable nucleus formation. The observed increase in nucleation probability with aging for small organics (observation \ref{aging}) is also in line with the proposed CNT model. The more time available for the solution to progress towards equilibrium before the laser irradiation, the higher the likelihood of the formation of large pre-nucleating clusters. Since the equilibrium timescale for a solution would depend on the solute's diffusivity, the reported higher diffusion coefficient for metal halides\cite{10.1016/S0167-7322(98)00088-9} compared to small organics\cite{10.1063/1.5099069} could be a possible reason why aging of metal halides is less significant for NPLIN probability.\par

Within a solution, in addition to the intended solute and solvent, there are often impurities that a laser can interact with. The majority of authors have overlooked the composition of the impurities present in their solution. The role of impurities in NPLIN has been demonstrated by Ward \etal\cite{10.1021/acs.cgd.6b00882} who observed a decrease in NPLIN probability with filtration (observation \ref{filtration}). Thus, impurity composition and quantity may have led to the contradicting results on polarization switching (observation \ref{switching}). The reported increase in the nucleation probability with laser peak intensity and pulse duration (observations \ref{Peak_intensity_dep} and \ref{pulsewidth}, respectively) can be rationalized using the increase in the energy available to heat up a nano-impurity for triggering nucleation. From theory, the current hypotheses based on both the OKE and DP mechanism depend only on the laser peak power (\Cref{order parameters,CNTeqn}) and not on the wavelength. However, the absorption spectra of impurities based on their composition could be a deciding factor. The composition of the solution impurities could vary depending on the solute manufacturer and solvent purity.\par

The reported laser peak-intensity threshold (observation \ref{threshold}) in NPLIN could be reasoned as the minimum energy required to form a significant vapor bubble surrounding a nano-impurity. Although the majority of observations can be explained using the nano-impurity heating hypothesis, there exist only a few theoretical works that explore the mechanism\cite{10.1021/acs.cgd.0c00942, 10.1063/1.5002002}. These theoretical works are largely phenomenological and lack the ability to quantify and predict the experimentally observed correlation between laser peak intensity, crystallization probability, and laser peak-intensity threshold. Nonetheless, predicted qualitative trends from simulations offer a means to compare experiments and theory. Hopefully, further efforts from both the theoretical and the experimental side can answer questions on the exact nature of these nano-impurities. This extended understanding can answer questions such as what are the physical properties of a given nano-impurity (physicochemical state- soluble or colloidal-, chemical structure, and more) that make it NPLIN active? \par

While employing multiple nanosecond laser pulses of 532 and 1064 nm, the reported alignment of urea crystals with a laser by Garetz \etal\cite{10.1103/PhysRevLett.77.3475} could not be reproduced by Liu \etal\cite{10.1039/c6cp07997k} (observation \ref{alignment}). The observation by Garetz \etal could be attributed to the possible polarization-dependent ablation of the crystals formed and secondary nucleation. The observed influence of irradiation pathway on nucleation probability (observation \ref{pathway}) by Clair \etal\cite{10.1107/S160057671401098X} could be a result of increased irradiated volume while directing the laser from the top of the vial. With an increased irradiated volume, the likelihood of the laser encountering a nano-impurity or a pre-critical cluster (possibly containing impurities) is larger. However, one cannot deny the possibility of the air-liquid interface playing an important role. Moreover, the reported zero nucleation probability by Kacker \etal\cite{10.1021/acs.cgd.7b01277} for samples that masked the laser entry within the solution further strengthens the argument on the role of direct light-solution interaction in NPLIN (observation \ref{contact}). Thus, clear evidence on the type of light-solution interaction occurring in NPLIN would help in establishing a concrete theory for predicting the NPLIN activity of a solution.

\section{High-Intensity Laser-induced nucleation (HILIN)}
\label{sec:HILIN}

\subsection{Phenomenology}

\begin{figure}[tbp]
	\centering
	\includegraphics[width=0.4\textwidth]{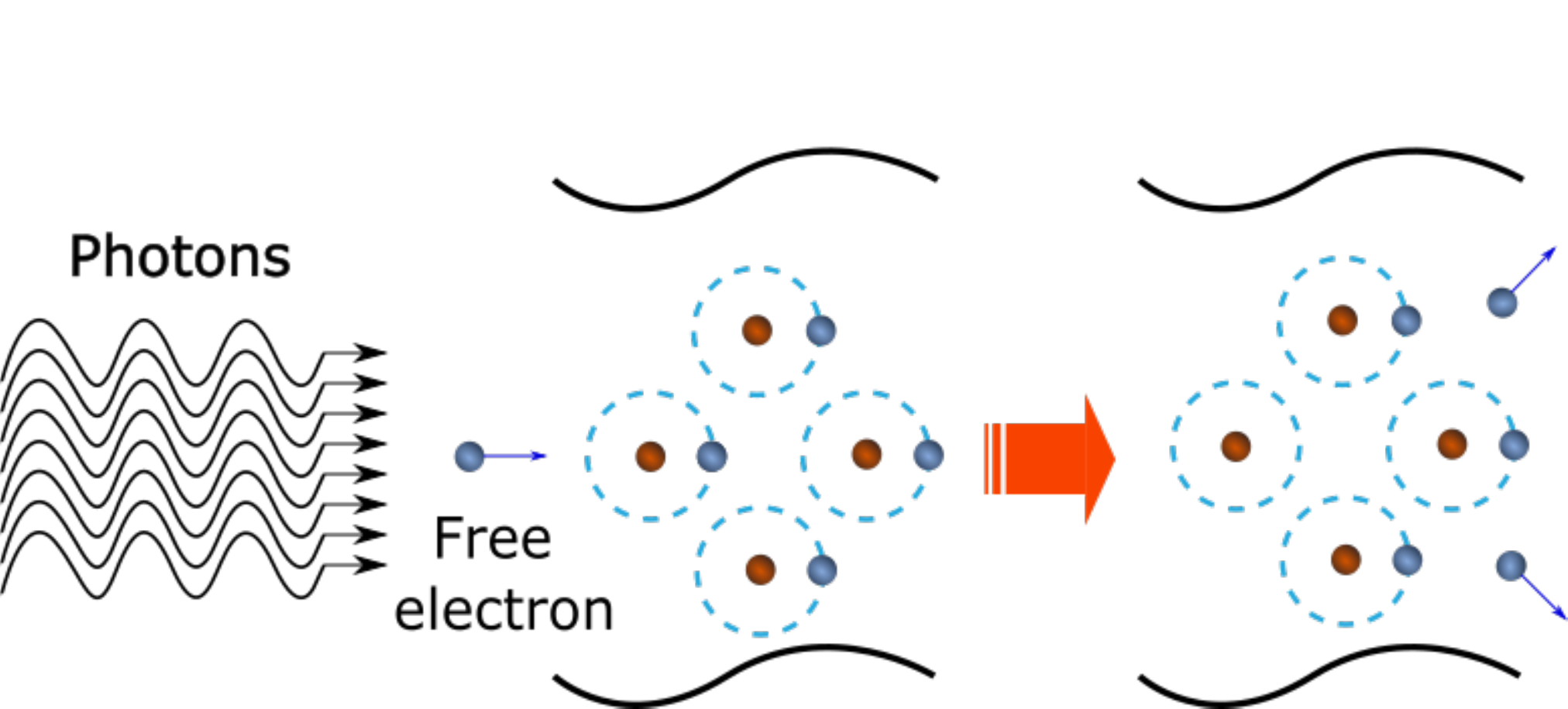}
	\caption{Avalanche ionization leading to optical breakdown and plasma formation\cite{10.5772/63892}.}
	\label{fig1100}
\end{figure}

The interaction of laser light with condensed matter 
at higher laser intensities has generated substantial attention since the invention of the laser in the 1960s \cite{10.1155/2015/463874}.  The interaction of a high-intensity laser pulse with condensed matter leads to optical breakdown and cavitation\cite{10.1007/s00340-005-2036-6}. Particularly for liquids, the literature contains many definitions of optical breakdown, but the most common definition considers it as the event connected to the minimum threshold energy needed to generate a plasma within the liquid\cite{10.1002/ppsc.200500918}. To generate such a breakdown event, the experiment needs to be operated at very high laser intensities ($\geq$GW/cm$^2$) using pulsed laser beams typically ranging from femtosecond to nanosecond pulse width \cite{10.1002/ppsc.200500918}.
Earlier literature work\cite{10.1063/1.1754840, 10.1063/1.1654204} reports optical breakdown in pure substances, \eg water, resulting from a multiphoton absorption avalanche mechanism \cite{10.5772/63892}, as illustrated in Figure \ref{fig1100}.
In this avalanche mechanism, the quasi-free electrons that are initially released from atoms and molecules gain additional kinetic energy as they are accelerated by the electric field of the laser. These electrons cause impact ionization of other atoms and molecules, thereby producing more free electrons. This event causes an electron avalanche, upon which a critical free electron density is achieved, leading to the formation of plasma within the liquid medium. The plasma generally heats up to several thousand Kelvins, upon which its volume expands leading to the emission of shockwaves\cite{10.1155/2015/463874}.\par

A considerable number of HILIN experiments have been reported in the literature with a diverse set of experimental setups and solute/solvent systems. The overarching goal of these experiments characterized with high laser intensities (GW/cm${^2}$) and short laser pulses (ns to ps) range from scientific exploration to improving the current understanding of laser-induced nucleation\cite{10.1103/PhysRevLett.77.3475}.  The results of these experiments have opened the door to new discussions concerning the control of nucleation mechanism as well as properties of resulting crystal in high-intensity laser-induced nucleation (HILIN)\cite{10.1063/1.5079328,doi.org/10.1016/j.jphotochemrev.2022.100530}. 

\subsection{Proposed mechanism}

\begin{figure}[tbp]
	\centering
	\includegraphics[width=0.5\textwidth]{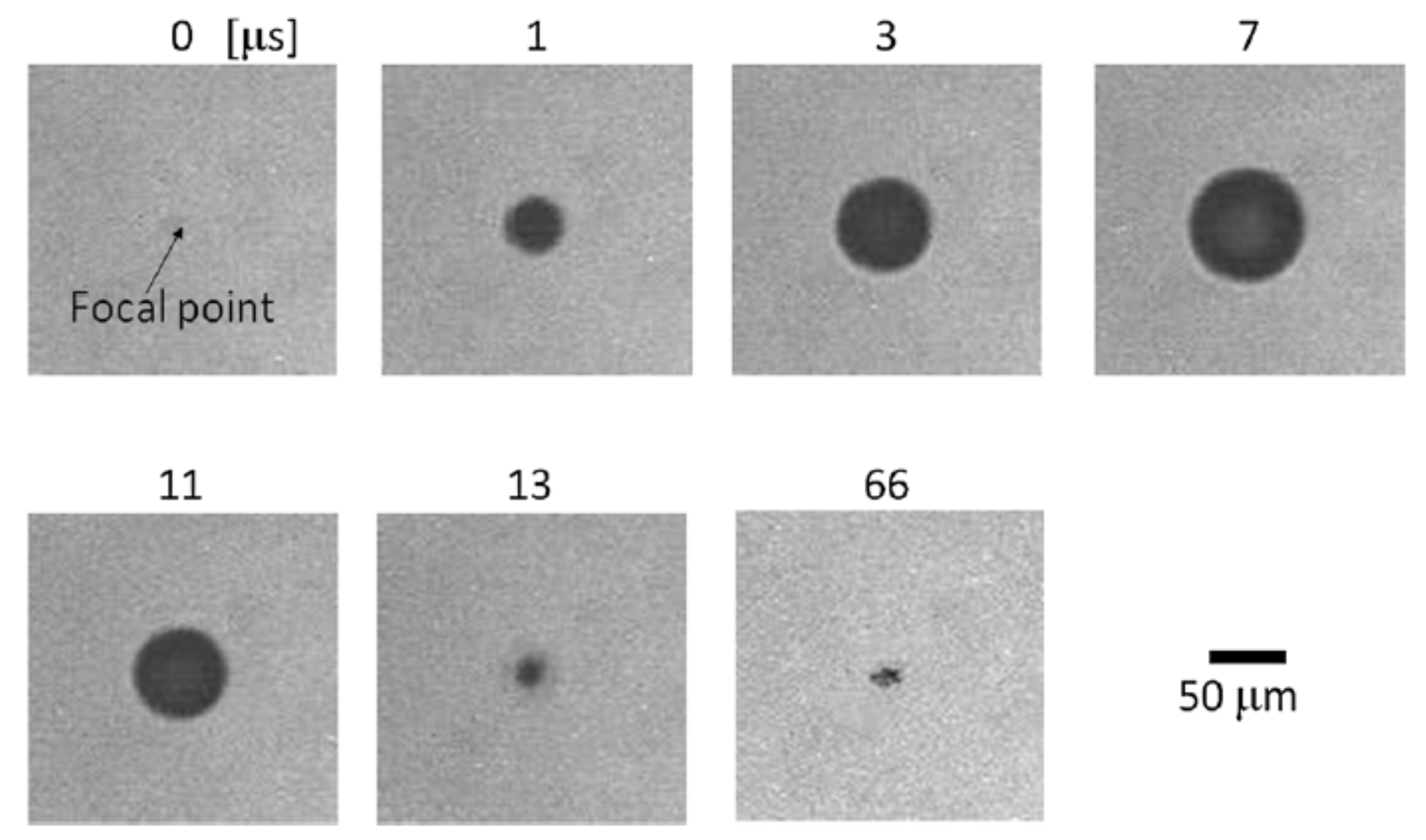}
	\caption{HEWL protein supersaturated solutions observed by a high-speed camera operating at $10^{6}$ frames/s after irradiation by a 200 fs pulse. The arrow shown at 0 $\mu$s corresponds to the focal point of the laser. The black sphere is the cavitation bubble expanding and collapsing after laser irradiation. \cite{10.1007/s00339-008-4790-x}. }
	\label{fig2}
\end{figure}

Understanding how HILIN  works in various solutions is intimately related to how plasma, shock waves, and emerging cavitation bubble upon focused laser irradiation interacts with solute and solvent.
Vogel \etal\cite{10.1021/cr010379n} suggested that a near-infrared short femtosecond laser pulse, when tightly focused into 
a solution, causes multiphoton absorption and subsequent ionization of solute and solvent molecules. This fast conversion of energy results in thermoelastic pressure along with the accumulation of heat, thereby increasing the vapor pressure of the solution. Cavitation bubbles are finally produced when the vapor pressure of the solution surpasses the atmospheric pressure\cite{10.1021/cr010379n}. As an example, the process of cavitation has been visualized in the experiments conducted by Yoshikawa \etal\cite{10.1007/s00339-008-4790-x} on a supersaturated solution of HEWL protein using a near-infrared femtosecond laser.
Microscopy images reveal that the cavitation bubble expands and collapses in microseconds after laser irradiation, as shown in Figure \ref{fig2}. The proposed nucleation mechanism among various solutions proposed in the literature is discussed below. The experimental conditions and sample compositions are summarized in Table \ref{table_LIN}.\par

\begin{figure}[tbp]
	\centering
	\includegraphics[width=0.5\textwidth]{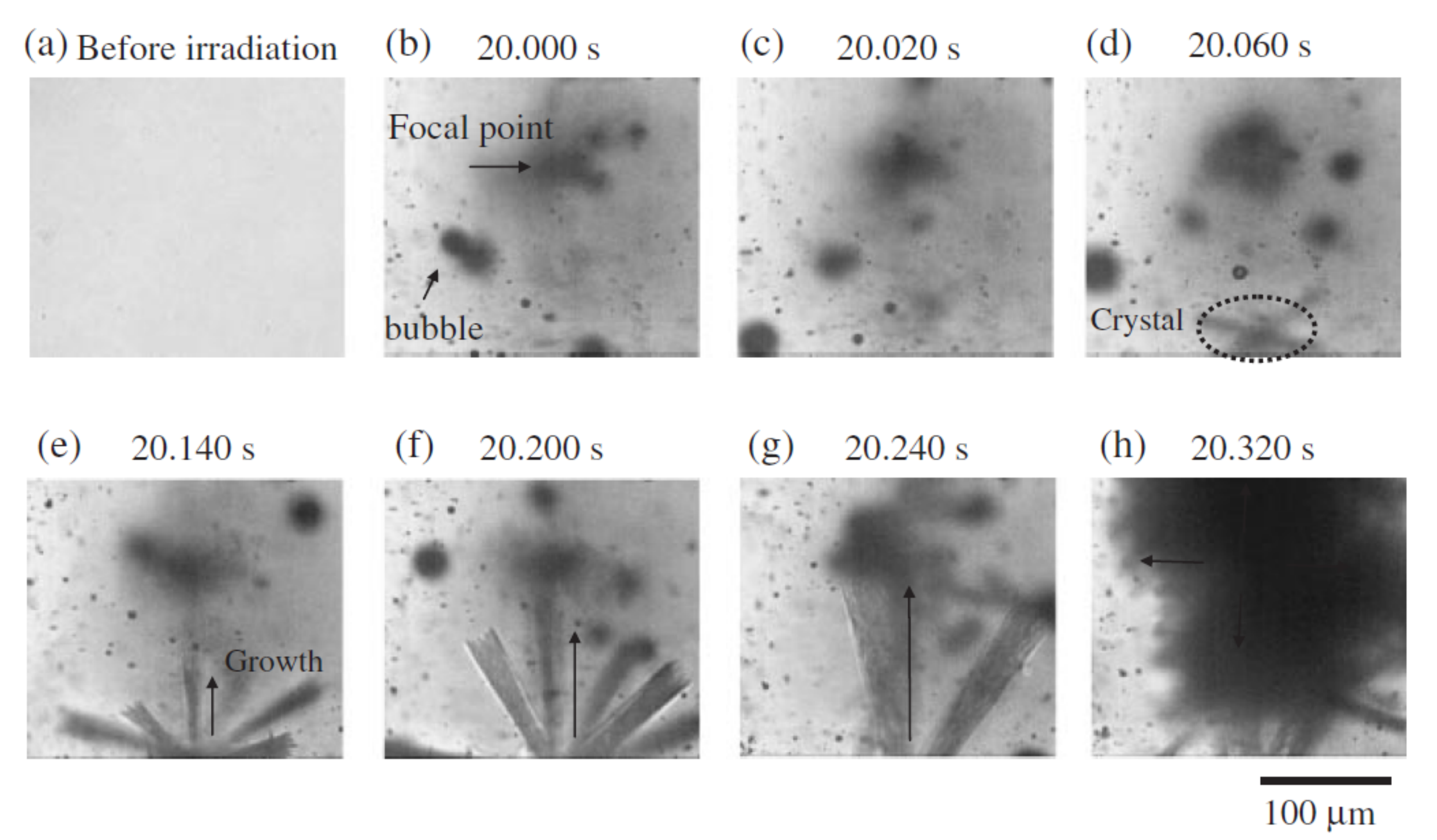}
	\caption{Urea crystallization observed by a high-speed camera, obtained after femtosecond laser irradiation at an energy of 340 $\mu$J/pulse for 12 M concentration. The arrow mark at 20 s indicates the focal point of the laser\cite{10.1143/JJAP.45.L23}.}
	\label{fig5}
\end{figure}

Yoshikawa \etal\cite{10.1143/JJAP.45.L23} performed experiments with different supersaturated solutions of urea in water (10.5 M - 12 M) using a Ti:sapphire femtosecond laser system (800 nm, 120 fs, 1 kHz) in 16 mm diameter pyrex glass tubes.  The laser was focused into the supersaturated solution through an objective lens (10X, N.A. = 0.4) at approximately 5 mm from the bottom of the glass tube. The generated crystals upon laser irradiation in the vicinity of the focal point are shown in Figure \ref{fig5}. Microscopy images revealed the formation of many cavitation bubbles of size ranging between 100 nm and 10 $\mu$m, subject to each laser pulse, that later diffused and collapsed. It was also observed that needle-shaped objects formed  with a standard CCD camera twenty seconds after irradiation, as shown in Figures \ref{fig5}d-\ref{fig5}f.
The crystallization threshold energies for urea concentrations of 11 M, 11.5 M, and 12 M were found to be 50 $\mu$J, 50 $\mu$J, and 240$\mu$J respectively. The corresponding laser energy density
for these threshold energies was measured to be approximately 10 J/cm$^{2}$, much higher than the optical breakdown of the solution (0.2 J/cm$^{2}$). At this laser energy density,
the authors predicted that the urea solution will undergo multiphoton absorption at the focal point of the laser,
leading to shockwaves and cavitation bubble formation. The shockwaves resulting from the expansion and collapse of cavitation bubbles\cite{10.1063/1.367512},\cite{10.1063/1.368906} generate transient pressures of the order of MPa-GPa, and these variations in pressure could  trigger nucleation\cite{10.1103/physrevlett.73.2853}. On the basis of these observations, the most plausible mechanism for crystal formation was explained in two steps. The first step is the nucleation and growth of urea crystals at the focal point of the laser due to the laser-induced shock wave and cavitation bubbles and the second step is the laser ablation of the already generated crystal leading to subsequent crystal growth due to multiple laser pulses\cite{10.1143/JJAP.45.L23}. \par

\begin{figure}[tbp]
	\centering
	\includegraphics[width=0.4\textwidth]{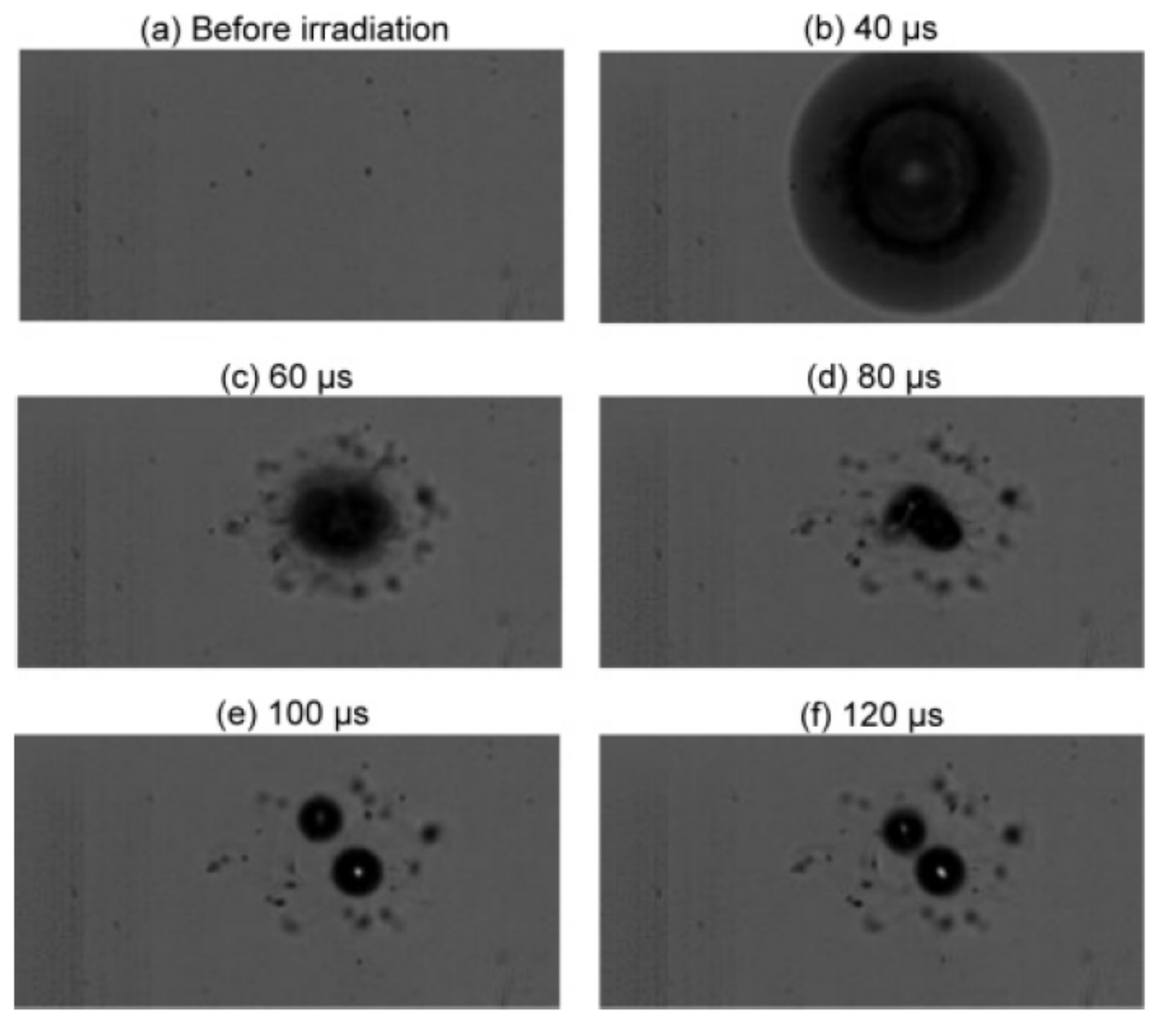}
	\caption{Bubble formation process before anthracene crystallization observed by a high-speed camera, obtained after single-shot femtosecond laser irradiation into its supersaturated solution at an energy of 12.7 $\mu$J/pulse with solution temperature at 25$^{\circ}$C\cite{10.1021/cg060631q}.}
	\label{fig7}
\end{figure}

\begin{figure}[tbp]
	\centering
	\includegraphics[width=0.4\textwidth]{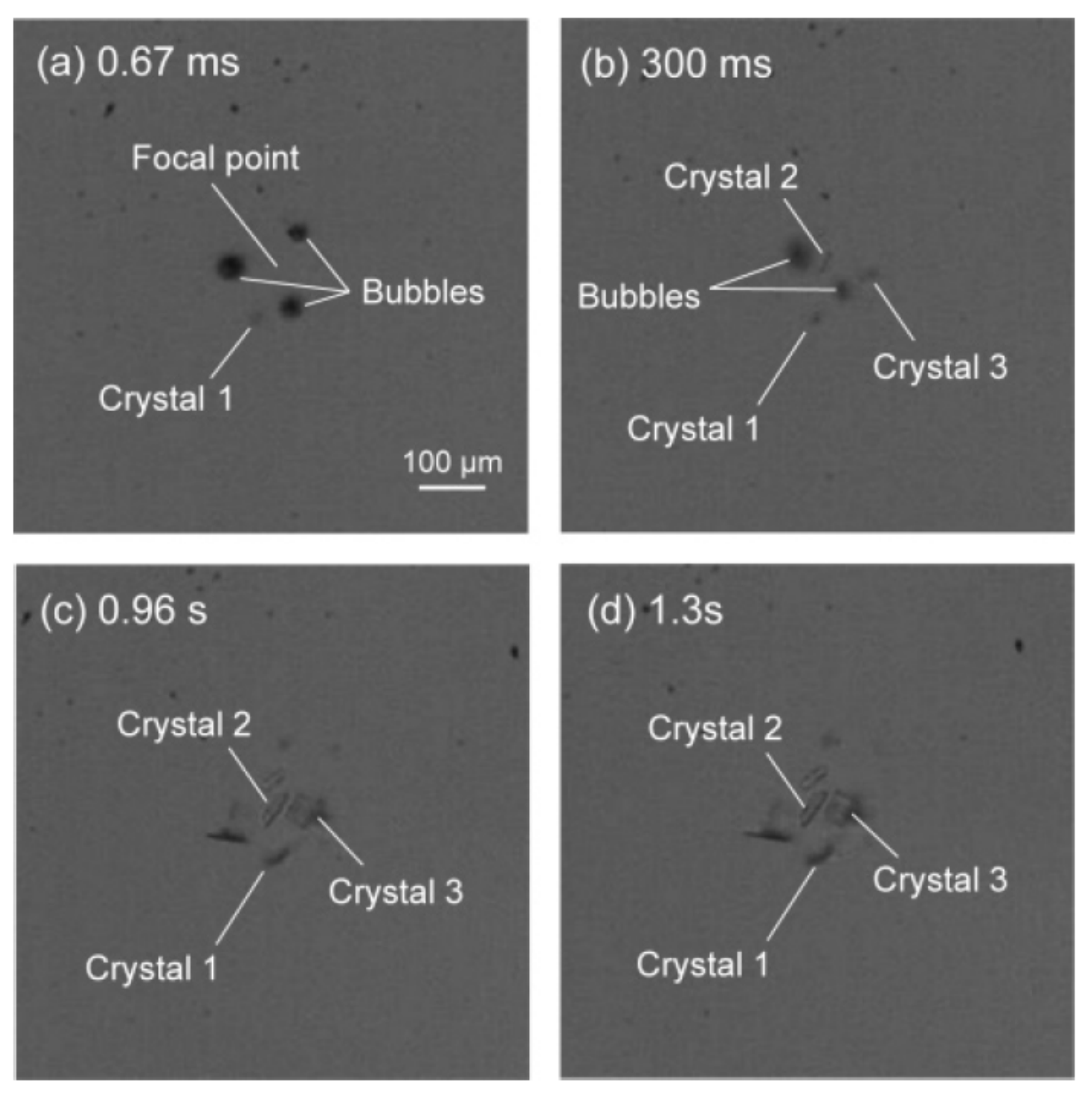}
	\caption{Anthracene crystallization observed by a high-speed camera, obtained after single-shot femtosecond laser irradiation into its supersaturated solution at an energy of 12.7 $\mu$J/pulse with solution temperature at 25$^{\circ}$C  \cite{10.1021/cg060631q}. }
	\label{fig6}
\end{figure}

Similarly, Nakamura \etal\cite{10.1021/cg060631q} conducted laser-induced nucleation experiments with supersaturated solutions of anthracene in cyclohexane (supersaturation 1.1-1.9) using a femtosecond Ti:sapphire laser pulse (800 nm, 120 fs, 1 KHz) in a 10 mm rectangular optical cell. The laser was focused into the supersaturated solution through an objective (10X, N.A. = 0.25) a few millimeters from the bottom of the cell, as shown in Figure \ref{fig8}d, and generated polyhedral-shaped crystals.
The cavitation bubble formation and subsequent crystal growth follows from the optical breakdown in pure water\cite{10.1007/s00340-005-2036-6}. According to Nakamura \etal\cite{10.1021/cg060631q}, the temperature of the supersaturated solution of anthracene increases rapidly at the laser focal point as the energy of the femtosecond laser pulse is consumed effectively through multiphoton absorption. This rapid temperature rise subsequently leads to shockwave emission along with the formation of a cavitation bubble\cite{10.1021/cr010379n}. However, as time progresses, the cavitation bubble further splits into small bubbles due to asymmetrical convection caused by jet flow during the asymmetric collapse of the cavitation bubble, as shown in Figure \ref{fig7}.
Finally, after some seconds, it was observed that polyhedral-shaped crystals of anthracene were formed at the laser focal point, as shown in Figure \ref{fig6}. From these observations, the authors hypothesized that the crystallization of anthracene might have taken place at the interface between the cavitation bubble and solution. This hypothesis was further verified based on the observation that the threshold energy (3.1 $\mu$J) reported for laser-induced bubble formation corresponds exactly with the minimum laser pulse energy needed to observe crystallization of anthracene\cite{10.1021/cg060631q}.\par

\begin{figure*}[tbp]
	\centering
	\includegraphics[width=0.45\textwidth]{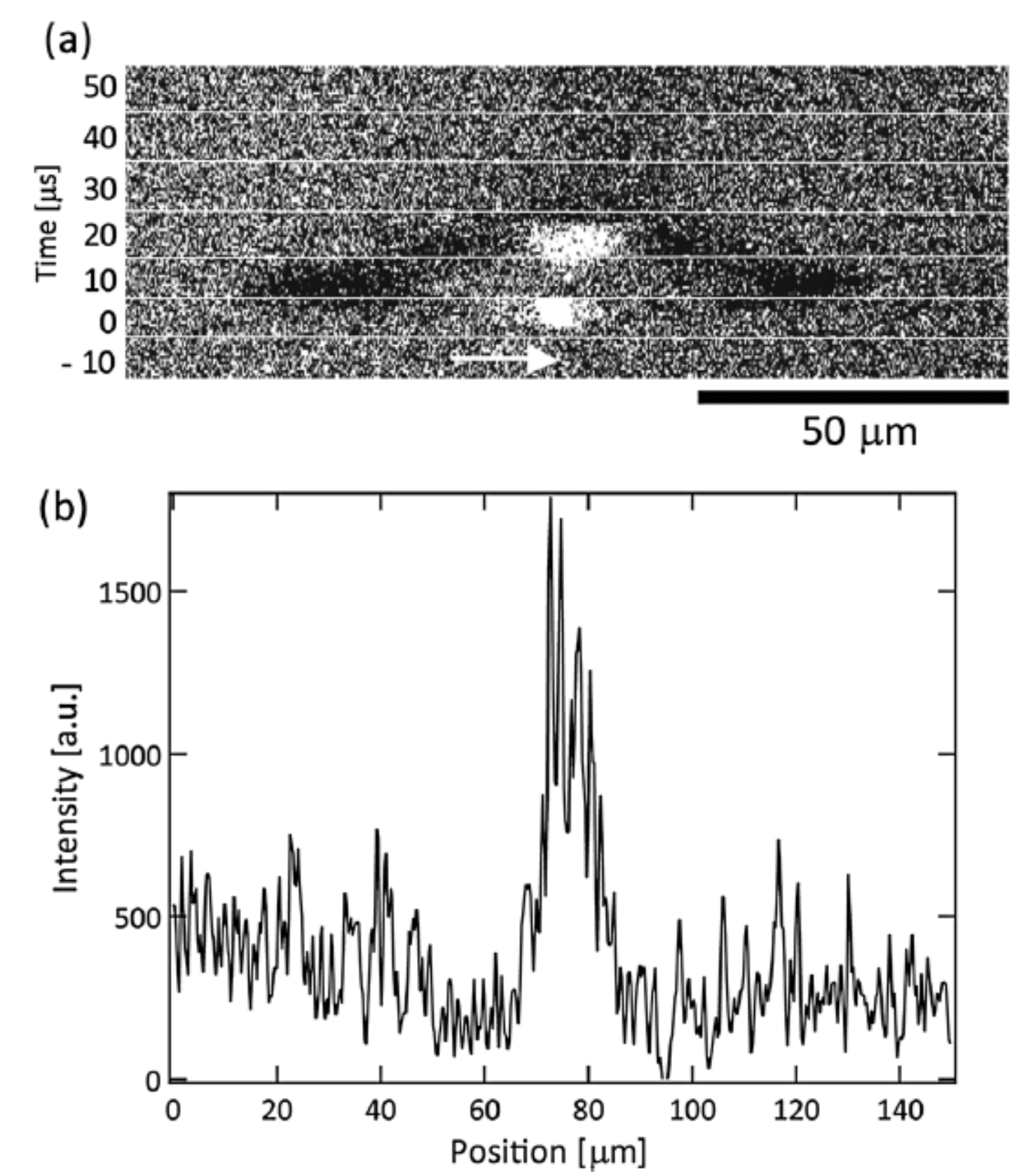}
	\caption{(a) Fluorescence images of F-lysozyme in agarose gel irradiated with a femtosecond laser pulse at 0 $\mu s$. The arrow mark at -10 $\mu$s indicates the focal point of the laser, (b) Fluorescence intensity profile of F-lysozyme in agarose gel at 20 $\mu$s\cite{10.1016/j.jcrysgro.2008.09.137}.}
	\label{fig3}
\end{figure*}

Cavitation bubbles were also observed in proteins subjected to laser pulse irradiation. To judge whether the cavitation bubbles trigger the nucleation mechanism, high-speed imaging experiments at an interval of 10 $\mu s$ have been performed with the help of fluorescent dye-labeled protein F-lysozyme in 2\% agarose gel using a femtosecond laser (780 nm, 100 fs, 1 KHz) in a capillary (0.7 mm ID, 50 mm length) by Yoshikawa \etal\cite{10.1016/j.jcrysgro.2008.09.137}. The laser was focused into the center of the capillary, containing supersaturated solution, through an objective (10X, N.A. = 0.4), as shown in Figure \ref{fig8}f.
The fluorescence images and fluorescence intensity profile of supersaturated HEWL protein solution in 2\% agarose gel are shown in Figure \ref{fig3}. The bright spot at the center of the fluorescence image directly after laser irradiation corresponds to plasma emission. The plasma emission was subsequently followed by a cavitation bubble that expanded and collapsed within 30 $\mu$s. However, the peak fluorescence intensity at 20 $\mu$s was found to be three times larger than the average peak intensity of the corresponding agarose gel medium, visible as an intense bright spot. This bright spot was attributed to a local high concentration of F-lysozyme at the focal point of the laser beam\cite{10.1016/j.jcrysgro.2008.09.137}. On the other hand, there were no bright spots observed during the expansion and collapse of the cavitation bubble after laser irradiation in the absence of agarose gel within the solution. Previously Nakamura \etal\cite{10.1021/cg060631q} hypothesized that the surface of the cavitation bubble acts as a preferential location for the accumulation of protein molecules generated after laser irradiation\cite{10.1021/cg060631q}. Therefore, the protein molecules that initially adsorbed onto the bubble surface probably are gathered together at the focal point during the rapid collapse of the cavitation bubble. The local increase in this protein concentration not only leads to an increase in fluorescence intensity but also to a nucleation event. \par

\begin{figure}[tbp]
	\centering
	\includegraphics[width=0.5\textwidth]{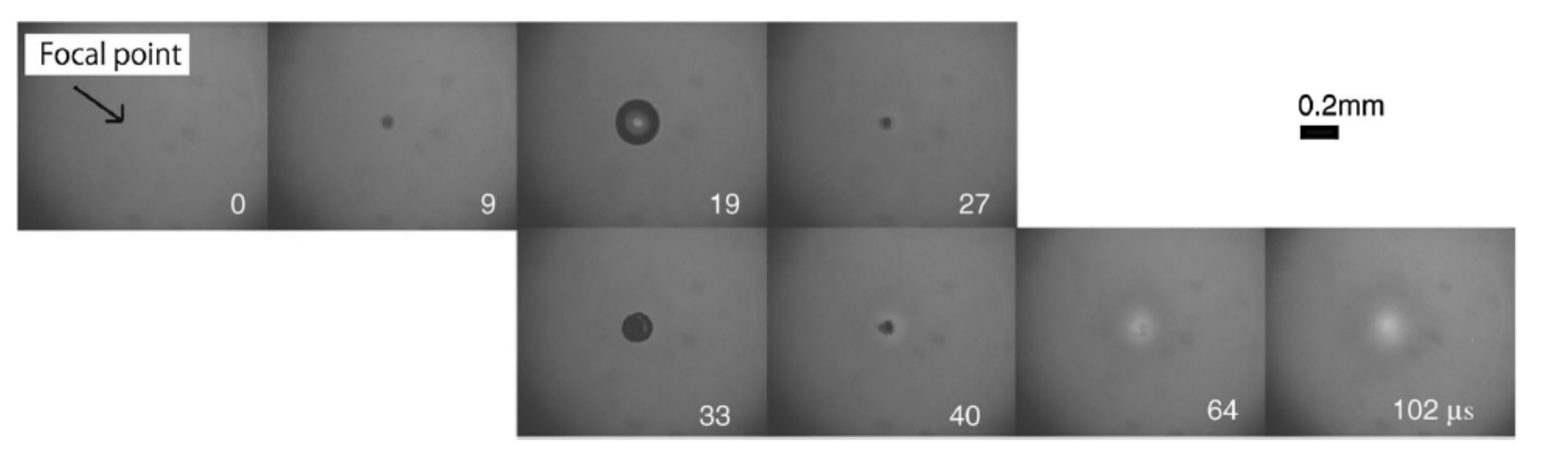}
	\caption{Cavitation bubbles observed by a high-speed camera operating at $10^{6}$ frames/s after irradiation with a 200 fs pulse into supersaturated cytochrome c solution\cite{10.1016/j.jcrysgro.2010.10.068}.}
	\label{fig4}
\end{figure}

Later, Iefuji \etal\cite{10.1016/j.jcrysgro.2010.10.068} performed laser-induced nucleation experiments with a supersaturated solution of protein cytochrome c using a femtosecond laser setup (780 nm, 200 fs, 1 KHz) and high-speed imaging in microbatch plates containing wells. The laser was focused at the center of the microbatch well, containing a supersaturated solution, through an objective (10X, N.A. = 0.25), as shown in Figure \ref{fig8}c.
Upon laser irradiation to the supersaturated solution, the cavitation bubble expanded and collapsed in 40 $\mu$s, leaving behind bright and dark areas, as shown in Figure \ref{fig4}. The bright area was observed at the focal point of the laser beam and the dark area surrounding the cavitation. These bright and dark regions were attributed to low and high protein concentration regions, respectively. The high concentration area could probably give rise to protein nuclei\cite{10.1016/j.jcrysgro.2010.10.068}. Despite the fact that the findings of these studies suggest that bright and dark spots indicate low and high protein concentration regions respectively, it is difficult to quantify this claim simply by looking at Figure \ref{fig4}. Further quantification in terms of the protein concentration as well as local temperature  surrounding the cavitation bubble after laser irradiation in a future study will be most informative in that regard.\par

The experimental techniques proposed by Yoshikawa \etal\cite{10.1016/j.jcrysgro.2008.09.137} and Iefuji \etal\cite{10.1016/j.jcrysgro.2010.10.068} collectively contributed to the development of a mechanistic view of laser-induced nucleation. On the application side, obtaining high-quality protein crystal using a femtosecond laser can be helpful for X-ray crystallographic structural studies\cite{10.1021/cg049709y}.\par



\begin{figure}[tbp]
	\centering
	\includegraphics[width=0.45\textwidth]{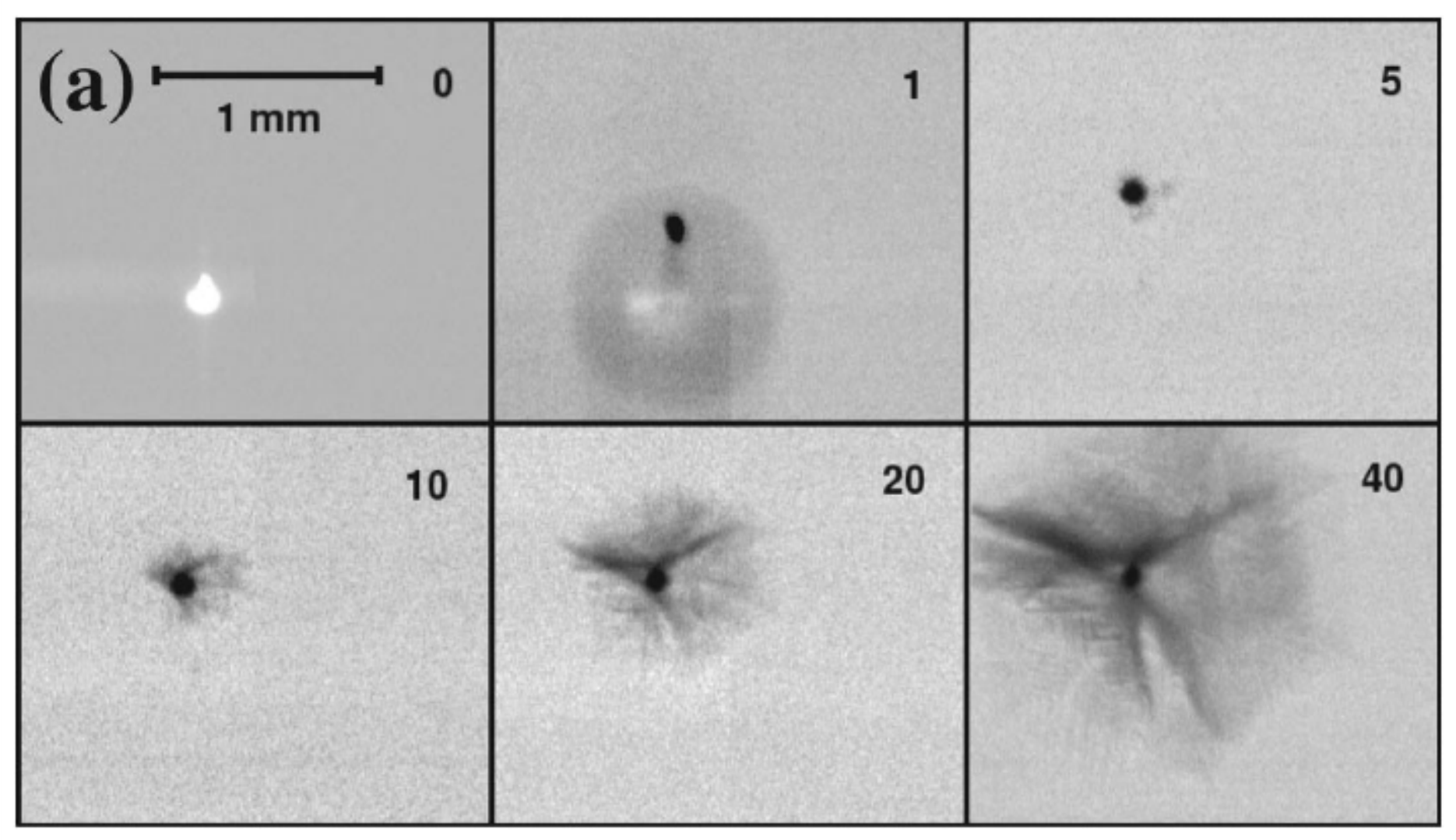}
	\caption{Ice crystallization observed by a high-speed CCD camera, induced by optical breakdown using 8 ns laser irradiation at an energy of 2 mJ.\cite{10.1103/PhysRevLett.99.045701} }
	\label{fig8000}
\end{figure}

Also for a single component system, focused Nd:YAG nanosecond laser irradiation (1064 nm, 8 ns, 20 Hz) into a rectangular cell containing supercooled water resulted in the nucleation of ice at the location of noncondensable gas bubbles, as shown in Figure \ref{fig8000}.  The laser was focused into the center of the rectangular cell containing supercooled water through a focusing lens, as shown in Figure \ref{fig8}k.
Lindinger \etal\cite{10.1103/PhysRevLett.99.045701} claims that homogeneous nucleation in the compressed liquid phase could be the possible mechanism for crystal formation. This was supported by experiments showing optical breakdown of supercooled water following laser irradiation along with cavitation bubble formation and its subsequent collapse into numerous small bubbles. It was also observed that when multiple laser shots were fired, the preexisting microbubbles from previous laser pulses acted as a heteronucleant for ice nucleation, if they were very close to the cavitation events happening from new laser pulses\cite{10.1103/PhysRevLett.99.045701}.\par

\begin{figure}[tbp]
	\centering
	\includegraphics[width=0.4\textwidth]{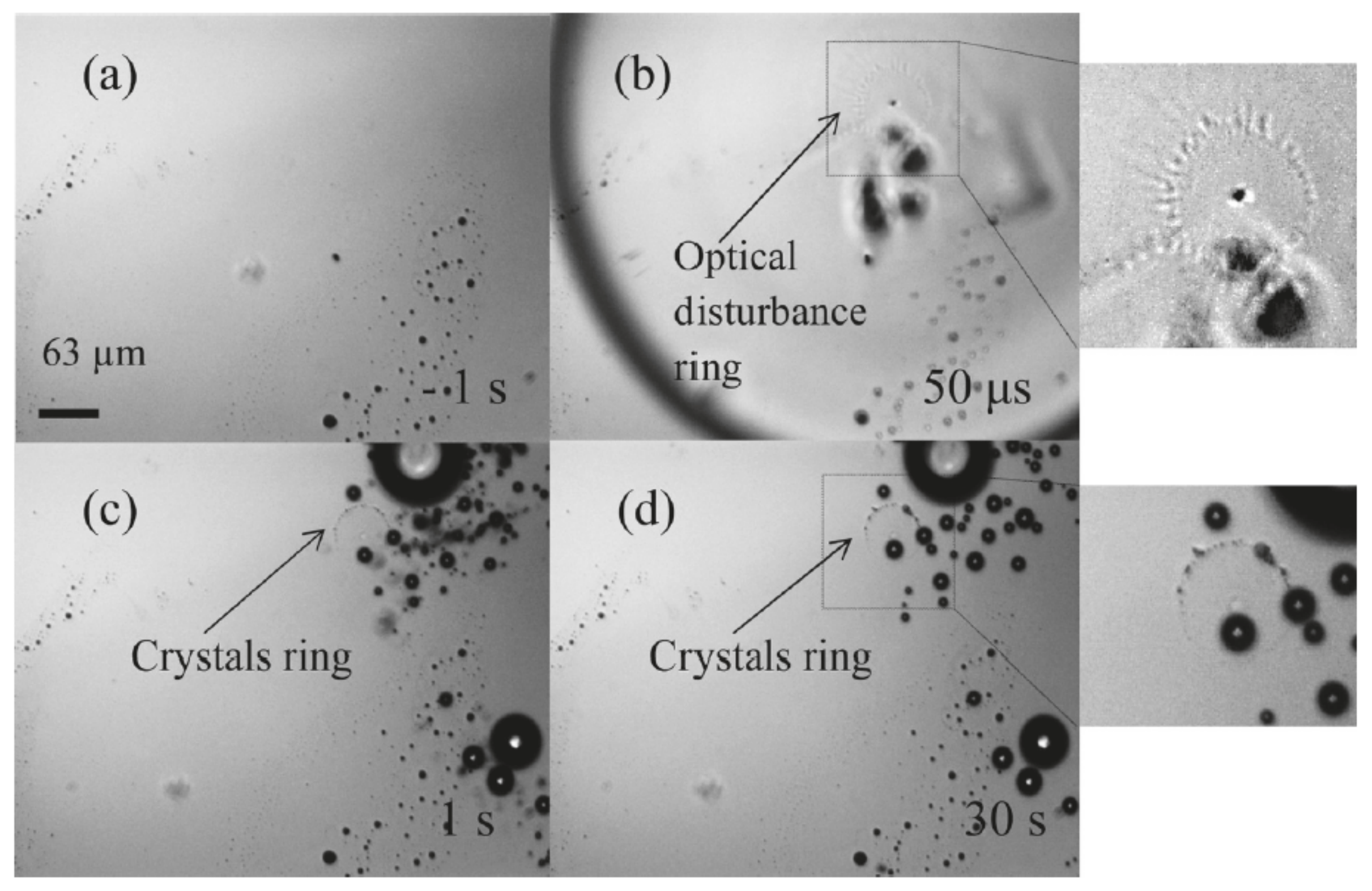}
	\caption{Cavity evolution and formation of (NH$_4$)$_{2}$SO$_{4}$ crystals observed by a high speed camera\cite{10.1021/cg2000014}.}
	\label{fig9000}
\end{figure}

Laser-induced nucleation experiments were performed with supersaturated solutions involving simple salts like (NH$_4$)$_{2}$SO$_{4}$ (0.2$\%$ and 0.4$\%$ relative supersaturations) and KMnO$_4$ (7$\%$, 14$\%$ and 21$\%$ relative supersaturations) by Soare \etal\cite{10.1021/cg2000014} using Nd:YAG nanosecond laser pulses (6 ns, 532 nm, 0.05-0.5 mJ). The laser was focused at the center of the two glass plates (placed 50-100 $\mu$m apart) containing supersaturated solution through a 20x objective, as shown in Figure \ref{fig8}f.
In the case of the (NH$_4$)$_{2}$SO$_{4}$ solution, it was observed that crystals were formed in the vicinity of the optical disturbance a few seconds after the cavitation bubble collapse, as shown in Figure \ref{fig9000}. Although a mechanistic route by which the cavitation bubble leads to crystal nucleation was not explicitly given, it was argued that crystal nucleation took place at the bubble interface, based on changes in the refractive index induced by the formation of nuclei at the maximum evaporation rate. This hypothesis was supported by recently performed direct numerical simulations of a laser-induced thermocavitation bubble by Hidman \etal\cite{10.1021/acs.cgd.0c00942}. To estimate the degree of supersaturation of the solution surrounding the vapor bubble, the initial growth stage of the bubble was modeled using numerical simulations. Hidman \etal\cite{10.1021/acs.cgd.0c00942} proposed that the evaporation of solvent around the bubble interface causes an increase in the local supersaturation and triggers nucleation in laser-induced nucleation experiments. This would be the case provided the predicted supersaturation of the solution in the vicinity of the bubble exceeds the solubility limit for a given threshold laser energy in the experiments. Also, it was found that 10 to 30 crystals of (NH$_4$)$_{2}$SO$_{4}$ (0.4$\%$ relative supersaturation) were formed per laser pulse, whereas with KMnO$_4$ higher supersaturations were needed for nucleation to occur, and a smaller number of crystals were observed. In some cases,  more laser pulses were needed to induce crystal formation. Soare \etal\cite{10.1021/cg2000014} interpreted this observation to be based on an insufficient evaporation rate emerging from a single laser pulse to induce nucleation, without actually quantifying it.
\par

Barber \etal\cite{10.1021/acs.cgd.9b00951} performed laser-induced nucleation experiments on supersaturated solutions of NaClO$_{3}$ at high energy densities (420 kJ cm$^{-2}$) using Nd$^{3+}$:YAG laser at two different wavelengths (532 nm and 1064 nm) in 3 ml vials. Laser irradiation in the form of circularly polarized light was focused at the center of the vial with the help of a planoconvex lens, as shown in Figure \ref{fig8}j, to understand the effect of helicity of light on the nucleation of NaClO$_{3}$  enantiomorphs. NaClO$_{3}$ as a compound exists in two enantiomorphic forms i.e. left-handed (l) and right-handed (r) enantiomorphs. Upon single laser pulse irradiation to the supersaturated solution, only one or two crystals were generated per vial at this high energy density. In addition, no significant correlation between helicity of light and NaClO$_{3}$ enantiomorphs was observed. Also, the number of d and l NaClO$_{3}$ crystals formed from the total number of samples irradiated with a laser at both wavelengths was found to be about identical for both right circular polarized light (RCP) and left circular polarized light (LCP). Barber \etal\cite{10.1021/acs.cgd.9b00951} proposed that crystallization takes place by initially favoring nucleation of monoclinic phase III NaClO$_{3}$ molecular clusters that later convert to cubic phase I NaClO$_{3}$, following Ostwald's rule of stages\cite{10.1021/cg401324f}\cite{10.1021/cg500527t}. The selectivity of monoclinic NaClO$_{3}$ phase III crystal nucleation over NaClO$_{3}$ phase I crystal was further attributed to its higher solubility values at room temperature and higher Gibbs free energy values below its melting point. Based on these observations, the plausible mechanism was discussed on the basis of local high supersaturation surrounding cavitation bubbles leading to monoclinic NaClO$_{3}$ phase III crystal nucleation upon laser irradiation. This was supported by a general idea that due to evaporation, there is an increased solute concentration at the bubble interface leading to nucleation after laser irradiation, as shown by the direct numerical simulations by Hidman \etal\cite{10.1021/acs.cgd.0c00942}. \par

\subsection{Reported solutions}

High-intensity Laser-induced nucleation (HILIN) experiments have been performed on different types of solutes, including small organic molecules and proteins. A classification of the sample holders, based on their geometry, volume, and position of the laser for HILIN experiments is schematically shown in Figure \ref{fig8}.

\subsubsection{Small organic molecules}

Concerning small organic molecules, HILIN experiments have been performed with supersaturated solutions consisting of 4-(Dimethylamino)-N-methyl-4-stilbazolium Tosylate (DAST) in methanol \cite{10.1021/cg049709y}, urea in water \cite{10.1143/JJAP.45.L23}, anthracene in cyclohexane \cite{10.1021/cg060631q} and glycine in water \cite{10.1016/j.jcrysgro.2012.11.018}.\par

Tsunesada \etal\cite{10.1016/s0022-0248(01)02266-7}, studied the nucleation of DAST crystals using Nd:YAG nanosecond laser (1064 nm, 23 ns) and later Hosokawa \etal\cite{10.1021/cg049709y} studied the nucleation of the same crystals using focused short pulse femtosecond Ti:sapphire laser (800 nm, 120 fs). It was found that the nucleation probability of DAST crystals obtained by a femtosecond laser setup at a fixed time interval is larger (10\%) than by a nanosecond laser setup (3\%), owing to its large peak intensity values and multiphoton absorption in a very short time. It was also observed that the nucleation probability of DAST crystals increased with a decrease in the repetition rate of the laser pulses. This was explained in terms of molecular cluster nuclei destruction by the train of laser pulses when the time interval between them is larger\cite{10.1021/cg049709y}. \par

Liu \etal\cite{10.1016/j.jcrysgro.2012.11.018} performed laser-induced crystallization experiments on supersaturated solutions of glycine in water using femtosecond laser (800 nm, 160 fs) at both air/solution interface and glass/solution interface and found that the crystallization probability at a fixed time interval is higher at the air/solution interface than at the glass/solution interface. It was found that the nucleation probability increased with an increase in the repetition rate of laser pulses, laser pulse energy, and exposure time of the laser. Two of the three above-mentioned parameters were kept constant while evaluating the nucleation probability with respect to the changing parameter\cite{10.1016/j.jcrysgro.2012.11.018}.

\subsubsection{Proteins}

The first HILIN experiments focusing on supersaturated solutions of proteins were performed in the early 2000's with the purpose of obtaining high-quality protein crystals. In these experiments, Adachi \etal\cite{10.1143/jjap.42.l798} studied the crystallization of proteins including HEWL, glucose isomerase, and ribonuclease H using a Ti:sapphire femtosecond laser. The important observation of these experiments was that the number of HEWL crystals generated upon laser irradiation increased with the number of irradiation pulses\cite{10.1143/jjap.42.l798}. The glucose isomerase protein crystals were also found to nucleate from a low supersaturated solution upon focused laser pulse irradiation \cite{10.1143/jjap.42.l798}. The induction time of Trypanosoma brucei prostaglandin F2$\alpha$ synthase (TbPGFS) was decreased from many months to a few days by focused femtosecond laser pulse\cite{10.1143/jjap.42.l798}. 
Apart from these water-soluble proteins, focused femtosecond laser pulse experiments were also carried out with membrane proteins. 
Adachi \etal\cite{10.1143/jjap.43.l1376} studied HILIN of membrane proteins like Aerobic respiration control sensor protein (AcrB) by irradiating focused femtosecond laser pulses to three different supersaturated solutions with various precipitant concentrations of 12\%, 13\% and 15\% polyethylene glycol (PEG) 2000. It was observed that upon laser irradiation, a supersaturated solution containing 13\% PEG (meta-stable region) generated large crystals (200 $\mu$m) of AcrB. However, for 15\% PEG spontaneous nucleation resulted in small crystals ($\le$100 $\mu$m)\cite{10.1143/jjap.43.l1376}. 


\par\vskip 1em

\begin{figure*}[tbp]
	\centering
	\includegraphics[width=0.8\textwidth]{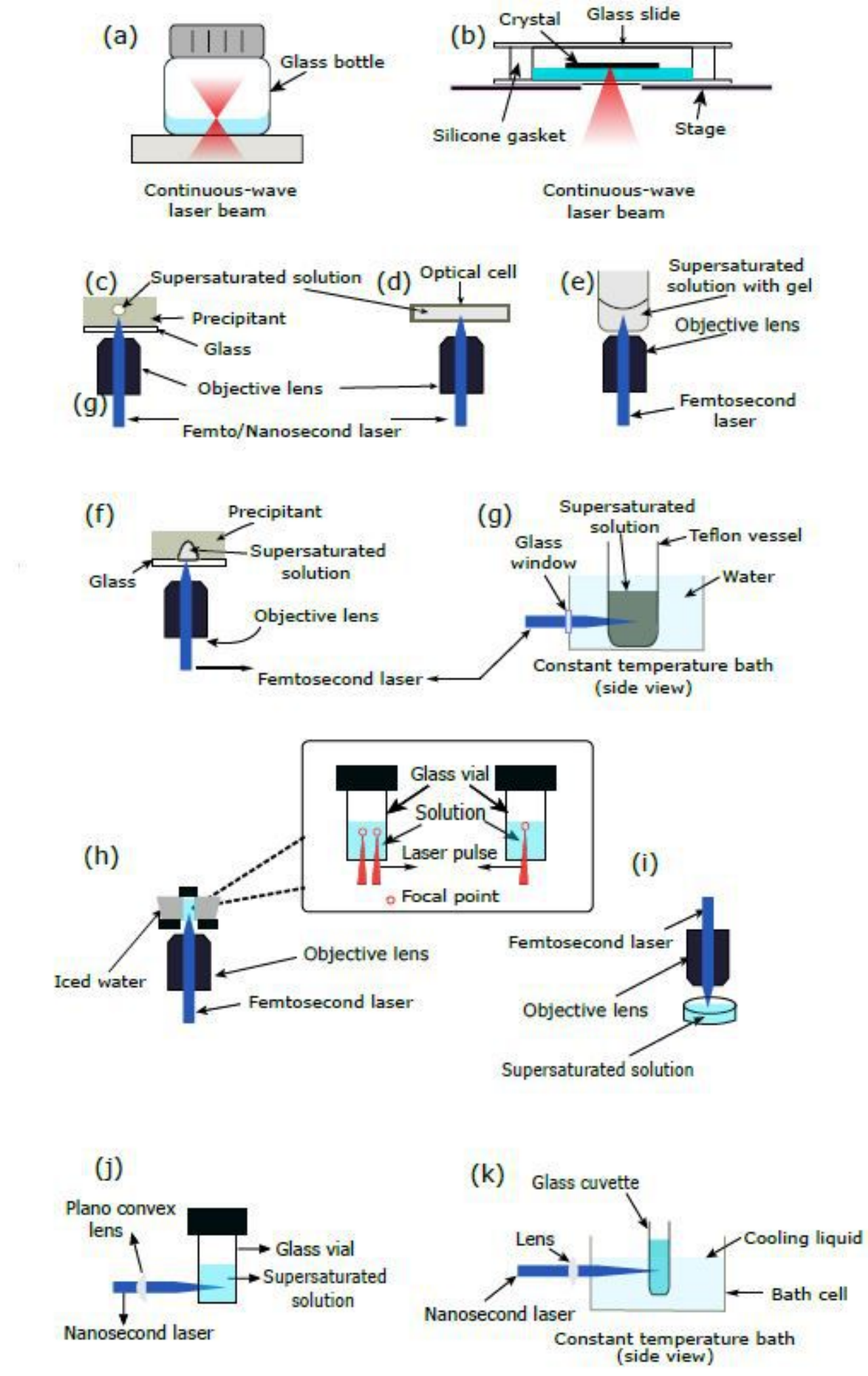}
	\caption{Schematic setups used in HILIN experiments.}
	\label{fig8}
\end{figure*}

\subsection{Potential for controlling polymorphic form}

Recently, focused femtosecond lasers have been reported to control the polymorphic form of compounds. Ikeda \etal\cite{10.7567/apex.8.045501} reported the first study on polymorph control of indomethacin in acetonitrile supersaturated solution ($\sigma$ = 3.5) by focusing a femtosecond laser through a 10x objective at 800 nm wavelength in 1 ml glass vials. In these experiments, the laser was focused in the center of the solution, close to the glass vial's side wall and at the air-liquid interface, as shown in Figure \ref{fig8}h. \par

At the location close to the glass vial side wall or at the air-liquid interface, nucleation occurred within a day after laser irradiation, giving rise to needle-shaped crystals with an approximate size of 1 mm. These crystals were found to be of metastable $\alpha$-form, having at least 8 months of temporal stability at room temperature in air and at 0$^{o}$C in solution. However, laser irradiation focused at the center of the solution resulted in plate-shaped crystals with an approximate size of 500-600 $\mu$m in 48 hours, corresponding to stable $\gamma$-forms. \par

These observations were explained by the fact that the laser irradiation at the glass vial's side wall and at the air-liquid interface resulted in an asymmetric bubble expansion and collapse with many long-lasting bubbles reaching the region of the solution meniscus along the glass wall.
It was also observed that the nucleation of meta-stable $\alpha$-form crystals occurred at the location of long-lasting bubbles. Ikeda \etal\cite{10.7567/apex.8.045501}  attributed this observation to the higher evaporation efficiency of the lasting bubbles along the solution meniscus, leading to higher local supersaturation and thus preferentially enabling the nucleation of the meta-stable $\alpha$ phase due to its lower interface energy, as described by the Ostwald rule of stages. Further experiments quantifying this observation will be most useful. Conversely, laser irradiation at the center of the solution resulted in a symmetrical bubble expansion and collapse, with long-lasting bubbles present along the region of the air/solution interface\cite{10.7567/apex.8.045501}. 
Ikeda \etal\cite{10.7567/apex.8.045501} proposed that the evaporation efficiency of long-lasting bubbles at the air/solution interface would be lower than at the solution meniscus, resulting in mild local supersaturation, without quantifying evaporation efficiency. Future efforts on quantification of this hypothesis point will be essential to predicatively control polymorphic form, a topic of primary interest for the industry. The mild local supersaturation values were argued to be responsible for the nucleation of the stable $\gamma$ phase. In summary, Ikeda \etal\cite{10.7567/apex.8.045501} suggests that controlling crystal polymorphs  requires precise control of the laser focal point.\par

Shubo \etal\cite{10.1021/acs.cgd.9b00123} performed laser-induced crystallization experiments on supersaturated solutions of acetaminophen in water ($\sigma$ = 1.5) by focusing femtosecond laser light (800 nm) through a 10x objective at the air-solution interface of the solution, as shown in Figure \ref{fig8}i. The experiments were performed to enhance the proportion of metastable crystals, \ie Form II and Form III polymorphs, by changing the laser pulse power and double-pulse delay.\par

During irradiation and immediately after cavitation, bubbles expanded and collapsed, and laser-induced crystallization of the acetaminophen solution resulted in needle-shaped crystals. In contrast, spontaneous nucleation of the same solution resulted in rhombic crystals. It was reported that laser irradiation to the solution for 2 minutes resulted in statistically reliable results
and prevented the transition of metastable Form II and III crystals to stable Form I crystals for several minutes. Furthermore, it was found that the percentage of needle-shaped crystals decreased with increasing laser pulse energy. 
Also, as laser pulse energies increased from 25 $\mu$J to 65 $\mu$J per pulse, the percentage of Form II and Form III needle crystals with an aspect ratio of 3 or higher increased from 5$\%$ to 20$\%$, then dropped to 5$\%$ as the laser pulse energies further increased to 95 $\mu$J per pulse. It was also noticed that the percentage of acetaminophen needle-shaped crystals increased in double-pulse experiments as compared to single-pulse experiments for the same laser energy. Also, it was observed that the length of the crystals decreased with an increase in the double pulse delay time. Finally, it was reported that the aspect ratio distribution of crystals was much wider for laser irradiation experiments, along with a 40$\%$ increase in Form II and Form III crystals as compared to spontaneous nucleation experiments.\par

Yu \etal\cite{10.1021/acs.cgd.0c01476} demonstrated polymorph control of sulfathiazole in a mixture of water and ethanol supersaturated solution by focusing femtosecond laser light (800 nm) through a 5x objective for a period of 1 minute, 500 $\mu$m from the air-liquid interface, as shown in Figure \ref{fig8}i. In other experiments, the laser source was combined with ultrasound, using a 40 kHz, 50 W ultrasonic cleaning machine to check the influence of ultrasound exposure along with laser-induced nucleation. The three main sulfathiazole polymorphs, II, III, and IV, were all observed to crystallize due to laser-induced nucleation. Of the three polymorphs, III is the most stable, making it the dominant polymorph for spontaneous nucleation. At 130$\%$ supersaturation and at low laser power, the III-polymorph was also observed to be the dominant nucleation product, but as the laser power increased from 10 to 80 $\mu$J per pulse, the proportion of metastable form II- and IV-polymorphs was seen to increase.\par

\begin{figure}[tbp]
	\centering
	\includegraphics[width=0.45\textwidth]{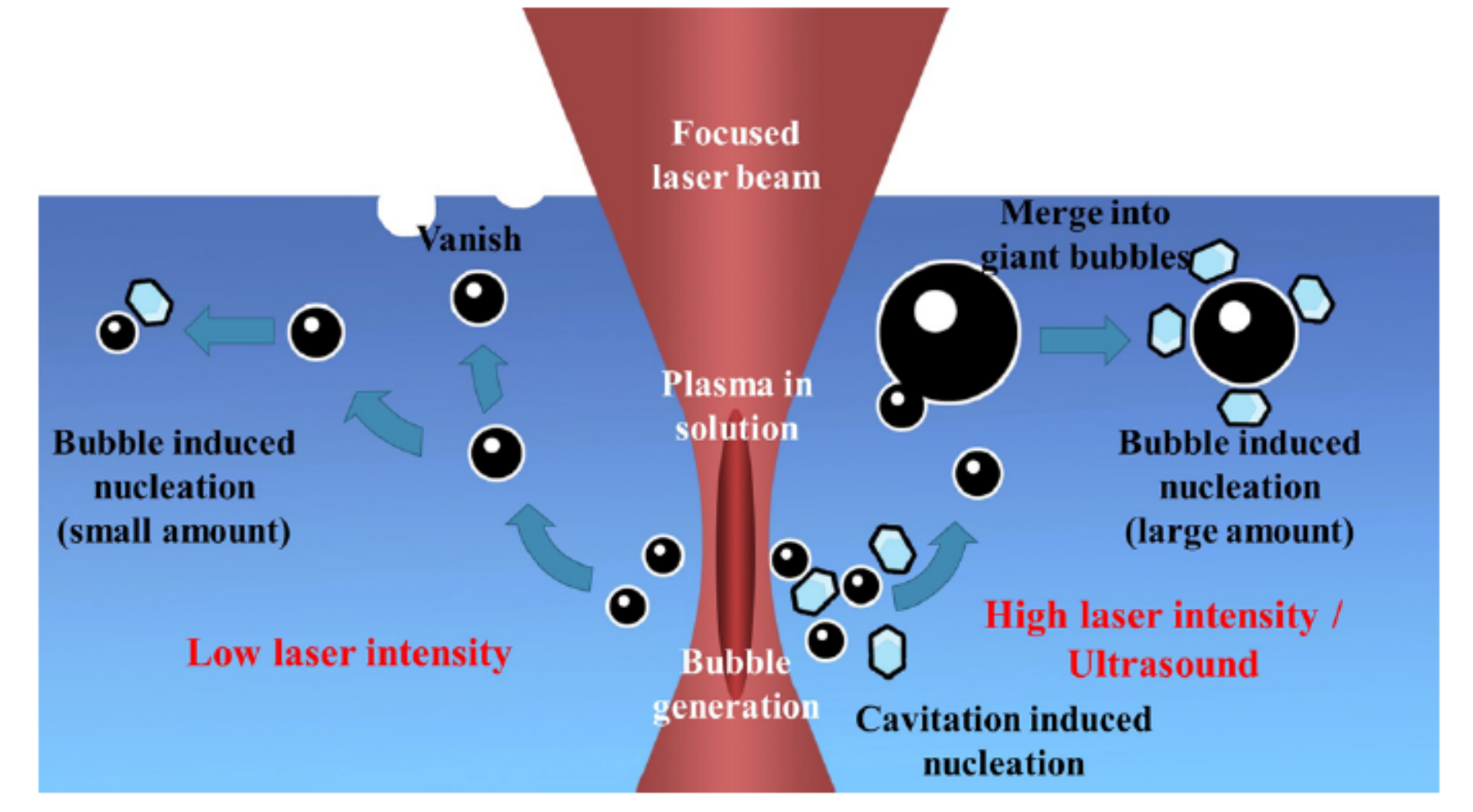}
	\caption{Schematic of sulfathiazole nucleation, comparing low intensity (left) with high intensity (right) femtosecond laser exposure\cite{10.1021/acs.cgd.0c01476}.}
	\label{fig200}
\end{figure}

The plausible mechanism for sulfathiazole crystal formation is shown in Figure \ref{fig200}. When an intense laser pulse is focused at the supersaturated solution, it causes solution cavitation leading to bubble formation. Low laser powers lead to lower number of crystals along with lower number of less-stable polymorphs. This is primarily attributed to the lower quantity of cavitation bubbles formation at low laser powers. However, a high laser power will create a large number of bubbles, which then merge into stable and long-lasting bigger bubbles. At this stage, it has been speculated that the interface of the cavitation bubble decreases the nucleation barrier, increasing the nucleation rate and thereby facilitating the crystallization of the less stable II- and IV-polymorphs in higher quantity as compared to low laser powers. It has also been reported that additional ultrasound will boost the merging process of cavitation bubbles into further larger and more stable long-lasting bubbles, thereby aiding in a faster nucleation process of less stable II- and IV-polymorphs\cite{10.1021/acs.cgd.0c00942}.


\subsection{Summary}

We first summarize common observations in HILIN experiments. We then discuss the consequences of these observations in terms of the proposed mechanism and finally, the open questions are highlighted.

\begin{enumerate}
\item \label{Labc} A broad range of solute-solvent systems, as well as pure compounds, undergo HILIN: Various small organic molecules (DAST in methanol\cite{10.1021/cg049709y}, urea in water\cite{10.1143/JJAP.45.L23}, anthracene in cyclohexane\cite{10.1021/cg060631q}), inorganics ((NaClO$_{3}$\cite{10.1021/acs.cgd.9b00951}, NH$_4$)$_{2}$SO$_{4}$ and KMnO$_4$ in water\cite{10.1021/cg2000014}) and proteins (HEWL\cite{10.1007/s00339-008-4790-x}, glucose isomerase, and ribonuclease H in water) have been crystallized upon exposure to high-intensity laser pulses with a pulse duration ranging from nano- to femtoseconds. Even undercooled pure water (single component) has been reported to crystallize\cite{10.1103/PhysRevLett.99.045701}. 

\item \label{L_PonS_E}  HILIN probability depends on supersaturation and laser pulse energy: Nakamura \etal\cite{10.1021/cg060631q} reported that the nucleation probability measured at a fixed time after laser exposure increases with increasing the supersaturation of anthracene in cyclohexane. Increasing laser pulse energy also increases the measured HILIN probability at a fixed time, while increasing solution temperature is reported to decrease it.


\item \label{Lpulsewidth} HILIN probability dependence of pulse duration and wavelength: Reported HILIN experiments often focus on the cavitation phenomenon and characterizing the resulting crystal properties. Despite a broad range of solute-solvent systems and lasers utilized, reported HILIN experiments rarely report the dependence of nucleation probability on pulse duration and wavelength, unlike NPLIN experiments. 

\item \label{Lthreshold} Minimum HILIN threshold depends on supersaturation: A minimum laser threshold energy is required to generate a cavitation bubble prior to crystal formation in supersaturated solutions. Yoshikawa \etal observed that this threshold depends on supersaturation\cite{10.1143/JJAP.45.L23}.

\item \label{L_optdisturbance} Physical changes observed around the cavitation bubble following laser irradiation: Soare \etal\cite{10.1021/cg2000014} reported ring-shaped optical disturbances in the vicinity of a cavitation bubble that was visible at least up to 30 seconds after laser exposure (Figure \ref{fig9000}). These optical disturbances have been interpreted as high-density regions forming around cavitation bubbles due to the local evaporation of the solvent. Moreover, long-lasting gas bubbles with lifetimes (at least tens of seconds) much longer than the lifetime of cavitation bubbles have been reported by various groups\cite{10.1021/cg060631q, 10.1021/cg2000014,10.1021/acs.cgd.0c01476}.

\item \label{Lpoly} Polymorphic control: Control of polymorphic form crystallizing from solution has been reported for indomethacin in acetonitrile \cite{10.7567/apex.8.045501}, acetaminophen in water\cite{10.1021/acs.cgd.9b00123} and  sulfathiazole in a mixture of water and ethanol supersaturated solution\cite{10.1021/acs.cgd.0c01476}. Experiments of Ikeda \etal\cite{10.7567/apex.8.045501} showed that controlling indomethacin polymorphs necessitates fine control of the laser focal point. Ikeda \etal\cite{10.7567/apex.8.045501} proposed that the focal point of the laser (at air-solution or at the center of solution) alters the local evaporation rate. Shubo \etal\cite{10.1021/acs.cgd.9b00123} observed crystallization of distinct acetaminophen polymorphs depending on the laser power.

\end{enumerate} 

HILIN experiments are characterized by high-intensity (GW/cm$^{2}$) femtosecond or nanosecond laser pulses interacting with supersaturated solutions. This material-light interaction results in an optical breakdown, plasma formation, emission of shockwaves, and thermocavitation along with nucleation of solute from solution. The exact mechanism of how plasma formation, shockwaves, and thermocavitation trigger nucleation and how these phenomena collectively influence the physical properties of emerging crystals is still an open question. The large collection of solute/solvent systems undergoing HILIN (observation \ref{Labc}) provides a good basis for formulating a working mechanism. Yet no information is available on the solute/solvent systems that do not undergo HILIN. Hence practical questions such as ``Can any supersaturated solution undergo HILIN regardless of the chemical identity of the solute?''  and  ``What are the optimum laser parameters (pulse energy, wavelength) required?'' remains unanswered. Answering these questions requires a complete mechanistic understanding explaining the role of plasma formation, shockwaves, and cavitation in crystallization from solution. The mechanism proposed by Vogel \etal\cite{10.1021/cr010379n} emphasizes cavitation bubble formation due to multiphoton absorption and subsequent ionization followed by fast conversion of energy into thermoelastic pressure and formation of heat.  As laser intensity increases, one may expect the formation of plasma, shockwaves, and thermocavitation bubbles to intensify and to have a more pronounced effect on crystallization. In experiments with simple salts, Soare \etal\cite{10.1021/cg2000014} observed fluctuations in the refractive index of material surrounding the laser focus. The authors hypothesized that these optical effects were caused by increased solute concentration around the cavitation bubble.  This observation was reproduced by direct numerical simulations of a laser-induced thermocavitation bubble by Hidman \etal\cite{10.1021/acs.cgd.0c00942}. Moreover, such enhanced concentration regions are also measured with fluorescent labeled lysozyme with high-speed imaging\cite{10.1021/cg060631q}.  These studies are certainly a step in the right direction. However, the simulations by Hidman \etal\cite{10.1021/acs.cgd.0c00942} did not take into account the formation of plasma and shockwaves. Furthermore, the experiments\cite{10.1021/cg2000014,10.1021/cg060631q} did not quantify the concentration and temperature increase during thermocavitation simultaneously making it difficult to calculate exact supersaturation, the driving force behind nucleation and growth. Following only the enhanced solute concentration due to the evaporation hypothesis proposed by Soare \etal\cite{10.1021/cg2000014} indicates that any solution can be crystallized by simply increasing the laser intensity. The limits of this postulate is another open question central to the adaptation of HILIN in scientific and industrial settings.


Nakamura \etal\cite{10.1021/cg060631q} and Yu \etal\cite{10.1021/acs.cgd.0c01476} proposed that the surface of the cavitation bubble could act as a preferential location for nucleation of solute molecules upon laser irradiation. This is a logical reasoning to explain the observations. However, at the early stages of plasma and bubble formation, high temperatures at the focal point are expected as most reported solutions have increasing solubility with increasing temperature. It is not sure that high supersaturations values required for observed nucleation rates will be created at the vicinity of the bubble interface during bubble expansion. 

 At this stage, combining experiments and simulations rationally is essential to answer such questions. For instance, experimentally measurable quantities such as bubble expansion dynamics or local temperature and concentration measurements can guide simulations to pinpoint a working mechanism. A working mechanism should explain all the observations listed above, particularly ones related to physical disturbances around the thermocaviation bubble (observation \ref{L_optdisturbance}). Also, observations \ref{L_PonS_E}, \ref{Lthreshold} are shared between NPLIN and HILIN. Moreover, it should shed light on questions such as ``Is it possible to compare NPLIN and HILIN directly and provide a common theory to explain these phenomena?'' An overarching physical reasoning supported by experiments is required to answer this question. Furthermore, wider implementation of the HILIN hinges on a complete mechanistic understanding.

\section{Laser trapping-induced crystallization (LTIC)} \label{sec:trapping}

\subsection{Phenomenology}
Laser trapping, also known as optical trapping, is a technique that allows the controlled manipulation of particles and molecules using a tightly focused continuous-wave laser beam, which exerts light-induced momentum on matter. The fine manipulation ability can be used, for instance, in the assembly of composite nanostructures\cite{10.1038/nnano.2013.208} and cell sorting\cite{10.1186/s40486-018-0064-3}. It is a versatile laser-induced phenomenon that can be applied to chemistry, physics, and bioscience research as it is contactless, non-photochemical and non-destructive \cite{10.1002/asia.201100105, 10.1103/PhysRevE.73.021911, 10.1016/j.jqsrt.2018.07.013}.
The technique was first demonstrated by Ashkin \etal\cite{10.1364/ol.11.000288} using particles ranging from 25 nm to 10 $\mu m$. Later on, it was seen that besides enabling the manipulation of molecules and molecular clusters, the electric field induced by the laser trapping also promoted the local increase of supersaturation in solution, consequently inducing nucleation events. Thus, the technique shows potential for spatiotemporal control of crystallization\cite{10.1021/cg100830x, 10.1039/c7cp06990a}, crystal growth \cite{10.1246/cl.2009.482} and polymorph control\cite{10.1021/jz900370x, 10.1021/cg300065x, 10.1039/c3pp50276g}.\par

\begin{figure}[tbp]
	\centering
	\includegraphics[width=0.45\textwidth]{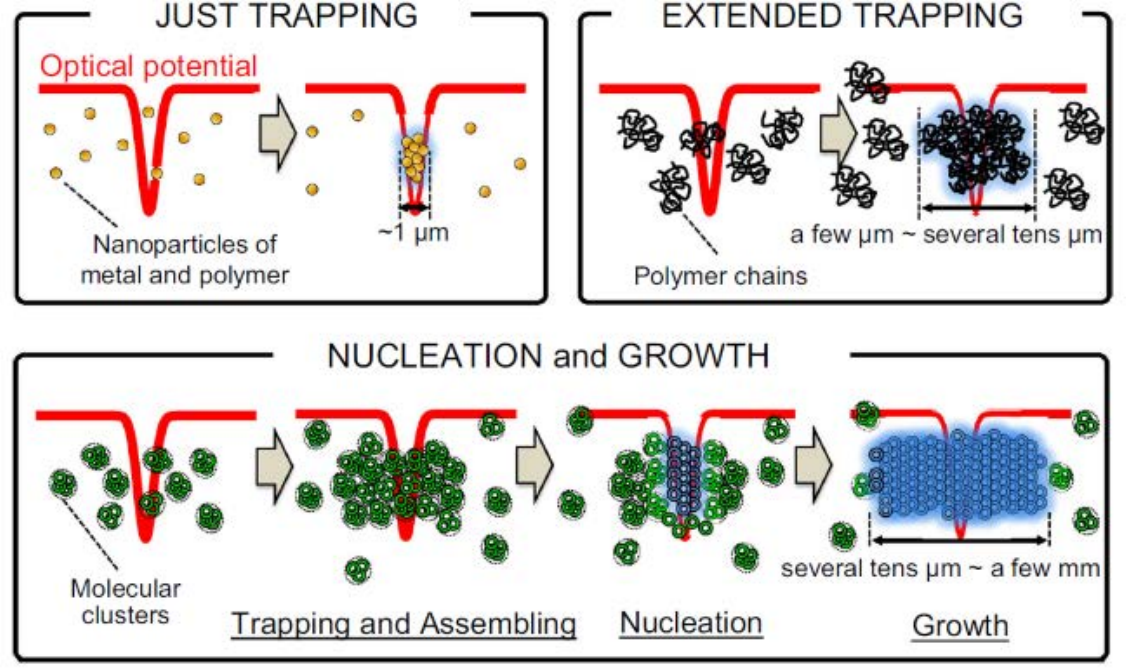}
	\caption{Categorization of the phenomena induced by laser trapping in an optical potential, according to Masuhara \etal\cite{10.1007/s10043-015-0029-1}}
	\label{categories}
\end{figure}

According to Masuhara \etal\cite{10.1007/s10043-015-0029-1}, laser trapping effects can be divided into three categories: `just trapping', `extended trapping', and `nucleation and growth', see Figure \ref{categories}. The first takes place when the optical force or optical potential induces the gathering of nano-objects or the formation of clusters around the focal volume - these clusters disassemble when the laser is turned off. If the gathering of molecules and molecular assemblage grows outside the focal volume, possibly due to the combined effects of laser trapping with heat and mass convection or intermolecular interactions, then `extended trapping' is taking place. Finally, one or both of the previous effects can induce nucleation and, consequently, crystal growth.\par

While both NPLIN and HILIN are performed with pulsed laser beams of short duration (femto- to nanoseconds), for the application of optical trapping, continuous lasers with varying exposure times (from tens of seconds to hours) are used. Long exposures can promote the assembly of solutes, inducing both phase separation and nucleation. Long exposure times using 1064 nm laser have been reported to elevate solution temperature by around 23 K/W in H$_2$O and 2 K/W in D$_2$O \cite{10.1021/jp065156l}. Yuyama et al. reported that temperature elevation also varies with the concentration of the solute\cite{10.1021/cg300065x}. Upon trapping, the authors estimated temperature increases of 6.8, 6.0, and 5.4 K/W for supersaturated, saturated, and unsaturated solutions, respectively. Thus, most reported studies have used D$_2$O as a solvent to suppress local temperature elevation caused by light absorption
(Table \ref{table_laser_trap}).

\subsection{Proposed mechanisms}

Despite being widely studied, the development of an accurate theoretical model for optical trapping is still based on approximations\cite{10.1016/j.jqsrt.2018.07.013}. A plausible mechanism was proposed by Ashkin et al.\cite{10.1364/ol.11.000288}, based on photon forces using two different cases, depending on the size of the object/molecule ($d$) and the wavelength of the incident light ($\lambda$). When $d \gg \lambda$, the photon force is explained using Mie scattering theory (ray optics approximation) - i.e. the refraction of light by the object changes the light's momentum which generates a force. If $d \ll \lambda$, for molecules or clusters foregoing nucleation, the optical trapping of an object can be explained using Rayleigh’s theory (dipole approximation), in which the potential energy $U$ for laser trapping is given by

{\footnotesize \begin{gather}
U=  \frac{(-\alpha|E|^2)}{2}
\end{gather}}

{\footnotesize \begin{gather}
\alpha=4\pi\epsilon_2 r^3\frac{(\frac{n_1}{n_2})^2-1}{(\frac{n_1}{n_2})^2+2}
\end{gather}}

\noindent where 
$E$ is the electric field vector of the incident light. $\alpha$ is the ``polarizability'' of the particle to be trapped, $r$ is the radius of the particle, and $\epsilon_2$ is the dielectric constant of the surrounding medium. $n_1$ and $n_2$ are the refractive indices of the particle and the surrounding medium, respectively. Note that, for the dipole approximation, the electromagnetic fields inside the particle under trapping are assumed to be homogeneous. \par

The optical forces are proportional to particle polarizability, which in turn depends on the particle volume and laser power. Since polarizability decreases with lower particle volume, trapping single molecules is particularly difficult as the optical forces cannot compete with Brownian motion. Therefore, the condition for stable trapping of a particle is $U\gg k_BT$, where $k_B$ and $T$ are Boltzmann's constant and the absolute temperature of the system, respectively\cite{10.1021/acs.jpcc.9b11651, 10.1039/c4cp04008b, 10.1021/acs.jpcc.9b11663}. The dipole gradient force can optically trap a wide range of small molecules as long as its refractive index ($n_1$) is larger than the refractive index of the solution ($n_2$). This includes small metallic nanoparticles ($d<$50 nm) and biological molecules, as previously reported by several authors\cite{10.1364/ol.19.000930, 10.1103/PhysRevE.70.061406, 10.1039/c4cp04008b, 10.1021/ar300161g, 10.1021/ja910010b}. \par

Once the target molecule or cluster is trapped, more molecules start to aggregate, resulting in an increase in volume and stronger radiation pressure. Thus, a nonlinear increase in concentration with time can be seen, which may induce liquid-liquid phase separation (LLPS) or crystal nucleation\cite{10.1021/jp210576k}.
Since the energy condition necessary for stable laser trapping to occur is ($U \gg k_BT$), the radius of the trapped particle should be much larger than the radius of a single molecule. Considering the solutes reported so far - inorganic salts, small organic molecules, and proteins - this implies that only a cluster of molecules or molecular aggregates can be trapped.  \par

In addition to laser trapping forces, aggregation of amino acid molecules (which undertake zwitterionic structure with a large dipole moment in aqueous solutions) is accompanied by Coulombic interactions and hydrogen bonding\cite{10.1021/jp9072334}. \par

Thermodynamically, an explanation using the classical nucleation theory (CNT) relies on the fact that growing nuclei have an energetically unfavorable surface-to-volume ratio that hinders the formation of the crystalline state. As a result, the nucleation depends uniquely on random fluctuations of the cluster, until a critical radius is reached. However, the two-step nucleation theory challenges CNT by claiming the existence of pre-nucleating clusters since nucleation can be enhanced by choosing a concentration near its ‘oiling out’ point. In this condition, the critical concentration fluctuations would give rise to liquid-like crystal droplets with very high concentrations and a high probability of nucleus formation. Under such circumstances, it would be easier for an external force to manipulate concentration.\cite{10.1039/c9sm01297d, 10.1038/s41557-018-0009-8}\par

Walton and Wynne\cite{10.1039/c9sm01297d, 10.1038/s41557-018-0009-8} presented a theoretical model based on the solution model of mixing. The authors showed that the concentration-dependent free energy can be manipulated near a critical point, and that optical tweezers can pull the high-refractive index liquid out of the mixture.
The basic idea behind the model by Walton and Wynne is that the nucleation process can be compared to a chemical reaction where the nucleus is the product and the solute in the supersaturated solution is the reactant. When the focused laser is turned on, the product state has its free energy lowered, even if the nuclei are not yet formed. For this reason, the equilibrium would be displaced towards the products, thus increasing the driving force for nucleation to happen by reducing the energy necessary. Hence, the nucleation (``reaction'') rate is increased. They called this process laser-induced phase separation (LIPS), which later can be followed by nucleation (LIPSAN - laser-induced phase separation and nucleation). The solution model of mixing was used to demonstrate how easily the free energy can get around the critical molar concentration. To account for the effect of the optical tweezing, the total electromagnetic energy stored in the laser volume was incorporated into the total free energy. Experiments performed to validate the model used a 785~nm continuous-wave diode laser (maximum incident power 200~mW in elliptical mode with a beam radius of 2.4 $\mu$m) focused by a 10x objective.
Their model showed that the optical trapping creates a zone of enhanced concentration with a high refractive index within the exposed volume and, consequently, a depletion zone around it, thus indicating phase separation. They also showed that increases in the laser power provided deeper laser traps which enhanced phase separation.

\begin{figure}[tbp]
	\centering
	\includegraphics[width=0.45\textwidth]{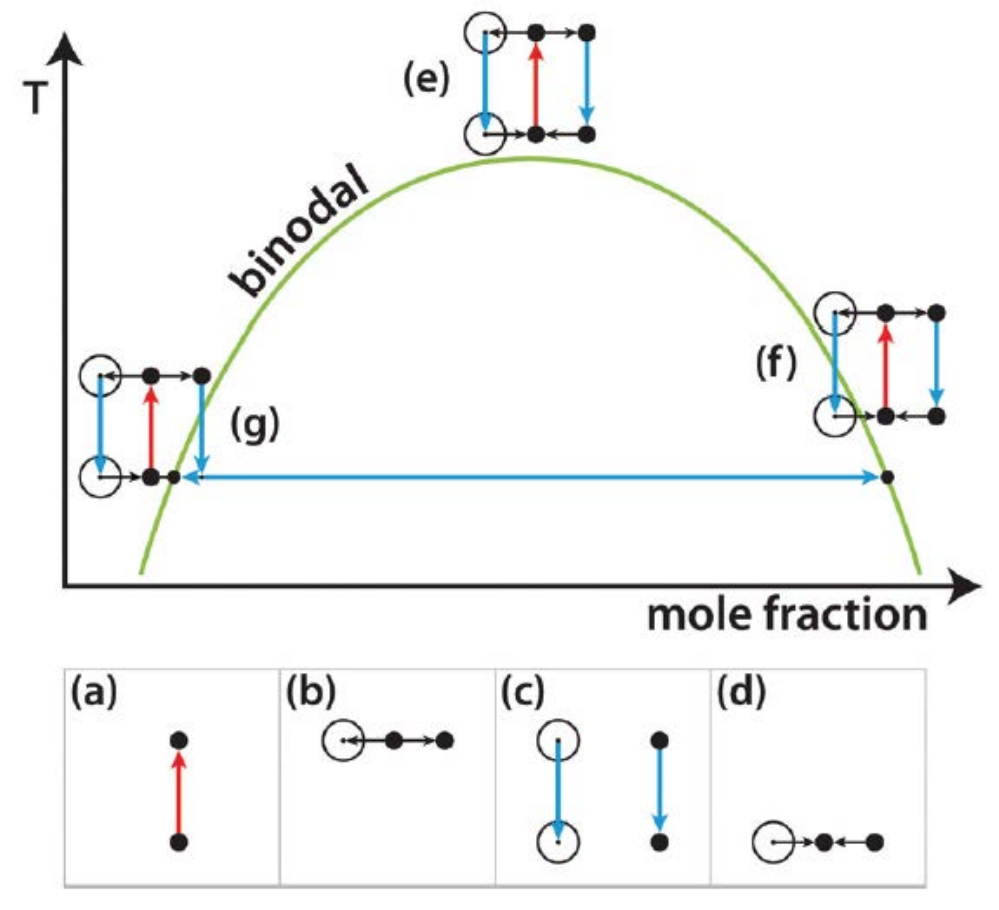}
	\caption{Depiction of a generic liquid-liquid phase diagram displaying the effects of both  heating and laser-induced phase separation\cite{10.1039/c9sm01297d}}
	\label{binodal}
\end{figure}

Considering the schematic phase diagram provided by Walton and Wynne\cite{10.1039/c9sm01297d} (Figure \ref{binodal}), which displays the combined effects of LIPS and heating, three possible scenarios are possible, namely (e), (f), and (g). In the figure, panel (a) represents the laser heating effect; (b) represents the enriched volume surrounded by a depleted region, using the dot and the circle, respectively; (c) the cooling after the laser is switched off and (d) the exposed volume going back to equilibrium. In both situations (f) and (g) depleted and enriched droplets will, due to fast cooling, fall into the unstable region after the laser is off. For the enriched droplet, according to the lever rule, phase separation will increase its concentration while reducing its size, until it eventually fades. Thus, it is the depleted region that triggers phase separation, which explains why the reported nucleation happens outside the laser-exposed volume. For LIPSAN, LIPS will first generate the region enriched with the compound with a high refractive index, and a depletion zone will form around it. Upon switching off the laser, both regions, rich and poor, will rapidly cool. The nuclei will then be formed in the depleted region. The authors have correlated their model with the results described for previous bulk NPLIN studies because the product, in general, shows a higher refractive index than the initial state (some exceptions were also reported, e.g. laser-induced gas bubble nucleation\cite{10.1063/1.3574010} and pure glacial acetic acid nucleation\cite{10.1039/c1cp22774b} in which the refractive index of the products was lower). Most importantly, in this model, nucleation (phase separation, if liquid-liquid) does not rely on pre-existing clusters to be trapped and aggregated but on the laser creating a potential that lowers the free energy of the solid (phase-separated) state.

\subsection{Experimental setups}

\begin{figure}[tbp]
	\centering
	\includegraphics[width=0.45\textwidth]{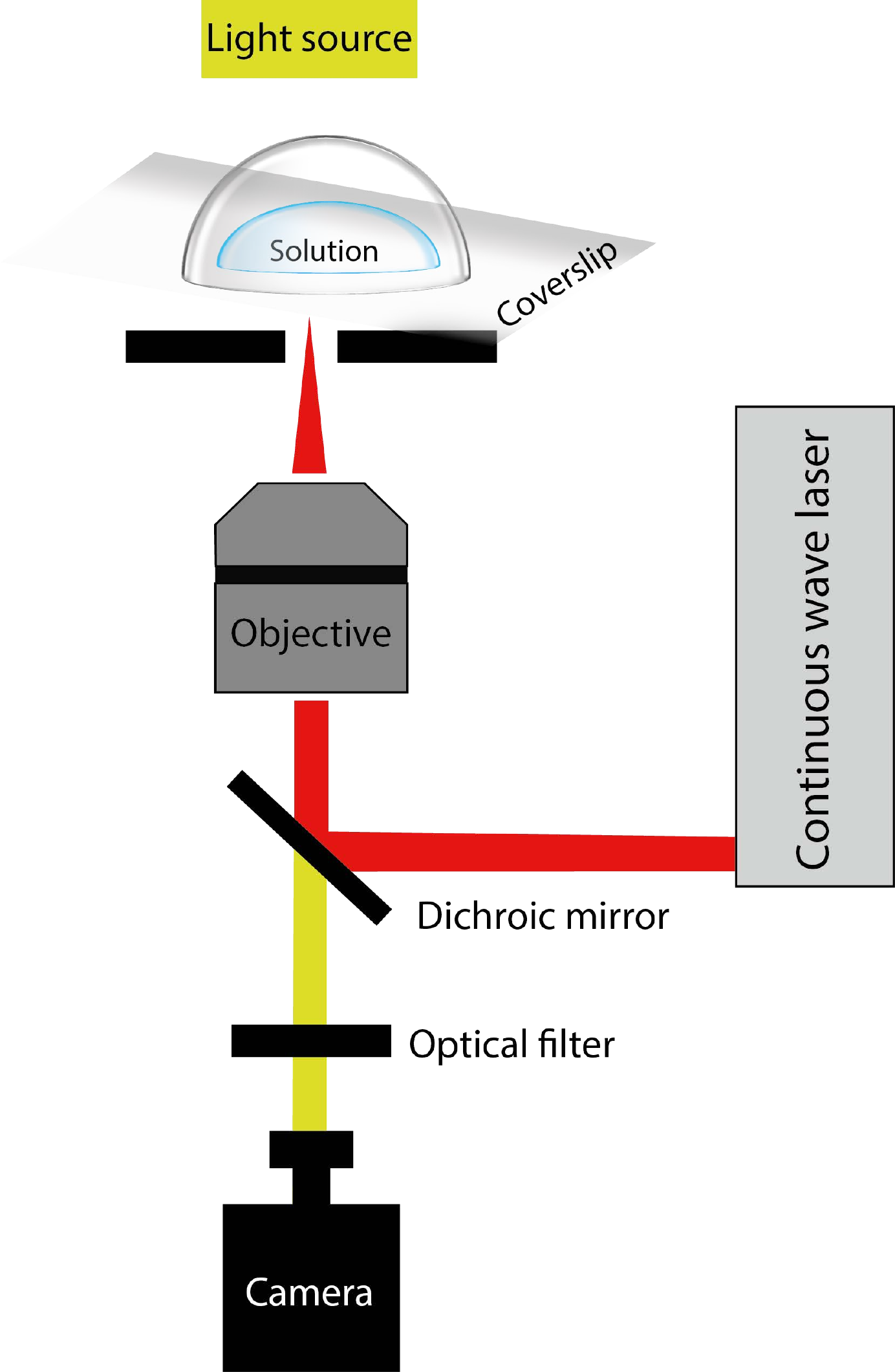}
	\caption{Example of a generic basic laser trapping setup. Adapted from Shilpa et al.\cite{10.1038/s41598-018-34356-0}}
	\label{laser_trapping_setup}
\end{figure}
Laser trapping can be achieved by focusing a laser beam through an objective lens with a high magnification, \ie concentrating the laser power in a very small area (typical diameter of a few microns). To achieve this, a simple experimental setup can be built by coupling the laser source with an optical microscope. This way, the objective can be used to both focus the laser beam and image the sample using a camera. An example of a generic laser trapping setup is shown in Figure~\ref{laser_trapping_setup}. The figure shows a collimated laser beam that passes by a dichroic mirror, which reflects the laser light toward the objective. This allows the light coming from the source in a different direction to reach the camera \cite{10.1038/s41598-018-34356-0}. Typical sample holders are similar to those in Figure~\ref{fig8}a and b, where (b) has been reported to be used both in the absence and presence of preexisting crystals. The type of sample holder used in each study is compiled in Table \ref{table_laser_trap}, along with other experimental conditions.

\subsection{Reported solutions}
\subsubsection{Small organic molecules}

The first works on laser trapping used an intense continuous wave Nd${^{3+}}$-YVO$^4$ laser beam (Neodymium-doped yttrium orthovanadate - 1064 nm) and a 40x objective at the air solution interface of a supersaturated D$_2$O glycine solution\cite{10.1246/cl.2007.1480}. For these experiments, supersaturated glycine solutions were placed over a glass slide covered with a glass dish. Glycine crystals were seen after \SI{16}{s} of laser irradiation which kept growing with further irradiation. The authors observed second harmonic generation (SHG) during the trapping of a growing crystal. This was related to the formation of $\gamma$-glycine polymorph that presents the chiral space group (P31 or P32) required for SHG. The formation of large glycine clusters within the solution was related to the large dipole moments of glycine. After being trapped by the intense focused laser beam due to increased interactions between clusters at the focal spot, these clusters would grow into critical nuclei. \par

Sugiyama et al.\cite{10.1246/cl.2009.482} showed the effect of laser trapping on glycine crystal growth rate. For the experiments, a droplet of glycine solution was placed in a covered glass slide (Figure \ref{fig8}b) until crystals were spontaneously formed. Then, the laser beam was focused near the seed crystal. The crystal growth rate was observed to be enhanced from the seed crystal toward the focal point and completely stopped when the crystal reached the focal spot. When more than one crystal was present, only one crystal showed growth while others dissolved until they eventually vanished. This was explained in terms of Ostwald ripening, using the differences in the surface free energy of the crystals. The results also showed that a local higher concentration of glycine is created by the laser beam not only at the focal point, but also in its surroundings. This is attributed to the suppression of glycine diffusion by the photon pressure at the focus of the beam as the laser traps large liquid-like clusters.\cite{10.1246/cl.2009.482}.\par

Laser trapping-induced crystallization was also reported for undersaturated solutions of glycine in D$_2$O\cite{10.1021/cg100830x}. In this work, solutions were prepared with concentrations of 50 and 68 \% of the solubility limit. The authors considered that glycine forms liquid-like clusters in solution (as reported by Chattopadhyay \etal\cite{10.1021/cg0497344}). Since the solutions are undersaturated, the liquid-like clusters were considered to be locally formed and transient. Exposure times in these experiments were around 400 s. The photon pressure traps some of the clusters at the focal point, promoting increased interaction within these clusters which then leads to nucleation. However, when crystals grow bigger than the focal spot, dissolution takes place as there is no photon pressure to bring together more clusters. Dissolution continues until the crystal reaches a size smaller than the focal spot, where it then grows again and repeats the cycle.\par

In a different study, also using glycine in heavy water, Yuyama \etal\cite{10.1021/jz100266t} describe the formation of a phase-separated glycine droplet of millimeter dimensions (5 mm), much larger than the laser focal point. This was achieved by focusing a 1064 nm CW laser through a 60x objective to the glass-solution interface over \SI{200}{s}. They also reported a reduction in the sample thickness along the laser light (from glass to air-solution interface), from 130 to approximately 5 $\mu$m, along with the formation of a liquid-liquid phase-separated droplet. The lowering of the sample thickness was attributed to the heterogeneous distribution of the surface tension due to temperature gradients. Further laser irradiation led to an increase in surface thickness to approximately 145 $\mu$m and a droplet size increase. These phenomena were attributed to an increase in glycine concentration around the focal point due to the trapping of clusters, which presents a higher refractive index. These droplets, consisting of high concentrations of glycine, were seen to last a few tens of seconds even after the laser was turned off. By focusing the laser at the air-solution interface, glycine crystallization took place. Similar changes in the height of the solution - from 130 to 5 $\mu$m, then increasing to 30 $\mu$m - were also observed when nucleation occurred for approximately \SI{20}{s} of laser irradiation. Since both nucleation and the highly concentrated droplet were seen after the changes in surface height (depression followed by elevation), it was argued that a phase-separated droplet is formed right before crystallization takes place.\par

While the above studies were made using glycine solutions in D$_2$O, Masuhara and others\cite{10.1351/PAC-CON-10-09-32} have investigated laser trapping-induced nucleation of saturated glycine solutions in H$_2$O, using a 1064 nm laser beam focused through a 60x objective. After 30 s of laser irradiation at the air-solution interface, a liquid-like domain was observed. The domain was not stable at the focal point but it was seen to grow continuously and float away, probably due to convection induced by a local increase in temperature. When this growing and floating process stopped, the domain immediately turned into a well-defined crystal. The experiments were repeated 10 times, showing the exact same behavior. It was hypothesized that in water, the high-concentration domain immediately turns into a crystal when the laser is stopped because the temperature increase is suppressed, leading to a supersaturation spike. Time-wise, the temperature elevation affects the rate of formation of the concentrated domain for glycine in H$_2$O so that it is comparable to the nucleation time in the D$_2$O solutions. Therefore, the domain is only visible because the crystallization is hindered by the temperature increase, otherwise, nucleation would take place faster due to the concentration increase.\par

Yuyama \etal\cite{10.1117/12.860241} studied laser trapping of urea in heavy water for different saturations (0.28 and 1.36) at room temperature, with a 1064 nm continuous-wave linearly polarized laser beam at 1.1 W power. By focusing the laser at the glass-solution interface, for the undersaturated solution, the formation of concentrated droplets was observed, as previously reported for glycine\cite{10.1246/cl.2009.482, 10.1021/cg100830x}. However, no crystallization was observed, regardless of the focal point position. This was explained by the increase in temperature that sharply reduces the supersaturation in urea-heavy water solutions. For urea, only crystal growth and dissolution were observed when the laser focal point was close to the surface of an existing crystal.\par    

In their work, Tsuboi \etal\cite{10.1021/jp9072334} used a CW Nd${^{3+}}$:YAG laser ($\lambda$ = 1064 nm) to show that the photon forces of a near-infrared (NIR) focused laser beam are able to manipulate molecular clusters of several amino acids via optical trapping. They described the trapping of glycine, D,L-proline, D,L-serine, and L-arginine in D$_2$O with different laser power thresholds, based on the compound. The laser power was varied from 0.6 W to 2.0 W with irradiation times up to 30 min.\par

Yuyama et al.\cite{10.1021/jz401122v} showed laser trapping-induced nucleation of L-phenylalanine using a continuous wave Nd$^{3+}$:YVO$_4$ laser focused through a 60x objective even for an undersaturated aqueous solution at \SI{25}{\celsius}. Crystals were seen after 170 s of laser exposure (1.1 W). This phenomenon was explained by the increase of local supersaturation due to laser trapping of liquid-like clusters. Plate-like crystals were formed, which corresponds to the anhydrous polymorph of L-phenylalanine. This polymorph is only expected to form at temperatures higher than \SI{37}{\celsius}, thus its formation confirmed the temperature increase due to the laser trapping - which the authors estimated to be around \SI{25}{\celsius}. Further continuation of the laser trapping led to crystal growth. However, contrary to what was observed by Rungsimanon \etal\cite{10.1021/cg100830x}, L-phenylalanine always showed continuous growth and no dissolution. This was hypothesized to be due to either heating, optical or electrostatic effects. The heating effect would create convection and constantly supply molecules/clusters to the growing surface. The optical effect would be promoted by the propagation of incident light through the crystal, generating optical potential at the growing surface. Lastly, the electrostatic effect would indicate that the laser locally charged the crystal surface, attracting loose clusters from the solution to the surface of the crystal. To investigate which of the hypothesized effects was responsible for the phenomenon, an experiment was performed where polystyrene particles (1$\mu $m) were added to the solution right after crystal nucleation. With the laser off, these particles were observed to exhibit Brownian motion around the crystal. When the laser was turned on, these particles immediately aligned themselves toward the crystal edges. The immediacy of the response and the observation of some optical trails within the crystals pointed to the optical effect. Thus, trapping was considered to be achieved by light propagation inside the crystal, confirmed by crystal growth, and even tunable by changing laser power. Optical trails were attributed to the optical anisotropy of the crystal.\par

In the same context, Yuyama \etal\cite{10.1021/acs.cgd.5b01505} have also measured the 2D growth rate of the single plate-like anhydrous crystal of L-phenylalanine, which was seen to directly depend on the laser power. In a later paper from the same group\cite{10.7567/1882-0786/ab4a9e}, L-phenylalanine three-dimensional crystal growth was studied, aiming at the changes in the thickness of the nucleated crystals, by applying reflection imaging - introduction of white light along with the optical trap so that 
constructive interference of the white light at the upper and lower faces of the crystal can be observed. The growth rate in the direction related to thickness was seen to be constant during laser trapping, while it decreased in the direction parallel to the solution surface. This behavior was explained by the thickness growth being directly promoted by the optical force from the laser, while in the other directions it is promoted by the optical force generated by the light propagation within the crystal.\par

In an attempt to couple the effects of enhanced laser trapping and laser ablation, Wu \etal\cite{10.35848/1347-4065} reported the nucleation and polymorph conversion of L-phenylalanine by using femtosecond laser trapping (800 nm, 100 fs, 80 MHz). Both undersaturated and supersaturated solutions were tested by focusing the laser at the air-solution interface. In unsaturated solutions, no immediate change was seen at the start of irradiation. After 3 minutes, a bright light emission, indicating optical breakdown and the formation of several bubbles at the focal point, was seen. After 12 minutes, plate-like crystals were observed, compatible with the anhydrous polymorph of L-phenylalanine, which continuously grew upon further irradiation. Interestingly, after some time, whisker-like crystals (monohydrate) started to form at the surface of the plate-like growing crystal. Both crystals dissolved when the laser was turned off. The threshold for crystallization under these conditions was identified to be 150 – 300 mW, which is estimated to be 3-4 times lower than with continuous lasers (1064 nm) with a 2 times shorter crystallization time. The crystallization was attributed to a combined effect of optical pressure and cavitation bubbles. The supersaturated solutions were used to prepare spontaneous whisker-like crystals which were then irradiated with the femtosecond laser. After crystal formation, the remaining solution was considered to be saturated and the laser was now focused near the glass-solution interface, at a densely populated whisker-like crystal region. A single bubble (a few tens of microns in size) was seen to form immediately after irradiation. Near the surface of the bubble, plate-like L-phenylalanine crystals were observed, despite not being the most stable polymorph at room temperature. Laser ablation of the mother crystal by the femtosecond laser was discarded as a cause of polymorph change after experiments. However, cavitation bubbles caused by the ablation of the solution surrounding the crystal is still a likely mechanism, where the increase in the concentration at the bubble surface determined the polymorphic outcome. \par

In another study involving chiral amino acid crystallization, Yuyama \etal \cite{10.1117/12.929381} reported  using a 1064 nm CW Nd$^{3+}$-YVO$_4$ laser to irradiate a supersaturated sample of L-alanine for 30 min on a focal spot at the air-solution interface. Nucleation was proven to be spatiotemporally controlled, as it was consistently seen within 10 minutes at the focal point. When a right-handed circularly polarized beam was used, faster crystallization was observed compared to left-handed circular polarization at laser power around 1.3 W, which was attributed to a higher increase in supersaturation. In laser power 1.0 W, an optical torque was induced in the direction of the beam polarization, as a rotation on the nucleated crystal in the same direction was observed. \par

Laser trapping-induced nucleation was also reported to selectively nucleate and grow organolead halide perovskite (methylammonium lead trihalide – MAPbBr$_3$) crystals from undersaturated precursor solutions\cite{10.1002/anie.201806079}. Experiments were performed using MAX/PbX2 (where X = Br, Cl, and I) precursors in N,N-dimethylformamide (DMF). Concentrations were varied from 0.1 to 1.3 M while a 1064 nm laser beam was applied, in power ranging from 0.1 to 0.6 W. Thresholds for inducing crystallization were identified as 1.2 M and 0.2 W, under trapping for around 100 s at \SI{18}{\celsius}. The crystal grew after nucleation while the laser was on. Once the laser was switched off, dissolution took place. Optical trapping was then cleverly combined with the retrograde solubility of the compound by heating the sample holder (\SI{100}{\celsius}). Not only dissolution was avoided but the nucleated crystal kept on growing, even after the laser was turned off. It was suggested that the precursors formed complexes in solution, which were trapped at the focal spot, along with some solvent molecules. Since the refractive index of the precursor is higher than that of the solvent, the concentration increased until supersaturation was reached, inducing nucleus formation and excluding solvent molecules from the crystal lattice. Slow crystal growth was reported for MAPbBr$_3$ and MAPbCl$_3$ while explosive growth was observed for MAPbI$_3$, related to two-photon absorption where the excess energy dissipation caused temperature increase of the surrounding solution. As the first two perovskite crystals do not significantly absorb laser light at 800-1064 nm, no temperature increase was expected whilst for the iodide-based solution light absorption was expected at around 820 nm, thus leading to a higher temperature in the surrounding area. Due to the inverse solubility of perovskite, this leads to fast supersaturation increase promoting explosive growth. Results suggest the use of laser trapping-induced nucleation to produce single crystals of perovskite with tailored optical and electronic properties. \par

Recently, Liao and Wynne\cite{10.1021/jacs.1c11154} reported experiments using high-power lasers (1040 nm, 1Hz, 4 ps pulse, and $\leq$8 W; CW 1064 nm, $\leq$10 W; and a 532 nm 0.05 W beam for Raman excitation) focused using an objective (NA = 0.7). 10 $\mu$L solutions were irradiated in a covered glass plate. The authors monitored the Raman shifts for droplets under evaporation in time (homogeneous nucleation experiments), to track crystal formation, and show the clear distinction between flattened Raman shifts to clear peaks indicating glycine $\gamma$-polymorph. Upon laser irradiation (20-30 min) of fresh glycine solutions, authors reported that no nucleation was induced. Instead, they found evidence of the formation of amorphous metastable microscopic particles ($\approx$500 nm) said to be on-path or off-path intermediates from solution to crystal. When placed under the laser focus, these amorphous particles immediately triggered the formation of crystals that continued growing even after the laser was off (indicating that the solution had become supersaturated). It was noted that laser caused the formation of a few crystals in separated nucleation events, mostly needle-like, which were later fully replaced by a prismatic crystal, evidencing a polymorph shift from $\beta$-glycine to $\alpha$-glycine. After these observations, the solutions were either aged overnight or irradiated with the 1040 nm laser for 30 min to initiate the formation of the metastable particles. Once formed, these particles were trapped with the 532 nm laser to investigate their development while monitoring Raman shifts. They showed that the trapped particles were (meta)stable in solution for over 200 s, then it expanded to a larger cluster which finally nucleated into a crystal. The Raman shifts showed a number of different and clear spectroscopic stages, from trapping, expansion, and nucleation, but the peaks fluctuate over time in intensity before becoming stable at the corresponding peaks for the produced polymorph. The observations in these experiments are in clear agreement with the previously reported need for aging before nucleation experiments with glycine solutions - and other solutions of small organic molecules and proteins. Moreover, the authors have related the fluctuations in the Raman spectra with the formation and dissolution of clusters explained by the CNT, which added by the heating effects of the laser could be enhancing the reorganization of the molecules and thus facilitating nucleation. Authors also found some support for the Optical Kerr effect due to a preferred formation of $\gamma$-glycine in irradiated experiments while non-irradiated solutions formed mostly $\alpha$-glycine. \par

In parallel, Urquidi \etal\cite{10.1073/pnas.2122990119} have used LTIC coupled with single crystal nucleation spectroscopy (SCNS) of glycine in water solutions. A 532 nm depolarized CW laser was used simultaneously for the Raman excitation and the optical trapping, focused at the air-solution interface through an objective (60 X, NA = 1.2) with laser power $>$1 W. Authors reported tracking Raman spectra evolution with 46 ms resolution for the glycine formation in room temperature. 128 experiments were performed with fresh glycine solutions (no aging was reported) and the nucleation times were seen to vary from minutes to hours. Blurred objects, assumed to be regions with an increased local concentration of glycine, were visible under the microscope, until a crystal was clearly seen. The Raman spectra were seen to vary over time, coinciding with the appearance of the blurred spots and the crystals. A nonnegative matrix factorization (NMF) algorithm was used to analyze the evolution of the Raman spectra. No formation of $\gamma$-polymorph was seen, but a short-lived Raman spectrum for $\beta$-glycine was observed, which finally became $\alpha$-glycine. Through spectral evolution analysis, the authors found fluctuations in the Raman shifts, similar to those reported by Liao and Wynne\cite{10.1021/jacs.1c11154}, that implies the formation of pre-nucleation clusters that convert into crystals. Hence, the evidence points to glycine following a nonclassical nucleation pathway, even though CNT could not be completely ruled out, where laser trapping enhances the increase of the local concentration by trapping these pre-nucleation clusters. The experimental Raman spectra were compared with those obtained from molecular dynamics (MD) simulation results. The Raman shifts from MD show that glycine forms linear networks through hydrogen bonding, which are responsible for the short-lived $\beta$ polymorph seen in the experiments. \par 

Urquidi \etal reported a temperature increase of $\approx$20 mK W$^{-1}$ in glycine-water solution under laser irradiation, which contradicts previous literature. This was brought to light by Liao and Wynne\cite{10.1073/pnas.2207173119} in a subsequent paper, where they estimated a solution heating around the focal point of 400 K using the conditions reported by Urquidi \textit{et al.}, Liao and Wynne\cite{10.1073/pnas.2207173119} also questioned the alleged trapping of aggregates of glycine-water claiming that the radius of those ($\geq$ 1 $\mu$m) is not enough for the trapping to overcome Brownian motion. Despite the criticism, Liao and Wynne\cite{10.1073/pnas.2207173119} proposed that Urquidi \etal observations pointed to the same phenomena they observed in their previous paper\cite{10.1021/jacs.1c11154} and agreed with their observations for the amorphous particles in MD.    

\subsubsection{Proteins}

Tsuboi \etal\cite{10.1143/JJAP.46.L1234} used a CW laser (Nd$^{3+}$:YAG $\lambda$ = 1064 nm) with an oil-immersion objective lens (NA = 1.3) and a laser power threshold of 0.25 W for trapping and inducing nucleation of hen egg lysozyme (HEWL). A CW Ar$^+$ laser ($\lambda$ = 488 nm) was added to provide excitation for Raman scattering. Samples were exposed to the laser for 1-2 hours before a lysozyme crystal was identified by Raman spectroscopy. The authors found an enhanced nucleation probability for the laser-exposed samples. Laser trapping was argued to overcome protein Brownian motion and gather clusters (oligomers) larger than 20 nm comprised of more than ten lysozyme molecules. \par

Tu and collaborators\cite{10.1021/cg401065h, 10.1021/cg501860k} have also investigated the effects of laser trapping on the crystallization behavior of solutions of HEWL in D$_2$O, focusing on crystal growth. First, they positioned the focus 10 $\mu$m away from a pre-existing HEWL crystal and studied temporal changes in growth rates of crystal faces\cite{10.1021/cg401065h}. Growth was reported to be well-controlled by laser trapping, yet different from spontaneous crystal growth. An extension of the dense cluster region is promoted due to convection and mass transfer, also induced by laser trapping. Within the extended trapping area, the concentration and rigidity of the clusters varied significantly from that of the homogeneous solutions, which enhanced crystal growth significantly. Moreover, the authors claimed that the laser trapping-induced a time-dependent variation in crystal morphology, compared to the spontaneous one, particularly regarding the growth rate of the \{110\} face, which showed a large decrease or increase according to the irradiation time. In their further work\cite{10.1021/cg501860k}, Tu \etal systematically studied the influence of important parameters - laser power, polarization, and focus point - on the growth behavior of HEWL crystals. Crystal growth is shown to be dependent on two kinds of cluster-concentrated domains with different concentrations, rigidity, and order. At first, trapping led to the formation of the so-called Domain 1 and, further irradiation led to Domain 2. Rigidity and ordering are both higher for Domain 1 than for 2 due to the concentration of clusters, which in its turn is higher for Domain 2 than for 1. Thus, the existence of Domains 1 and 2, respectively, inhibits or enhances crystal growth when compared to spontaneous phenomena. Different chemical and physical properties for the domains are reported to be obtained by tuning laser parameters - power, polarization, and focal position.\par

HEWL was also studied by Yuyama et al.\cite{10.1039/c7cp06990a} regarding the nucleation position and the morphological characteristics of these crystals. The authors showed that when irradiating the solution for 1 hour with a 1.1 W laser power, HEWL crystals were seen 30 minutes after the laser was switched off but not during trapping. Crystals were seen in similar numbers, and considerably larger sizes, as the spontaneous nucleation that happened only after 4-6 hours. Even so, nucleation time decreased by 20 times with the use of laser trapping. The crystals were reported to nucleate a few millimeters outside of the focal volume, instead of forming everywhere in the solution. Lower laser power (0.5 W) was reported to have no effect on nucleation time or location. Changes in the external shape of the crystals were also observed as an effect of laser trapping. Square-shaped crystals were seen to increase and become predominant with increasing laser power (0, 0.5, 0.8, and 1.0 W), whilst spontaneous crystals were preferentially hexagonal or with a tilted shape.

\subsubsection{Inorganic molecules}

Laser trapping-induced nucleation of KCl crystals was reported by Cheng \etal\cite{10.1021/acs.jpcc.9b11651} to happen in three different morphologies, depending on the laser power and polarization used. In their work, the laser power was varied from 0.4 to 1.4 W in linearly and circularly polarized (respectively, LP and CP) laser beams, focused with a 60x objective, and the samples were irradiated for a maximum of 30 minutes. Only needle-like crystals were seen under low laser intensities, while increases in laser intensities increased the probability of cubic crystals. Also, different light polarization generated different KCl morphologies – linearly polarized light led to needle-shaped, rectangular, and cubic crystals, whilst circularly polarized gave only needle and rectangular crystals, but never cubic crystals.
This points out that the light polarization dictates the physical state of the dense clusters, which would enable polymorph control: i) the same laser power gave lower nucleation probabilities for CP than LP; ii) no cubic crystals were formed under CP; and iii) only under CP, at 0.9 W, needle-like KCl crystals were formed. CP was also reported to induce rotation of the generated crystal and its dissolution, after which the crystal would recrystallize, yet in a smaller size, repeatedly with further laser exposure. \par

Cheng \etal\cite{10.1021/acs.jpcc.9b11651} confirmed the critical role of surfaces or interfaces for the optical trapping-induced nucleation experiments: crystallization only happens if the laser is focused at the air/solution interface. They explained that gradient forces, as well as scattering forces, generated by the laser beam are proportional to the 3$^{rd}$ and 6$^{th}$ power of the targeted object size, respectively. Considering the target is a cluster, the scattering forces are dominant with an increase in cluster size upon laser irradiation at the solid-liquid interface. Thus, clusters are de-trapped from the focal point, inhibiting nucleation. On the other hand, at the air/solution interface, both forces lead to an increase in the concentration which increases cluster size up to nucleation.

\subsection{Potential for controlling polymorphic form}

\begin{figure}[tbp]
	\centering
	\includegraphics[width=0.45\textwidth]{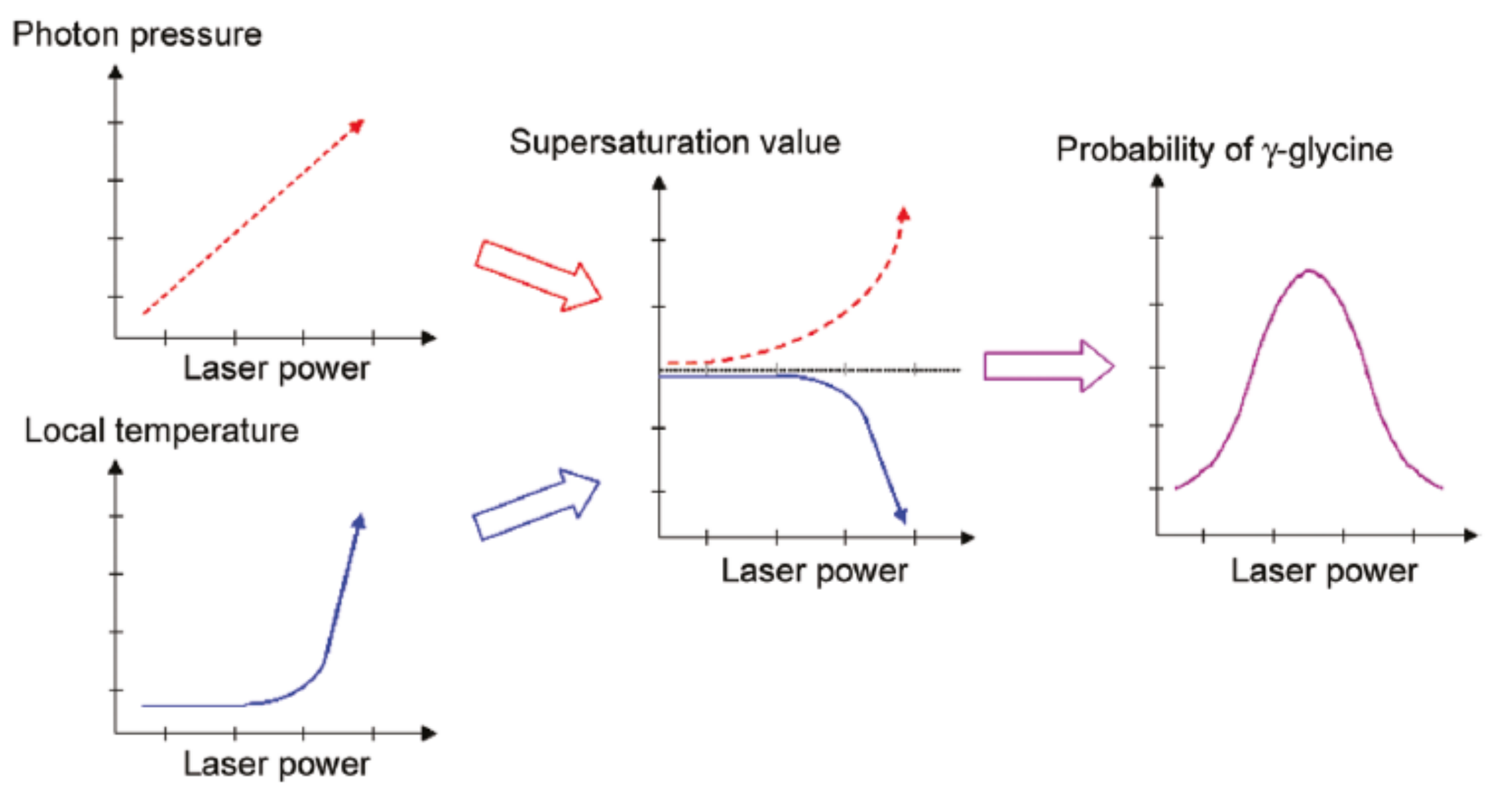}
	\caption{Effect of laser power on photon pressure and supersaturation and its influence on $\gamma$ glycine nucleation probability\cite{10.1021/jz900370x}.}
	\label{figX}
\end{figure}

Polymorph control has been reported as one of the abilities of laser trapping-induced crystallization. Rungsimanon \etal\cite{10.1021/jz900370x, 10.1021/cg100830x} reported the first study on polymorph control by photon pressure of a CW-NIR linearly polarized laser beam focused at the air-solution interface through a 60x objective using 1064~nm wavelength. The authors prepared supersaturated solutions of D$_2$O and glycine. The crystals produced were analyzed with FTIR and Single Crystal X-Ray Diffraction - SCXRD. They investigated 10 samples at each laser power - ranging from 0.8 to 1.4~W. For laser powers 0.8 and 1.0~W, only $\alpha$-glycine was obtained. With increasing laser power, the probability of getting $\gamma$-glycine gradually increased to a maximum of 40~\% at 1.3~W. The formation of the $\gamma$ polymorph was explained by laser power-dependent effects. Enhanced photon pressure around the focal spot results in increased supersaturation - supersaturation shows non-linear increases with the laser power - thus increasing the probability of $\gamma$-glycine nucleation. On the other hand, an increase in photon pressure also leads to local temperature elevation, which reduces the supersaturation value. This reduction of the supersaturation, above a threshold, would lower $\gamma$ polymorph nucleation probability, giving a bell-shaped nucleation probability curve, as shown in Figure~\ref{figX}.\par

\begin{figure}[tbp]
	\centering
	\includegraphics[width=0.5\textwidth]{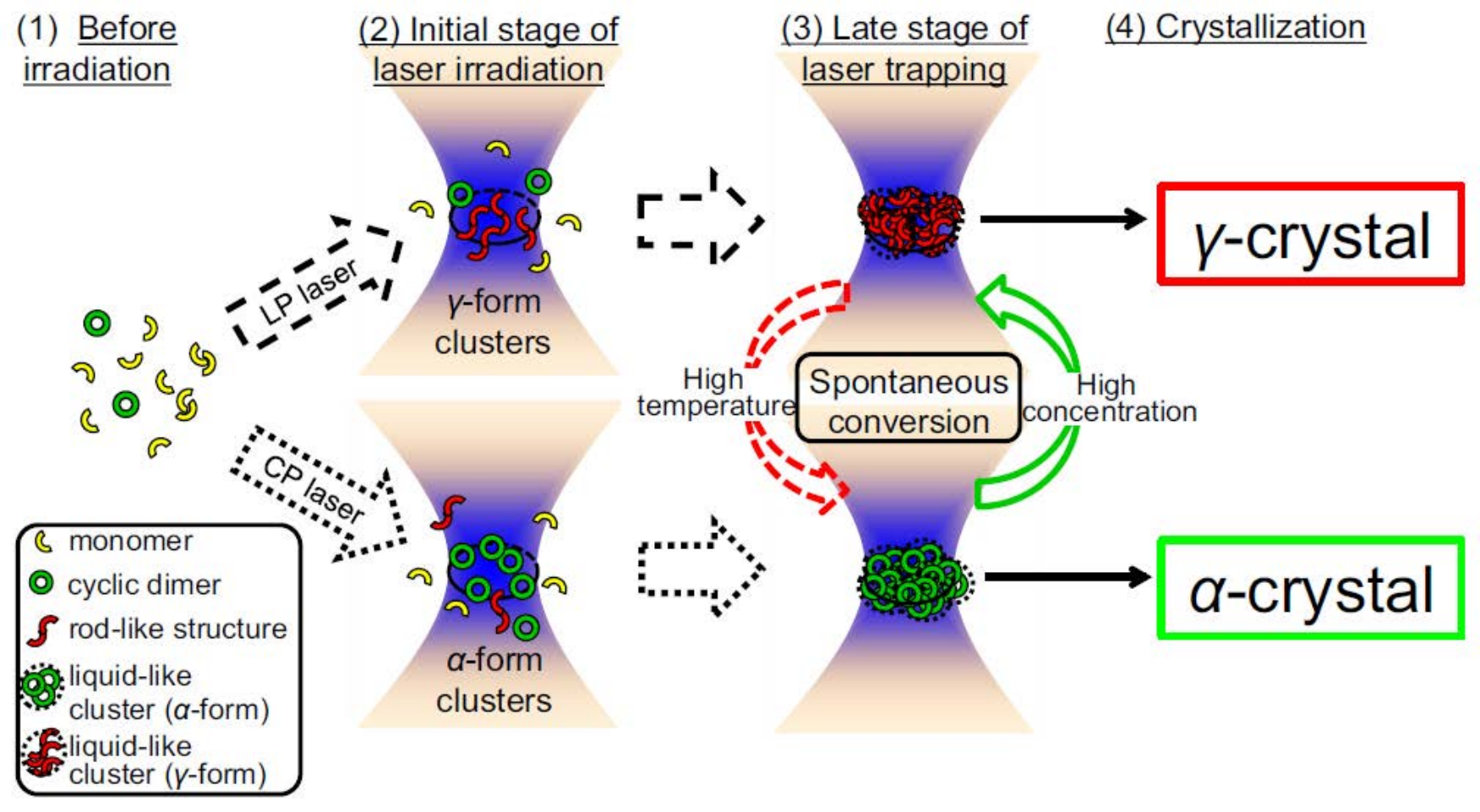}
	\caption{Effect of laser polarization on glycine dimers organization and polymorphic outcome.\cite{10.1007/s10043-015-0029-1}}
	\label{polymorph_selection_yuyama}
\end{figure}

The formation of $\alpha$- or $\gamma$-glycine was reported to depend strongly on laser power, polarization, and initial supersaturation. Overall, in saturated solutions, the formation of $\gamma$-glycine showed higher probability while the $\alpha$-form was obtained for supersaturated solutions. This was related to the metastability of the supersaturated solution, which would kinetically favor the formation of the least thermodynamically stable crystal $\alpha$-glycine. For both polymorphs, bell-shaped nucleation probability curves as a function of laser power were observed due to the competition between supersaturation increase by laser trapping and temperature increase (which lowers supersaturation). The highest nucleation probabilities – compared regarding $\gamma$-glycine – for CP lies at 1.1~W whilst for LP it lies at 1.3~W. In explaining polarization-dependent polymorph formation, the authors compared their results with the NPLIN experiments by Garetz \etal\cite{10.1021/cg050460}, which were also performed with glycine, in which a CP pulsed laser would form disk-like ($\alpha$-glycine) and LP would form rod-like ($\gamma$-glycine) polarizable clusters. Changing polarization and supersaturation resulted in the formation of different polymorphs, suggesting that polymorph formation depends on laser polarization and supersaturation. Yuyama \etal\cite{10.1021/cg300065x} also reported that CP laser efficiently led to $\alpha$-disk-like clusters and consequently $\alpha$-glycine, while LP gave mostly the $\gamma$-polymorph in saturated and supersaturated solutions (see Figure~\ref{polymorph_selection_yuyama}).
For undersaturated glycine solutions, the authors reported that LP laser irradiation at 1.3~W laser power drastically increased the nucleation probabilities to 90~\% of $\gamma$-polymorph crystallization, while CP showed no crystallization of $\gamma$-polymorph at all. Since solutions are undersaturated, the spontaneous formation of clusters in solutions is unlikely. Therefore, the authors assumed that the cluster formation was possible due to laser trapping effects, which consequently affected the formation of rod-like ($\gamma$ form) and disk-like ($\alpha$ form) clusters in CP and LP lasers, respectively, up to 1.3~W. Over 1.3~W, the formation of $\gamma$-glycine under CP exposure was related to the spontaneous conversion from $\alpha$ to $\gamma$, rather than a change in the molecular arrangement provided by the laser. \par

It is noteworthy that, in terms of polymorphs, Masuhara \etal\cite{10.1351/PAC-CON-10-09-32} have shown the nucleation and growth of the least stable $\beta$-phase at room temperature (although with only 10~\% probability). These results were reported for saturated solutions of glycine in H$_2$O at laser trapping powers of 1.1 and 1.2~W (1064~nm). Under the same conditions, but for glycine in D$_2$O, no $\beta$-form was found. This shows that not only the laser power but also the choice of solvent impacts the polymorph formed. \par

Selective crystallization of L-phenylalanine pseudopolymorphs from both water (H$_2$O) and heavy water (D$_2$O) solutions was also reported. For L-phenylalanine, monohydrate is the most stable form around \SI{25}{\celsius} but stability reverses to the anhydrous form around \SI{36}{\celsius}. Yuyama \etal\cite{10.1039/c3pp50276g} have shown that different pseudopolymorphic forms, namely anhydrous and monohydrate, were obtained by laser trapping of supersaturated solutions with H$_2$O and D$_2$O, respectively. They have considered the stability of each form under given temperatures and concluded that different crystals were obtained in water due to laser heating of the solution, which is known to have a lower effect on D$_2$O solutions. The same research group also reported laser trapping to control L-phenylalanine pseudopolymorphs by a judicious choice of laser power, polarization, and solution concentration. Wu \etal\cite{10.1021/acs.cgd.8b00796} reported, for an undersaturated D$_2$O-L-Phe solution, the formation of a single crystal of anhydrous L-Phe whilst in supersaturated solutions, multiple monohydrate crystals were generated in an area around the focal point. For saturated solutions, the authors observed that with increasing laser power, anhydrous and monohydrate L-Phe crystals were favored by, respectively, linearly and circularly polarized laser light. The proportions of the favored pseudopolymorph showed an increase with increases in laser power. \par

Niinomi et al.\cite{10.1021/acs.cgd.7b01116} reported the possibility of nucleating and controlling the growth as well as dissolution of a single short-lived metastable sodium chlorate crystal by means of optical trapping. Experiments were done with hemispheric microdroplets, ranging from 30 to 60~$\mu$m, confined between two glass slides and exposed to a circularly polarized laser beam (532~nm) focused through a 60x objective at the air-solution interface. Nucleation was seen after 10 to 40 minutes. The metastable form of the nucleated crystal was evidenced by its morphology - parallelogram-shaped with birefringence (only the metastable form of sodium chlorate exhibits birefringence) - and was confirmed via Raman spectroscopy. The crystal growth was seen to continue even after the crystal escaped the focal point until the crystal size exceeded the edges of the microdroplet. Contrary to previous observations, crystal growth also continued after the focus was shifted to the inside of the solution or glass-solution interface. Crystal dissolution was observed when the laser was switched off. The continued growth of the crystal out of the focal point was attributed to the trapping of crystalline clusters by surface optical potential due to the propagation of light within the nucleated crystal. In further work, Niinomi \etal\cite{10.1021/acs.cgd.9b00600} reported in-situ observation of step formation/dissolution and wetting transition on the metastable form of sodium chlorate. By switching the laser trap on and off, several two-dimensional micron-sized islands were formed and could be controlled by the focal spot location. This showed the influence of the electrical field gradient force on the surface kinetics. \par

In recent work, Shih \etal\cite{10.1021/acs.cgd.1c00822} reported a novel polymorphism of $\beta$-cyclodextrine ($\beta$-CD), obtained through optical trapping laser-induced crystallization using 1064 nm continuous wave laser beam. The authors first found evidence of the new polymorph when prism- and plate-like crystals were formed under the same conditions from undersaturated solutions of $\beta$-CD and D$_2$O. Even though X-ray diffraction of the plate-like crystals was not possible, major differences were found on the Raman spectra of the prism and plate crystals, where both X-ray and Raman for the prism crystal were compatible with literature data for $\beta$-CD.12H$_2$O. It is noteworthy that upon continuous laser irradiation, a prism crystal always formed on the surface of the plate crystal; the former continuously grew while the latter showed no changes as long as the laser was on. When the laser was turned off, the prism crystal immediately dissolved while the plate-like crystal remained intact longer, despite the solution being undersaturated, which showed higher thermodynamic stability of the latter. Results were interpreted as an indication that the formation of such a polymorph was only possible due to the specific pre-arrangement of molecules provided by the laser trapping. Authors have also systematically assessed the influence of initial concentration, laser power, and laser polarization on the polymorphic outcome (polymorphs were distinguished using the distinct morphologies and Raman spectra). High initial concentrations ($S>$0.7) yielded exclusively the prism $\beta$-CD while lower concentrations ($S<$0.28) showed only plate-like crystals, with both polymorphs forming for concentrations in between. Increasing laser intensity was shown to favor the prism form, while circularly polarized laser beam was shown to favor plate crystals. A mechanism was also proposed for the polymorph formation based on the packing structure of pre-clusters, molecular alignment, and concentration enhancement at the laser focus. Their results highlight the possibility of using laser trapping-induced crystallization to generate and investigate possible new polymorphs in solutions with concentrations below the saturation point. 

\subsection{Coupled LIPS and NPLIN}

Gowayed \etal\cite{10.1021/acs.cgd.9b01255} proposed to study the nature of a LIPS droplet by monitoring the nucleation of glycine in D$_2$O solutions occurring spontaneously after the laser was switched off and the nucleation induced by exposing the same droplet to a single unfocused 7~ns laser pulse (0.4-GW/cm\textsuperscript{2}, 1064~nm), \ie NPLIN of the LIPS droplet. Laser trapping was performed with a continuous-wave 1064~nm Nd:YAG laser beam focused by a 50x objective at the glass-solution interface of a 200~$\mu$L solution for 2 minutes. Four types of experiments were performed, two with LIPS, one with spontaneous nucleation, and one with NPLIN, at $S$=1.36; and two without LIPS but with $S$=1.50 (spontaneous and NPLIN). A laser pulse was applied in the solution immediately after the laser trapping was stopped, with an intensity of 0.4~GW/cm$^2$. When laser-trapping was performed alone, crystals were observed after around 30 minutes, whilst for NPLIN within the LIPS droplet, nucleation happened almost immediately. Only $\alpha$ crystals were formed within the LIPS droplet and both pure $\alpha$ and $\gamma$-glycine - as well as mixtures - formed around the phase-separated droplet.

\subsection{Summary}
We first itemize the observations from experiments and then discuss their consequences in terms of mechanism and application.

\begin{enumerate}
   
   \item \label{LTIC_concentrationregion}In LTIC, optical forces from a focused continuous-wave laser beam bring together solute clusters at the laser focal spot. Optical trapping causes the local concentration to increase, eventually leading to a spatiotemporally controlled nucleation event at the focal point. The liquid-like region with enhanced concentration can extend a few hundred microns away from the focus\cite{10.1007/s10043-015-0029-1}.
   \item \label{LTIC_Only}LTIC only happens when the focal point is at the air-solution interface\cite{10.1021/acs.jpcc.9b11651}. 
   \item \label{LTIC_Nucleation+Growth}Not only nucleation, but also crystal growth control, can be achieved using laser trapping-induced crystallization.
   \item \label{LTIC_LIPS}When the laser is focused inside the solution droplet or at the glass-/crystal-solution interface, only laser-induced phase separation (LIPS) is observed. LIPS droplets are highly concentrated with a higher refractive index compound, which may promote, and enhance, crystal growth\cite{10.1246/cl.2009.482}. 
   \item \label{LTICabc}Laser trapping-induced crystallization is demonstrated in various kinds of solutions of inorganic, small organic compounds and proteins, both in H$_2$O and D$_2$O super- and saturated solutions. 
   \item \label{LTIC_heating}Due to the long duration of laser irradiation (from a few seconds to hours), most researchers use heavy water to cope with the heating effects prominent in regular water.
   \item \label{LTIC_sat_undersat} LTIC has also been reported for undersaturated solutions of glycine in D$_2$O\cite{10.1021/cg100830x} and L-Phenylalanine in H$_2$O\cite{10.7567/1882-0786/ab4a9e}.
   \item \label{LTIC_liquidclusters} Phase-separated liquid-like droplets have been observed in LTIC experiments prior to crystallization\cite{10.1021/cg100830x,10.1021/jz100266t}. The reported size of these droplets can be larger than the laser spot size (a few microns to millimeters) and the lifetime varies based on experimental conditions.  
   \item \label{LTIC_Critical}Critical experimental parameters for laser trapping-induced crystallization are: the focal point of the laser, the laser power, the solvent used, and the solute concentration.
   \item \label{LTIC_Polymorph} Polymorph control was seen with LTIC by changing the polarization of the laser beam.
   \item \label{LTIC_Morphology_control}Different morphologies are also induced in non-polymorphic forming crystals by means of laser polarization\cite{10.1021/acs.jpcc.9b11651}.
    
\end{enumerate}

As LTIC has been reported for a broad range of solutes with both D$_2$O and water (observation \ref{LTICabc}) and a thermodynamic model has been proposed for LIPS\cite{10.1039/c9sm01297d, 10.1038/s41557-018-0009-8}, the next steps should be exploring and extending the limits of the current understanding. A promising direction is coupled heat, mass, and momentum modeling of the phenomena which will provide insights into kinetics. Such progress naturally requires quantitative measurements of the instantaneous temperature and concentration in the presence of laser irradiation and possibly thermally driven flow. These measured instantaneous fields will enable calculation of the driving force behind nucleation and growth, supersaturation. This approach is a challenging task experimentally as most LTIC experiments are conducted at the liquid-gas or liquid-glass interface. Yet it may offer explanations to open questions and observations not explained by the proposed thermodynamic model such as ``Why does the concentrated region extends a few hundred microns away from the focus (observation \ref{LTIC_concentrationregion})?'' and ``Why does laser trapping only induces nucleation at the air-solution interface (observation \ref{LTIC_Only})?''. Moreover, the role of heating (observation \ref{LTIC_heating}) and other critical experimental parameters (observation \ref{LTIC_Critical}) may be quantitatively elucidated opening a door to predictive models. Any developed model should also offer an explanation of observations not shared in NPLIN and HILIN, particularly the ability to observe crystallization in undersaturated solutions with LTIC (observation \ref{LTIC_sat_undersat}) and the emergence of phase-separated liquid-like droplets (observation \ref{LTIC_liquidclusters}). Recent approaches integrating low-intensity continuous lasers with pulsed lasers have recently shown significant promise as they allow following the time-lapse spectroscopic signature of the solute as it crystallizes and grows\cite{10.1021/jacs.1c11154,10.1073/pnas.2122990119,10.1021/acs.cgd.9b01255}. These approaches would also benefit from experiments quantifying time-dependent concentration and temperature fields.

\section{Indirect methods} \label{sec:indirect}
In this section, we explore indirect approaches where the laser does not irradiate the solution directly, but the energy is transferred through an intermediate medium. We classify the indirect methods influencing  crystallization in three broad categories. 
\begin{enumerate}
\item Nucleation through laser-induced shockwaves: the irradiation energy is first transferred by absorption of a solid object in contact with a solution and the resulting shockwave is thought to trigger nucleation.
\item Surface plasmon-assisted crystallization: plasmonic nanoparticles absorb laser energy, enhancing the electromagnetic field through localized surface plasmon resonance. Plasmonic effects respond to specific wavelengths and confine the electromagnetic field to very small hotspots where trapping forces are enhanced.  Additionally, due to plasmonic heating effects, the local temperature is elevated. This can initiate bubble formation as well as secondary phenomena such as capillary and Marangoni flows. This connected set of events collectively influences crystallization and the resulting crystal properties.
\item Pulsed laser ablation: in pulsed laser ablation of crystals, a laser is directly focused on a macroscopic crystal and induces photomechanical and photothermal stress, which leads to fragmentation and ejection of fragments. The resulting crystals then serve as seeds for secondary nucleation. Alternatively, the growth mode can be selectively altered at the laser focus. 
\end{enumerate}
It should be noted that in the last decade, studies focusing on indirect methods leveraging light fields to control nucleation have gained considerable popularity, but due to the wide variety in experimental setups and scope, proper classification is challenging. Nonetheless, we present our best effort to classify this diverse field.

\subsection{Laser-induced shock wave}
Mechanical agitation is commonly utilized in industrial practice to enhance effective nucleation rates and growth \cite{Myerson2002}. Approaches such as sonocrystallization \cite{10.1016/j.ultsonch.2010.06.016}, stirring \cite{10.1038/190251a0}, or even shaking or scratching sides walls of vials, all cause a disturbance in the solution leading to enhanced nucleation and growth compared to an undisturbed solution. The underlying mechanism for all these methods is suspected to involve a combination of heterogeneous nucleation, cavitation, and shockwaves. Yet, the exact mechanism is still debated in the literature.\cite{10.1039/c7cp03146g,10.1016/j.cep.2019.03.017}
 
Mirsaleh-Kohan \etal\cite{10.1021/acs.cgd.6b01437} experimented with saturated and undersaturated solutions in a setup in which a laser (average power of 800 mW) was focused into the solution or onto a thin metal plate floating at the air-solution interface. Whether the laser was focused on the solution directly or on the metal plate, similar crystallization behavior was observed. From these experiments, they deduced that nucleation was triggered by the shock/sound waves created through laser-material interactions. Connecting their previous characterization of shock/sound waves with pressures up to 20 MPa they assumed similar conditions were present in experiments where the laser was focused onto the plate. Compounds such as sodium chloride, sodium bromate, and tartaric acid, were all successfully crystallized.

The authors also sketched a thermodynamic framework in an attempt to explain their results. The framework formulates a nucleation enhancement factor within the CNT framework, that takes into account the effect of pressure and temperature fluctuations altering the chemical potential. A statistically significant number of repetitions are essential for nucleation studies yet only a limited number of experiments was reported in this study. 
We suspect this might be due to the labor-intensive and manual nature of these experiments. Previous work from the same group\cite{10.1063/1.1588750} describes how the velocity of laser-induced shock waves was quantified. A He:Ne laser beam, positioned at a known distance from the laser focus, was used to determine the time of transit of the sound pulse. It would have been useful if these measurements on the speed of shock/sound waves were accompanied by imaging of crystal formation. Yet, it is not hard to imagine that such simultaneous measurements can be experimentally challenging. Mirsaleh-Kohan \etal provide some scanning electron microscopy and petrographic images, which they use for size determination \cite{10.1021/acs.cgd.6b01437}. However, the authors provided no information on particle size distribution, which would have allowed the reader to judge the applicability of the proposed method in an industrial setting. 

In one of our earlier works \cite{10.1021/acs.cgd.7b01277}, inspired by results of Mirsaleh-Kohan \etal\cite{10.1021/acs.cgd.6b01437}, laser-induced shock waves were investigated as one of the possible mechanisms behind NPLIN. A statistically significant number of samples (more than a hundred independent samples) were irradiated with laser pulses. Control vials were masked by placing black tape on the vial containing the solution to block laser irradiation while measuring the pressure resulting from irradiation. The resulting nucleation probabilities for the unmasked and the masked control vial were $100\%$ and 0$\%$ respectively, while the measured pressure wave values were 20 mbar and 200 mbar, respectively. Although the masked vials experienced more intense shock waves quantified by a transducer, the nucleation was solely dependent on the laser passing through the solution. The employed laser energy density in this multiparameter investigation was 80 MW/cm$^2$ \cite{10.1021/acs.cgd.7b01277}. 

Mirsaleh-Kohan \etal reported a laser energy density of around 1.9 x $10^{12}$ W/cm$^2$ in their work, a number that is above the ionization threshold. Therefore, it should not be classified as NPLIN but LIN. Not only was the irradiation energy density in their work  several orders of magnitude greater, their setup might also be more efficient in transferring the laser energy into the solution. This can be due to the fact that the floating metal plate directly translates the light momentum into the solution, while in the work by Kacker \etal\cite{10.1021/acs.cgd.7b01277}, there is first the tape that absorbs the laser energy, which then needs to go through the glass before it reaches the solution.

Alexander \etal also discussed the shock wave mechanism \cite{10.1063/1.5079328} in the context of NPLIN \cite{10.1039/c7cp03146g}. The authors attempted to recreate the NPLIN polarization switching effect as achieved by Garetz \etal Additionally, an additional set of vials was exposed to ultrasound by placing them in a standard laboratory ultrasonic bath (37 kHz, effective power 80 W) for a duration of 120 seconds. Another set of vials was exposed to mechanical shock. Comparable nucleation probabilities for NPLIN, ultrasound, and mechanical shock were reported, with a similar dependence on supersaturation, hinting at a common mechanism. This mechanism may involve ultrasound-induced cavitation and pressure shock waves causing localized increases in pressure and supersaturation. The experimental results presented in Liu \etal along with others \cite{10.1039/c7cp03146g,10.1021/acs.cgd.7b01277} point to a common mechanism amongst NPLIN, ultrasound, and mechanically generated shocks, that is yet to be quantified.



\subsection{Surface plasmon-assisted nucleation}

In this approach, a continuous-wave (CW) laser is focused on closely packed nanoparticles or lithographically manufactured features that produce surface plasmon resonance (SPR) upon irradiation. Surface plasmon resonance is the collective oscillation of free electrons on surfaces of metallic nanoparticles upon light irradiation.  In this process, the electrons on metal surfaces oscillate collectively as they synchronize and interact with the electrical field oscillation of the incident light.

SPR has been utilized to trap a plethora of systems ranging from nanoparticles to DNA, with two-dimensional arrays of metallic nanostructures fabricated on a substrate\cite{10.1021/jz501231h}. This concept, also referred to as plasmonic trapping, enables trapping at sub-excitation irradiation wavelengths and hence offers superior control compared to traditional laser trapping approaches\cite{10.1021/acs.cgd.8b01361}. However, one should pay specific attention to  the contributions of molecular transport mechanisms triggered by plasmonic heating.

Notably, Niinomi \etal have shown particular interest in leveraging SPR to induce nucleation. For this technique to be more broadly utilized, other groups have to focus their efforts on SPR-induced nucleation. Niinomi \etal\cite{10.1039/C6CE01464J} demonstrated NaClO$_3$ chiral crystallization can be selectively induced by the optical trapping of Ag nano-aggregates using a continuous wave circularly polarized laser ($\lambda= 532$~nm emitted from a Spectra Physics Millennia eV laser with intensity 940 mW). In this study, the laser was focused at the air-solution interface of unsaturated NaClO$_3$ aqueous solution containing Ag nanoparticles. Chirality of the final crystal could be controlled by altering the handedness of the circularly polarized laser. It should be noted that this result is somewhat reminiscent of the polarization switching effect in NPLIN observed by Garetz \etal However, the NPLIN polarization effect could not be reproduced by other groups, so this connection should be further investigated. Niinomi \etal\cite{10.1039/C6CE01464J} provide thermodynamic arguments explaining chiral control, hypothesizing that plasmon-induced circular dichroism causes chiral bias as a consequence of an enantiomeric difference in chemical potential of l- and d-crystalline clusters.  
Combining numerical simulations and experiments, 
Niinomi \etal\cite{10.1021/acs.cgd.6b01657} explained their previous results\cite{10.1039/C6CE01464J} of crystallization induced by visible laser trapping of silver nanoparticles (AgNPs) in the context of plasmon heating. In this study, the authors placed the focal spot at the air/unsaturated mother solution interface. The numerical analysis of temperature distribution pointed out that the temperature reaches \SI{390}{\celsius} at the focal spot because of plasmonic heating. They concluded that crystallization occurred due to enhanced supersaturation caused by local solvent evaporation via plasmonic heating.

Despite their clear bias, the nanoparticle aggregate properties are not well-defined, and can only be controlled to a certain extent. In one of Niinomi's subsequent works, a plasmonic substrate of gold triangular trimers was produced. Not only does this allow for plasmonic hotspot design at the nanogap, but it is also possible to perform a finite element analysis of the electromagnetic field \cite{chiralbiasplasmonictriangles}. The energy density of the CW laser ($\lambda = 1064$ nm) was fixed at 1 MW/cm$^2$, and it was focused on one triangular trimer, submerged in saturated D$_2$O. A single crystal was grown from the focal spot, and was determined to be an achiral, metastable precursor crystal. Continued irradiation enabled polymorphic conversion to the final, chiral crystal. Again, chiral bias could be induced by changing the handedness of the laser, resulting in an enantiomeric excess of over 50\%. The authors hypothesized that the resulting bias was a consequence of the momentum transfer of the EM field by either the spin angular momentum (SAM) or orbital angular momentum (OAM). SAM transfer could possibly, taking into account a difference in refractive index in chiral pre-nucleation clusters, lead to the diffusion suppression of either the d-(l-CPL) or l-clusters (r-CPL), thus favoring one handedness over the other. The second mechanism, being somewhat reminiscent of an OKE mechanism, would involve the distortion of the achiral octahedral structure of the primary crystal by an optical torque affected by the orbital angular momentum. 

A follow-up numerical EM-field analysis study was performed, in which the triangular structure and laser were virtually mimicked to further investigate the spatial distribution of the enantioselective chiral optical potential \cite{doi:10.1021/cg500250e}.
The trapping potential was calculated for particles of various sizes and chirality. The calculated chiral optical potential well created by the plasmonic hotspot was not capable of trapping virtual particles with chirality values representative of NaClO$_3$ crystals as measured by conventional far-field ORD. However, the magnitude of the difference in chiral gradient force between l- en d-chiral virtual NaClO$_3$ nanospheres was found to be comparable to the dielectric gradient force calculated in previous experiments on laser-trapping from undersaturated solution.
As an alternative, it was investigated whether the chiral bias could be explained thermodynamically. The framework by Alexander and Camp \cite{doi:10.1021/cg8007415}, in which the Gibbs free energy for nucleation is reduced by the electrostatic potential, was extended with an additional term accounting for the reduction in free energy due to the chiral optical potential. However, calculating the nucleation rates for both the d- and l- enantiomorphs yielded equal rates, and it was therefore concluded that the giant crystal enantiomeric excess could not be explained by thermodynamic contributions.
Rather, it was hypothesized that the enantiomeric excess could be explained by a difference in the frequency of incidence of crystal nuclei at the nanogap, either due to the possibility of heterogenous nucleation or the local difference in chiral clusters due to enantioselectively biased diffusion.


Some additional effects have been reported utilizing plasmonic-assisted trapping (PAT) \cite{plasmoncontacttransfer}. Saturated aqueous NaClO$_3$ microdoplets were sprayed onto a gold gammadion nanostructure, and a CW circularly polarized Nd3+:YVO4 laser beam ($\lambda$ = 1064 nm, Spectra Physics, J20-BL-106C) was focused through a 60x objective on the periphery of a microdroplet. Authors noticed similar formation of achiral precursor and subsequent transition to chiral crystal as reported earlier. Additionally, moving the focal spot outside the microdroplet led to `creeping' of the crystal outside of the droplet towards the focal spot. When the creepage of a chiral crystal came in contact with the achiral precursor, polymorphic transformation propagated from the contact
point, transforming the achiral crystal.
Interestingly, the authors observed the formation and short lifetime of a liquid distinct from the solution, which could potentially be contributed to the existence of a liquid precursor. This would be in line with a two-step nucleation mechanism.

The plasmonic-assisted trapping has been demonstrated for compounds other than NaClO$_3$. The Niinomi group succeeded in PAT of acetaminophen, and found that after stopping laser irradiation, crystals first relaxed into a liquid domain, before diffusing out \cite{doi:10.1021/acs.cgd.8b01361}. A 20 mW CW laser was focused with a 60x objective onto the interface of a thin film of saturated acetaminophen solution and a plasmonic film consisting of repeated left-handed gammadion structures. Crystallization was observed in an annular pattern around the focal spot, at a radius between 19-20 microns. When the focal spot was moved, the annular distribution followed.

The crystallization at a very specific distance from the center of the focal spot, which resulted in the annular pattern, was rationalized to arise from a balance of thermophoretic and electric gradient forces. The thermophoretic force acts as a repulsive force from the focal spot. 

The electrical field enhancement in the plasmonic near-field was evaluated by FDTD. Electric field enhancement was observed at the nanogaps between the gold gammadion structures. Subsequent gradient forces perpendicular to the plane were calculated. The thermophoretic forces together with the gradient forces were used to calculate an optimal radius for crystal nucleation, which was reasonably close to the observed crystal radius value.

Stopping laser irradiation resulted first in the produced polycrystals to swiftly relax into concentrated microdroplets, before diffusing back into the solution. This prompt response suggests that the crystallization cannot be attributed solely to thermophoretic forces originating from the temperature increase at the focal spot and that the crystallization behavior is likely a result of the contribution of the plasmonic near-field-enhanced electric gradient force. The authors referred to Alexander's dielectric polarization hypothesis that an, in this case, plasmon-enhanced, electric field gradient force would decrease the Gibbs free energy barrier for nucleation \cite{AlexanderCampCNTtheory}. 

Interestingly, plasmonic nanoparticles have been considered in the discussion of a theoretical study focusing on the physical mechanism behind NPLIN by Nardone and Karpov \cite{10.1039/c2cp41880k}. 





\subsubsection{Optothermally generated bubble trapping}

As indicated in previous sections, bubble formation plays a critical role in LIN, and is suspected to be involved in NPLIN as well. The effects of bubble formation during optical trapping have been known for some time \cite{Opticaltrappingbubbleeffects}.

Various authors have shown that the effects of bubble formation can be utilized in microfluid flow and particle manipulation. Xie and Zhao have reviewed a wide variety of applications of optothermally generated surface bubbles \cite{surfacebubbleanditsapplications}. Several benefits of (gold) plasmonic substrates for bubble generation were highlighted, namely the convenience that plasmonic particles provide for fine-tuning optical properties. Their size also makes them ideal nano heating sources.
The fundamentals and applications of radiation-induced plasmonic nanobubbles have recently been reviewed by Zhang \etal\cite{plasmonicnanobubblereview}. Details on localized surface plasmon resonance and bubble formation can be found there.

Due to bubble formation, the plasmonic substrate is isolated from the fluid, resulting in a sudden temperature spike of the nanoparticle, and convection caused by strong temperature gradients.

Convection is conjectured to be twofold, consisting of natural and Marangoni convection. Natural convection is a result of density gradients induced by local heating. Marangoni convection is the result of a surface tension gradient/temperature-dependent shear force at the bubble surface. Although various authors reported that Marangoni convection can be considered to be the dominant effect, one should not overlook other phenomena such as thermophoresis and van der Waals forces.

Setoura, Ito, and Miyasaka studied stationary bubble formation around a single Au nanoparticle due to plasmonic heating, and evaluated the bubble-induced convection for parameters such as Au particle diameter and laser fluence \cite{AuNPBubbleFormationEffects}. It was found that the velocity of the convective flow could be increased by increasing laser power. There are several examples of this technique being utilized to trap quantum dots, polystyrene microparticles, and macromolecules such as DNA strings \cite{particletrappingandmanipulationbymicrobubble}. 
However, there is only a handful of examples of the induction of nucleation with this technique.


Fujii \etal\cite{10.1021/am201799b} employed a 1064 nm CW laser to irradiate a patterned gold nanoparticle thin film. The laser power was varied in the range of 0 - 100 mW. An aqueous glycine supersaturated solution at 3.6 M was used as a crystallization medium. The temperature at the focal point was calculated to scale linearly with laser power, and the laser power threshold for bubble formation (0.15 mW) roughly corresponded with gold thin film temperatures being at the boiling point of the medium (373 K). A microbubble was formed after several seconds of irradiation, and approximately 1 minute after bubble generation a dense liquid cluster of glycine was observed.
After switching off the laser, crystallization of the liquid precursor was observed from inside to outside. It was argued that the temperature drop due to laser shutdown increased the supersaturation to above the metastable zone, thereby inducing transformation from the liquid phase to the crystal. Yamamoto \etal used the technique to grow macroscopically anisotropic petal-like structures of diporphyrin \cite{10.1038/s41598-018-28311-2}.

Perhaps the most elaborate work on the potential role of a plasmonic-heating induced microbubble in primary nucleation belongs to an experiment of the Niinomi group \cite{Niinomi2018}. A 30 mW CW ($\lambda = 532$ nm) laser was focused with a 60x objective on a gold plasmonic nanolattice under a stagnant, saturated NaClO$_3$ droplet, and a microbubble was produced. The laser focal point was slightly off-center relative to the center of the microbubble. 
After microbubble formation due to the plasmonic heating effect, achiral crystals nucleated as a result of Marangoni convection. The local supersaturation near the bubble was estimated by measuring the growth rate. Assuming linear dependence on supersaturation for growth rate, the local supersaturation was calculated to be 365\%. Furthermore, a polymorphic transition to a chiral crystal was observed. These 'mother' crystals were seen to be fragmented, resulting in a large number of smaller crystals of the same handedness dispersing throughout the solution. It was hypothesized that this fragmentation was due to microfluidic shear stress originating from the Marangoni convection at the bubble/substrate interface. The authors referred to the work of Namura on different convective trapping modes by bubble formation \cite{doi:10.1063/1.4942601}, classifying the trapping behavior as the horizontal trapping mode.


It should be noted that a plasmonic substrate is not a prerequisite for bubble nucleation. However, observations for crystallization with alternate substrates are limited, and provide neither theoretical nor numerical analysis. Nonetheless, it is likely that similar effects are present.
Salam \etal\cite{Nucleanteffect} experimented with several nucleants for the crystallization of a number of simple biomolecules. CW irradiation from a ($\lambda = 1064$ nm) Nd:YAG laser was used for irradiation of the nucleants, with energy ranging between 120 and 600 mW. Focusing the laser on a copper wire submerged in a 3 M glycine solution, bubble formation was observed, followed immediately by glycine crystal formation around the bubble. It was hypothesized that the occurrence of the air-liquid interface increased local saturation thus leading to nucleation. 

A study by Wu et al. demonstrated femtosecond pulse laser trapping-induced nucleation \cite{Wu_2020}, as well as polymorphic conversion due to surface bubble effects.
The laser fluence ranged from $4*10^2-54*10^2$ J/cm$^2$, which corresponds to laser pulse powers in the range of 20-400 mW. Pulses with a repetition rate of 80 MHz were focused at the air-solution interface of L-Phe solutions at supersaturation 0.8 for several minutes. The laser wavelength was 800 nm instead of the 1064 nm commonly used in CW laser trapping experiments. After 3 minutes, a bright light was observed, which the authors ascribed to 'white light supercontinuum', a phenomenon that occurs by self-modulation of the laser in the medium. Additionally, bubbles were formed at the focal spot, which quickly diffused out. After approximately 12 minutes, several plate-like crystals formed, which in the case of L-Phe, can be ascribed to the anhydrous polymorph. The results are in agreement with the CW laser trapping-induced nucleation of L-Phe crystals from undersaturation, as reported earlier \cite{doi:10.1021/acs.cgd.8b00796}. The threshold laser power for pulsed-induced nucleation, however, was estimated to be 3-4 times smaller than the CW trapping case, which highlights the efficiency of pulse trapping. At higher laser energy ($54 * 10^2$ J/cm$^2$) the nucleation probability dropped to zero, most likely caused by the flow induced by the cavitation bubbles and the temperature increase. Additionally, by continuing the pulsed irradiation onto the plate crystals, whisker-shaped crystals, the monohydrate polymorph, appeared on the plate crystals. This is remarkable, as the formation of the monohydrate form had previously only been observed to grow from supersaturated solution. The authors furthermore succeeded in converting whisker-like crystals, produced by evaporative crystallization, to the anhydrous polymorph. Although the polymorph conversion was bidirectional, it was argued that the mechanisms were not. Crystal fragmentation induced by laser ablation was considered, but the laser energy was calculated to be far below the ablation threshold. Instead, the conversion was initiated by bubble formation. Polymorphism was determined by the degree of concentration increase at the surface of the bubble.

\subsection{Pulsed Laser Ablation}


In this section, we will briefly touch upon the extensive literature on pulsed laser ablation of crystal surfaces and pulsed laser ablation in liquid (PLAL). The laser ablation mechanism is fundamentally different than the putative (NP)LIN mechanisms. While in LIN, the laser is focused into the solution, evoking cavitation, in PLAL the laser pulse is focused onto a solid face, often belonging to a crystal, inducing a (thermoelastic) stress wave, which causes the material under focus to ablate and break, releasing particles into the solution.

Paltauf and Dyer provided an extensive review of photomechanical effects in laser ablation \cite{laserablationreview}. They indicated that a ``stress confinement'' condition is satisfied when heating yields minimal expansion and strain, which is achieved when the laser pulse duration is much smaller than the acoustic relaxation time of the solid medium.
Zhigilei and Garrison performed large-scale molecular dynamics simulations on the laser ablation of organic solids and found that the ejection process is heavily dependent on the rate of laser energy deposition \cite{MDablation}. Stress confinement occurred for shorter laser pulses and corresponds to a lower ablation threshold. For laser fluences close to the ablation threshold, the mechanical fracture is localized under the ablation site, while for higher fluences the stress affects the material over larger distances.

Femtosecond or deep-UV laser processing of crystals for the purpose of micro- or macroseeding has been reviewed previously by Yoshikawa \etal\cite{C3CS60226E}

The use of a femtosecond laser for the ablation of crystal surfaces was reported to promote and control the crystal growth of hen egg-white lysozyme (HEWL) and glycine, respectively. Tominaga \etal\cite{10.1038/nphoton.2016.202} irradiated the (110) face of a tetragonal crystal with about 10,000 pulses of 0.25 $\mu$J (slightly above the ablation threshold for HEWL, $\sim 0.2 \mu$J) using a near-infrared laser. At energies close to the threshold, the ablation is known to be induced photomechanically, which implies minimum heat generation and minimum damage to the bulk crystal, compared to longer laser pulse durations. Crystal growth was monitored by laser confocal microscopy combined with differential interference contrast microscopy (LCM-DIM) and X-ray diffraction for 32 days. The results clearly showed the enhancement of crystal growth of the ablated face in single crystallinity without any deterioration of crystal lattice parameters. Image analysis revealed that before laser ablation, the crystal surface was covered by microcrystals, most likely originating by 2D-nucleation. When an area of around 6 micron diameter was etched by the femtosecond laser ablation, a single rectangular microcrystal was observed, initiating spiral growth on that face. After 100 hours, a smooth face had fully covered the ablated face and no additional microcrystals were visible. The authors explained the phenomenon through the ablation of the (110) face by the femtosecond laser which ejected tiny crystal fragments from the bulk (with minimal damage to the surface), which induced screw dislocation and thus initiating the spiral growth. Alternatively, they hypothesized mechanical disruption by the laser might directly induce lattice misalignment. Further work by Suzuki \etal\cite{10.1021/acs.cgd.8b00697} goes beyond growth enhancement and demonstrated control of organic crystal shape, using $\alpha$-glycine as a model crystal and varying laser energy, observed under similar LCM-DIM. In their work, when a single femtosecond laser pulse with an energy of 0.28 $\mu$J, which is below the threshold ablation for glycine (0.35 $\mu$J), was focused (focal radius $\sim 1\mu$m) at the (010) face of a plate-like crystal, no growth was observed after 2 hours. However, for energies at the exact ablation threshold for glycine, 0.35 $\mu$J and 1.2 $\mu$J in the same crystal shape and focal radius, immediate growth was observed, forming a full pyramid-like crystal after only 6 minutes for the lower laser energy. Higher energies (1.8 $\mu$J) were observed to break the bulk crystal. Similarly to the study with HEWL, in this work, crystal growth steps were also visible after laser ablation, originating spiral growth that covered the whole surface in a matter of seconds. Moreover, the authors observed that when irradiating the faces of a hexagonal and lozenge-like initial crystal, the irradiated faces immediately grew to form sharp edges. Sequential laser shots to the same lozenge-like crystal showed that even `tailored' star-shaped glycine, not possible to be formed spontaneously, could be obtained. These studies highlight the possibility of using femtosecond laser ablation to control crystal growth that can be applied in obtaining functional crystals. 

In a recent paper by the Yoshikawa group, the effects of pulse duration on the ablation mechanism of L-Phe crystals were investigated \cite{fspsnspulseinvestigation}. Video recording and atomic force microscope imaging of the ablated crystal surface showed protruding material around the ablated area for ps and ns pulses, pointing at the occurrence of a photothermal mechanism, while for fs pulse exposure, the ablation appeared to be sharp without thermal deformation around the impact site of the laser. This is in line with theoretical considerations of photomechanical and photothermal regions\cite{MDablation}. For fs irradiation, single-crystalline growth was observed, while for ps and ns pulses polycrystallinity occurred. Furthermore, spatial control of the crystal growth was investigated by visualizing growth velocity by LCM-DIM monitoring of the growing crystal faces. The average growth rate of individual crystal faces could be increased by selectively ablating them. The growth rate increased varying from 2.5 - 4 times relative to the growth rate pre-exposure. Some LCM-DIM images clearly showed the occurrence of spiral hillocks on the ablated face, while others did not. The authors suspect this could mean that besides spiral growth, other growth modes could be induced.

Dell'aglio \etal\cite{10.1016/j.apsusc.2015.01.082} gave a solid overview of the PLAL mechanism, in which the formation of the plasma cloud is the central phenomenon, followed by a shockwave and bubble formation.  Nanoparticles are produced in the first stage of the PLAL process, \ie during the plasma cooling. The plasma cooling starts from the external shells of the plasma and proceeds to the interior, thus the temperature in the plasma core remains constant (4000-6000 K), allowing nanoparticles to be formed at thermodynamically constant conditions, which, according to the authors, results in a narrow size distribution. Experiments by Reich \etal\cite{C9NR01203F} on PLAL of gold and silver substrates suggest that the majority of the particles are of small diameter (8 nm) and are contained within the cavitation bubble, but a small portion of particles of larger diameter (15-20 nm) was seen to precede the cavitation bubble front, which explains the bimodal size distribution that is said to be typical of PLAL of noble metals. Variation of laser fluence and choice of liquid medium can significantly impact the resulting size distribution \cite{PLALeffectsofliquidandlaserfluence}. In the case of metal nanoparticles produced by PLAL, the laser melts the surface layer of the metal target, after which the liquid metal disperses through the solvent, and then solidifies as nanoparticles\cite{10.1016/j.apsusc.2015.01.082}.
In a similar set-up, focusing a laser on a graphite surface\cite{10.1063/1.2132069}, a high-density plasma region is formed and as a result of the elevated temperature and pressure in the plasma, the ablated graphite transforms into its diamond allotrope. It should be mentioned that this experiment employs high laser energy densities (10$^{11}$ W/cm$^2$). It is stated that at such high energy density, the ablated graphite experiences extreme pressure and temperatures: in the regions of 10-15 GPA, and 4000-5000 K, above the diamond formation line in the carbon phase diagram. Using this method, the authors have reported successfully synthesizing nanocrystals of a number of high-pressure phase materials, including C$_3$N$_4$ and BN crystals\cite{WANG200310, C3N4PLIIR}.

\section{Molecular Simulation}
\label{sec:MD}

The experimental techniques discussed above offer the possibility to induce nucleation under various conditions and infer possible nucleation mechanisms. However, obtaining direct and comprehensive insight into the molecular structure, local dynamics, and nucleation mechanisms is prohibited by the spatial and temporal resolution of measurement techniques. Simulations have proven to be a powerful way to complement experiments and help to interpret experimental measurements. In particular, a bottom-up simulation technique such as classical molecular dynamics (MD) simulation has been used in various studies to elucidate nucleation mechanisms or probe dynamics without relying on nucleation models, assumed equations of state, or known transport coefficients, since these properties are emergent in MD simulation. 

MD allows accurate control over system conditions and provides access to local properties at high spatial and temporal resolution. On the other hand, the high computational cost of atomistic simulations is prohibitive for covering the macroscopic length and time scales over which experimental measurements are performed. Furthermore, the vast majority of classical MD studies do not allow for the occurrence of chemical reactions. Therefore, any reference to laser-induced nucleation in this section refers to NPLIN. 

Although MD has been used to study crystal nucleation for over 40 years \cite{10.1063/1.432681}, current computing power and modern simulation techniques have served as an impetus in recent years\cite{10.1021/acs.chemrev.5b00744}. To date, only a few studies have considered the effect of laser irradiation on nucleation.  
Classical MD simulations of homogeneous nucleation have mostly considered systems of less than $10^5$ molecules. Observing a statistically significant number of nucleation events throughout a simulation time of up to hundreds of nanoseconds in such small systems is highly unlikely, even under conditions that are strongly favorable for nucleation. Alternatively, various simulation techniques have been employed to bias nucleation events, as will be briefly discussed in Section\,\ref{sec:sampling}.

\subsection{Equilibrium simulations} \label{sec:EMD}

NPLIN experiments have shown that nucleation in a supersaturated KCl solution can be induced with a laser pulse of at least 100 ps. However, no stable crystals were observed with pulses shorter than 5 ps.\cite{10.1021/jp5049937} 
This suggests a two-step nucleation mechanism in which the growth and restructuring of an amorphous cluster leads to the formation of a stable crystal. If this is the case, then the minimum required pulse duration for stable nucleation may depend on the ion pairing lifetime, their diffusivity, and the speed at which ionic clusters restructure \cite{10.1063/1.3268704}. To investigate this hypothesis, Sindt \etal\cite{10.1021/jp5049937} performed MD simulations to explore the growth and restructuring of spontaneously forming KCl clusters over a wide range of ionic concentrations. The authors found an increase in the ion pairing lifetime with increasing KCl concentration, with values beyond 40 ps in supersaturated KCl solutions. The average residence time of water in the first hydration shell of the ions also increased with ion concentration, up to 18 ps for a supersaturation of $S=1.96$. These findings suggest that a laser pulse of at least 100 ps is indeed long enough for ions to join or leave a cluster during the pulse, whereas a pulse of 5 ps would be too short. 

Although the fast dynamics of ions compared to the duration of the laser pulse might be important, it is not a sufficient condition for nucleation to occur. For example, far below the saturation limit of a KCl solution, the ion pairing time and the residence time of water in the ion hydration shell is also of the order of 10 ps \cite{10.1063/1.4896380}, yet NPLIN would not occur under such conditions. Moreover, another study has reported that ion desolvation, rather than diffusivity, is the determining factor for an ion to join a cluster.\cite{10.1021/jacs.5b08098} 
To identify how the dynamics in a cluster varies with concentration, Sindt \etal computed self-intermediate scattering functions. The authors found a difference between the relaxation behavior below and above the saturation limit, which suggests complex single-ion dynamics that might be affected by its correlation to the cluster. 
Pairing and relaxation time scales help to explain the required laser pulse duration in NPLIN, but it remains an open question how amorphous clusters turn into stable crystals. To shed light on this aspect, Lanaro and Patey \cite{10.1021/acs.jpcb.6b05291} developed a method to track the evolution of NaCl clusters in water. Their analysis showed that cluster size was not the only criterion for stability and nucleation. Instead, the nucleation probability and lifetime were found to be strongly correlated with the crystallinity of the pre-nucleation cluster. 

The findings of Sindt \etal\cite{10.1021/jp5049937} and Lanaro and Patey \cite{10.1021/acs.jpcb.6b05291} indicate that certain local conditions can lead to nucleation via a two-step mechanism. However, neither study has addressed how these local conditions are affected or caused by laser irradiation. As described in Section \ref{sec:NPLIN}, several mechanisms have been proposed that could explain the occurrence of NPLIN. Sindt \etal\cite{10.1063/1.5002002} used MD simulation to investigate a cavitation-induced nucleation mechanism. A carbon nanoparticle (4 nm in diameter) was simulated to represent an impurity in the solution, that was suggested to absorb the energy from the laser pulse. After equilibrating the particle in a 20\% supersaturated NaCl solution at room temperature and atmospheric pressure, it was instantaneously heated to several thousand Kelvin. The heat was then conducted to the surrounding fluid, resulting in cavitation at the particle surface. For the first 2 nanoseconds after the onset of cavitation, ions moved away from the vapor-liquid interface. This resulted in a temporary enhancement of the ion concentration a few nanometers away from the interface, where ion clusters with a high level of crystallinity formed. Stable nucleation was not observed within the 2.8 ns simulation time. 
 
The formation of ionic clusters in the study of Sindt \etal\cite{10.1021/jp5049937} supports the possibility of a cavitation-induced nucleation mechanism. However, it is unknown if nucleation in experiments would indeed initiate where the clusters formed in the simulations, or what the local supersaturation would be at the location of these clusters. Local supersaturation can be obtained from simulated temperature and density profiles, such as those presented by Sindt \etal The simulations of Sindt \etal would, however, not yield a representative supersaturation for two reasons. First, the cavitation and ion distribution would not be quantitative correct because the TIP3P water model that was used has a boiling point of 593 Kelvin and a self-diffusion coefficient that is more than twice the experimental value \cite{10.1134/S1990793110020065,10.1063/1.1329346}. Second, the initial heat of the nanoparticle remained present in the finite-size system, leaving no room for further dissipation. The solution absorbed heat from the nanoparticle until it reached a uniform temperature above the experimental boiling point, but below that of the TIP3P water model. The final fluid temperature in this system depends on the size of the simulation system. This finite-size effect could be avoided by calculating the minimum required system size based on the known heat diffusion rate and the intended simulation time. However, this would require a huge system or limit to a very short simulation time. A more feasible solution is to couple the fluid a few nanometers away from the nanoparticle to a heat sink representing the environment \cite{10.1021/acs.jpcc.8b04017}. The target temperature of the heat sink should then ideally be varied in time based on temperature evolution estimates from a continuum model, such as the model of Sun \etal\cite{10.1017/S0022112009007381}.

\subsection{Nonequilibrium simulations} \label{sec:NEMD}

Instead of reproducing local fluid conditions that laser irradiation could cause, the influence of a laser can be probed directly in a nonequilibrium MD simulation. Nonequilibrium MD simulations contain an external force that drives the system away from thermodynamic equilibrium. The effect of a laser can be most logically mimicked by imposing an oscillatory electromagnetic field. However, implementing the magnetic field contribution in combination with the popular velocity Verlet integrator is problematic and has therefore not been included in most MD packages, with some exceptions.\cite{10.1007/s00894-020-4349-0}
This inconvenience, combined with the fact that the Lorentz force is much smaller in magnitude than the electric force, has caused many researchers to neglect the magnetic component entirely.\cite{10.1039/C5CP00629E} Electric fields are easy to impose in MD and have been widely applied in the context of electrokinetic transport.\cite{10.1063/1.1543140,10.1039/c5cp03818a, 10.1103/PhysRevLett.128.056001} Few MD studies have investigated nucleation under the influence of an electric field. 

Parks \etal\cite{10.1021/acs.cgd.7b00356} investigated the effect of high-intensity static electric fields on a pre-formed paracetamol crystal in an aqueous solution. An electric field was found to suppress the crystal growth rate of a supersaturated solution by up to 40\% (for an electric field of 1.5 V/nm). The suppressed growth rate was explained based on the fact that the aligned liquid paracetamol molecules under influence of the electric field favored the liquid phase over the solid phase. This decrease in growth rate was found to be consistent with experimental observations of paracetamol nucleation under the influence of a magnetic field.\cite{10.1002/crat.201400432} Parks \etal also reported a solid-state transition into a new polymorph of paracetamol in which the dipoles of the molecules are more aligned with the direction of the electric field than in the previously known polymorphs. This transformation was induced by an electric field of 1.5 V/nm and was metastable in the absence of an electric field. Its metastable character, combined with high solubility in water, showed a large potential for the bioavailability of this polymorph. 

Inspired by the use of MD to discover new polymorphs, Bulutoglu \etal\cite{10.3390/pr7050268} explored the solid-state transformation of infinitely periodic $\alpha$, $\beta$ and $\gamma$-glycine crystals. A new polymorph was formed under the influence of a strong electric field (0.25-1.5 V/nm). Each of the three initial structures was found to transfer into this new polymorph for a sufficiently strong electric field oriented along the right direction. The polymorph remained stable for at least 125 ns once the electric field was removed. However, the long-term stability of structures cannot be established from direct MD simulation. To gain insight into the stability and rare events, one needs to perform enhanced sampling, discussed in Section \ref{sec:sampling}.

The fact that the electric field strengths used by Parks \etal\cite{10.1021/acs.cgd.7b00356} and Bulutoglu \etal\cite{10.3390/pr7050268} are several orders of magnitude larger than experimentally accessible field strengths is typical of nonequilibrium MD and is needed to induce a transition within the accessible simulation time. The transition probability and corresponding time scale under experimental conditions can then be extrapolated from the simulation results. 
In some cases, an observable effect under a large perturbation field might actually be negligible under experimental conditions. This was concluded by Knott \textit{et al.}, who investigated the influence of a laser-induced reorientation on the nucleation energy, using Monte Carlo simulations of a Potts lattice gas.\cite{10.1063/1.3574010} On the other hand, it is also possible that an experimentally observed phenomenon is not reproduced in simulation even when a large electric field is applied. This can happen for instance for phenomena that find their origin in a perturbation of the electron cloud of atoms.\cite{10.1039/C5CP00629E} Such phenomena are not reproduced by the commonly used non-polarizable force fields, but can be effectively accounted for with the much more computationally expensive polarizable models.

\subsection{Enhanced sampling and seeded simulations} \label{sec:sampling}

The timescale limitations of MD have led to the development of various techniques that alter the natural dynamics of a system to efficiently sample its free energy landscape (\eg metadynamics and umbrella sampling) or a transition path (\eg transition path sampling and forward flux sampling). Such techniques are widely used in research areas where rare events are important, such as the folding of macromolecules\cite{10.1073/pnas.1534924100}, biological reactions.\cite{10.1080/08927022.2014.923574}, and nucleation,\cite{10.1063/1.2888999,10.1073/pnas.1421192111,10.1021/acs.jctc.9b00795} The efficiency and the challenge in applying enhanced sampling methods hinge on the ability to express the problem in terms of a small number of reaction coordinates. Making a judicious choice of reaction coordinates typically requires some prior knowledge of the system. The free energy landscape or transition path is then determined in the reaction coordinate space. As such, the information provided by these simulations is also limited to this coordinate space.  
Such a sampling approach can be employed to investigate the effect of an electric field on nucleation. Note that the effects of an oscillatory or single pulse excitation cannot be directly considered in enhanced sampling techniques because time does not evolve naturally in enhanced sampling simulations. Although static electric fields have been used in combination with enhanced sampling,\cite{10.1063/1.4832383} we are not aware of any of such studies in the context of laser-induced nucleation. 

Another way that researchers have mitigated the time scale limitations of MD to study nucleation is by introducing a cluster seed, serving as a template for nucleation \cite{10.1063/1.1931661,10.1021/jacs.5b08098,10.1063/1.5024009}. This was done for instance in the above-discussed study of Parks \etal\cite{10.1021/acs.cgd.7b00356}. In fact, this is to the best of our knowledge the only study that exposes the seeded cluster to an electric field to study the effect of laser irradiation. 

The idea behind the seeding approach is that the free energy of the initial seeded configuration is close to, or larger than, that of the critical nucleus, such that nucleation or crystal growth can occur within the accessible simulation time. The flip side of this approach is that the resulting nucleus depends strongly on the seed, such that prior knowledge of the nucleus structure, shape, and size is essential. Sun \etal\cite{10.1103/PhysRevLett.120.085703} recently introduced a biased seeding method that allows a starting point further removed from the critical nucleus. In this `persistent embryo method', a harmonic spring force stimulates the growth of a small initial seed without prescribing the resulting cluster shape. The magnitude of the biasing force gradually vanishes as the cluster grows to subcritical size, allowing for spontaneous further growth to reach the critical size. The relatively small free energy increase from subcritical to critical size can be spontaneously overcome on the nanosecond time scale accessible in MD. By performing 50 independent simulations in which a subcritical cluster was allowed to spontaneously grow or shrink, Sun \etal extracted information about the critical nucleus size and the kinetic prefactor under the assumption of classical nucleation theory. 

Although the persistent embryo method is a great way to effectively reduce the free energy barrier and increase the corresponding nucleation probability, its potential for the study of laser-induced nucleation is not straightforward for two reasons. First, if this method is applied in the presence of a static electric field, it is hard to isolate the effects of the imposed field on nucleation, since nucleation is simultaneously driven by another biasing force. Second, the persistent embryo method assumes the nucleus size to be the only relevant reaction coordinate, whereas laser-induced nucleation is often believed to be a two-step process in which structural reordering of the pre-nucleation cluster plays a key role.\cite{10.1021/acs.jpcb.6b05291} To account for such a two-step process, the method would at least need to be adapted to allow the restructuring of the cluster. It would, however, not be straightforward how to define a subcritical cluster and if restructuring should be driven in the same way as growth.

\subsection{Current challenges}

The studies discussed above illustrate the wide-ranging applicability of MD simulation to investigate molecular structure and dynamics, local fluid properties, and molecular-level mechanisms. These features are instrumental to provide an explanation or prediction of experimentally observed phenomena, such as NPLIN. Yet, quantitatively predicting nucleation has proved challenging, with reported nucleation rates calculated from MD spanning tens of orders of magnitude.\cite{10.1063/1.5024009} The predicted rate is sensitive to various methodological factors, such that a quantitative prediction requires careful consideration of each aspect discussed in the following subsections. These apply both to the study of spontaneous homogeneous nucleation and laser-induced nucleation. 

\subsubsection{Sampling technique and interpretation} 

We briefly discussed the ability of enhanced sampling techniques to overcome the time scale limitations of MD by biasing its dynamics. However, since these biased simulations do not evolve naturally following Hamiltonian dynamics, the nucleation rate cannot simply be predicted from the number of nucleation occurrences during a finite simulation time. Instead, some form of transition state theory is typically applied, with the kinetic prefactor being determined from additional simulations. Thus, sampling and interpreting the free energy barrier typically requires assumptions on the nucleation mechanism. 

\subsubsection{Finite size effects}

The large computational cost of atomistic simulation limits the accessible length scales. On the other hand, coarse-grained MD simulations might not be sufficiently detailed to reproduce orientations of small molecules, such as water, or the correct dynamics. Atomistic simulations of $10^4$-$10^6$ atoms can be sufficient for the study of nucleation-related conditions or phenomena, but a few guidelines should be kept in mind. In the first place, the simulation box should always be sufficiently large to avoid the interaction of a particle with itself via periodic boundaries. Interaction of the nucleus with itself will lead to a too-high nucleation rate. However, satisfying this constraint on the simulation box size is not sufficient to avoid finite-size effects. For example, self-diffusivity is known to be sensitive to the size of the simulation box.\cite{10.1021/acs.jctc.8b00625} 

An issue that is specific to the study of nucleation, is the chemical potential variation when nucleation extracts solutes from the finite liquid solution.\cite{10.1073/pnas.1421192111,10.1021/acs.chemrev.5b00744,10.1021/acs.jpcc.8b04017} This results in an unrealistic drop in the supersaturation, which strongly affects the predicted nucleation rate and the nucleus size.  
The fluid reservoir either needs to be very large to ensure that the supersaturation is not depleted due to nucleation, or a correction needs to be applied to account for the effect of finite system size on supersaturation.\cite{10.1021/jacs.5b08098} The fact that many simulations in the literature have not satisfied either of these measures contributes greatly to the variety of reported nucleation rates. 

\subsubsection{Interaction force field}

The physical accuracy of MD data depends on the quality of the force field describing interactions between atoms. Most classical force fields are based on the assumption of pair interactions and point charges, with polarization effects often being ignored. Furthermore, force field parameters are most commonly optimized to reproduce a small set of thermodynamic and structural properties under specific conditions, such as an infinitely diluted aqueous solution at room temperature and atmospheric pressure. Consequently, such force fields are not guaranteed to be accurate under the local conditions (i.e., concentration, temperature, pressure) under which nucleation occurs. For example, many ion force fields perform poorly in terms of solubility in water at room temperature, as well as at the temperature dependence of solubility.\cite{10.1063/5.0012102,10.1063/1.5124448} 
If the solubility limit of a force field is incorrect, then so is the supersaturation and thus the nucleation rate. Indeed, nucleation rates have been shown to be extremely sensitive to the force field.\cite{10.1103/PhysRevE.76.061505}
Notably, existing force fields with an electronic charge correction yield better solubility, but their inaccurate description of the solid phase makes them unsuitable for the study of nucleation.\cite{10.1063/1.5121392} 

\subsubsection{Control over system conditions}

Thermostats and barostats are typically employed to maintain a constant homogeneous temperature and pressure throughout the system. This can interfere with spontaneous density fluctuations that can serve as a precursor for nuclei. Perhaps more importantly, nucleation and crystallization are exothermic processes, causing a local temperature elevation that should not be artificially (and homogeneously) removed by a thermostat. Similarly, barostats scale the system size to modulate the system pressure, assuming homogeneity. When the system contains interfaces between multiple phases, pressure fields vary locally and the homogeneous expression is no longer valid \cite{10.1063/1.4737927}. These issues with temperature and pressure control can be avoided by introducing a thermostatted piston that controls the nominal pressure far from the nucleus. 

\subsection{Summary}

Molecular dynamics simulation has proved a useful tool to provide  molecular-level insight that can ultimately be used to tailor and control nucleation processes.
\begin{itemize}
\item Many studies have provided insight not by aiming to mimic a NPLIN experiment, but rather by targeting specific conditions or mechanisms that may be crucial to NPLIN.
\item Nonequilibrium MD simulation studies have found that a laser field can induce new polymorphs of paracetamol and glycine.\cite{10.1021/acs.cgd.7b00356,10.3390/pr7050268}
\end{itemize}

Two things are needed in order to quantitatively bridge the atomic length and time scales to the experimentally observable scales. 
\begin{itemize}
\item Force fields are needed that are suitable for modeling supersaturated solutions at elevated temperatures and pressures. 
\item Limitations of accessible length and time scales should be mitigated with the help of enhanced sampling and biasing techniques. Various such techniques have already been used in the study of spontaneous homogeneous nucleation, whereas few studies thus far have investigated the effects of laser irradiation on a supersaturated solution.
\end{itemize}


\section{Concluding remarks}
\label{sec:concluding}

In this review article, we summarized the body of existing literature where light fields are used to manipulate crystallization from solution. We classified the literature on laser-induced crystallization that evolved into four fields, namely NPLIN, HILIN, LTIC, and indirect methods, while limiting our work to mostly non-photochemical methods with the exception of HILIN where the role of plasma formation on crystallization is still an open question. In essence, all contributions covered in this study originate from light-material interactions. Hence the underlying physicochemical mechanism intimately depends on the experimental configuration (for instance which interfaces the irradiation interacts with), solution thermodynamics shaped by the chemical identity of solute and solvent, as well as properties of the light field interacting with the solution. 

Scientists contributing to these four fields utilized properties of the light field, such as intensity, wavelength, polarization, and exposure time, as well as experimental configuration and solution thermodynamics to evoke a plethora of interesting mechanisms to alter properties of emerging crystals through laser-material interactions shaping nucleation, growth, and secondary phenomena. These mechanisms offer rich physics laden with open questions requiring an interdisciplinary approach spanning coupled heat, mass, and momentum transfer to laser ablation, electrokinetics, and spectroscopy. We hope that the detailed classification provided will highlight the shared fundamental understanding and experimental know-how and thus facilitate scientific exchange. 

\begin{acknowledgement}
This work was funded through the Open Technology Programme by Netherlands Science Foundation (NWO), project number 16714 (LightX). The authors thank the members of LightX user committee for their productive discussions; Dr. Jörn Gebauer (Bayer AG), Dr. Clemens Bothe (Bayer AG), Dr. Bart Zwijnenburg (Nobian B.V.), Dr. Rob Geertman (Janssen Pharmaceutica), Dr. Andreas Sieber (Lonza Group), Ir. John Nijenhuis (TU Delft), Prof. dr. Antoine E.D.M. van der Heijden (TU Delft) and Dr. Herman J.M. Kramer.
\end{acknowledgement}


\clearpage
\begin{table}[htbp!]
\rotatebox{90}{
\begin{minipage}{2.8\linewidth}
  \centering
  \caption{Compilation of experimental conditions for NPLIN}%
  \label{table_NPLIN}%
  \resizebox{\linewidth}{!}{%
    \begin{tabular}{cccccccc}
    \hline
    \multicolumn{1}{m{9em}}{\centering \textbf{Setup classification}} & \multicolumn{1}{m{10.43em}}{\centering \textbf{Laser specification}} & \multicolumn{1}{m{6.285em}}{\centering \textbf{Exposure frequency \newline{}(\SI{}{Hz})}} & \multicolumn{1}{m{8.785em}}{\centering \textbf{Exposure time\newline{}(\SI{}{s})}} & \multicolumn{1}{m{8.07em}}{\centering \textbf{Laser peak-intensity \newline{}(\SI{}{MW/cm^2})}} & \multicolumn{1}{m{14em}}{\centering \textbf{Solvent}} & \multicolumn{1}{m{8em}}{\centering \textbf{Solute}} & \multicolumn{1}{m{8em}}{\centering \textbf{Supersaturation}} \\
    \hline \hline
    \Cref{ExpSetups_NPLIN}a - sl\cite{10.1103/PhysRevLett.77.3475} & P\textsuperscript{A}-ns-nfoc-L-1064 & 10    &   -    & 50-250 & H$_2$O & CH$_4$N$_2$O  & 1.1-1.29 \\ \hline
    \Cref{ExpSetups_NPLIN}a - sl\cite{10.1021/cg0055171} & P\textsuperscript{A}-ns-nfoc-L-1064 & 10    &    -   & 350   & H$_2$O & C$_2$H$_5$NO$_2$ & 1.38-1.45 \\ \hline
    \Cref{ExpSetups_NPLIN}a - sl\cite{10.1103/PhysRevLett.89.175501} & P\textsuperscript{A}-ns-nfoc-L/C-1064 & 10    & 60    & 350   & H$_2$O & C$_2$H$_5$NO$_2$ & 1.38-1.45 \\ \hline
    \Cref{ExpSetups_NPLIN}a - sl\cite{10.1021/cg050041c} & P\textsuperscript{A}-ns-nfoc-L/C-532/1064 & 10    & 60    & 20-350 & H$_2$O & CH$_4$N$_2$O  & 1.23 \\ \hline
    \Cref{ExpSetups_NPLIN}a - sl\cite{10.1021/cg050460+} & P\textsuperscript{A}-ns-nfoc-L/C-532/1064 & 10    & 60    & 240, 460 & H$_2$O & C$_2$H$_5$NO$_2$ & 1.2-2 \\ \hline
    \Cref{ExpSetups_NPLIN}a - sl\cite{10.1021/cg800028v} & P\textsuperscript{A}-ns-nfoc-L/C-532 & 10    & 60    & 240   & H$_2$O & L-Histidine & 1.4-1.8 \\ \hline
    \multirow{2}{*}{\Cref{ExpSetups_NPLIN}b - sl/ll/al\cite{10.1021/cg800696u}} & P\textsuperscript{A}-ns-nfoc-L-532/1064 & 20    & 10    &   3.2-58, 125    &   \multirow{2}{*}{Buffer sol.}    & \multirow{2}{*}{HEWL}  & \multirow{2}{*}{-} \\ 
        & P\textsuperscript{B}-ps-nfoc-L-532 & 30    & 10    &  240     &       &   &  \\ \hline
    \Cref{ExpSetups_NPLIN}d - al\cite{10.1021/cg5004319} & P\textsuperscript{C}-ns-foc-L-532 & 11000 & 0.1   & $>$ 3   & H$_2$O & KCl   & $>$ 1.2  \\ \hline
    - \cite{10.1103/PhysRevE.79.021701} & P\textsuperscript{C}-ps-nfoc-L-532 & 10    &   -    & 3.9   & 5CB & 5CB & - \\ \hline
    \Cref{ExpSetups_NPLIN}e - sl\cite{10.1021/acs.cgd.9b00362} & P\textsuperscript{A}-ns-nfoc-L-1064 & 2, 10    & 1, 6 pulses & 2.5-100  & H$_2$O & KCl   & 1.06 - 1.1 \\ \hline
    \Cref{ExpSetups_NPLIN}a - sl/al\cite{10.1021/acs.cgd.8b00688} & P\textsuperscript{A}-ns-nfoc-L/C-1064 & -      & single pulse & 500-550 & agarose gel & C$_2$H$_5$NO$_2$ & 1.5 \\ \hline
    \Cref{ExpSetups_NPLIN}e - sl\cite{10.1021/acs.cgd.0c00669} & P\textsuperscript{A}-ns-nfoc-L-1064 & 10    & 1-2 pulses & 80-105 & H$_2$O & C$_2$H$_5$NO$_2$ & 1.4-1.6 \\ \hline
    \Cref{ExpSetups_NPLIN}a - sl\cite{10.1021/cg8007415} & P\textsuperscript{A}-ns-nfoc-L/C-1064 &   -    & single pulse & $>$ 6.4 & H$_2$O & KCl   & 1.053-1.102 \\ \hline
    \Cref{ExpSetups_NPLIN}a - sl\cite{10.1016/j.cplett.2009.09.049} & P\textsuperscript{A}-ns-nfoc-L-1064 & 10    & 10    & 2.14, 2.30 & H$_2$O & KCl   & 1.066-1.076 \\ \hline
    - sl/al\cite {10.1021/ja905232m} & P\textsuperscript{A}-ns-nfoc-L-1064 &   -    & single pulse &  7-55     & agarose gel & KCl   & 1.06 \\ \hline
    - sl\cite{10.1039/c1cp22774b} & P\textsuperscript{A}-ns-nfoc-L-1064 & 10    & 3     & 9-900 & CH$_3$COOH & CH$_3$COOH &  \\ \hline
    \Cref{ExpSetups_NPLIN}a - sl\cite{10.1021/cg300750c} & P\textsuperscript{A}-ns-nfoc-L-532/1064 &  -   & single pulse &    5-40     & H$_2$O & KCl/KBr   & 1.06 \\ \hline
    \Cref{ExpSetups_NPLIN}f - sl\cite{10.1021/acs.cgd.5b00854} & P\textsuperscript{A}-ns-nfoc-L-532 & 10    & 10 pulses & 17-70 & H$_2$O & KCl   & 1.08 \\ \hline
    \Cref{ExpSetups_NPLIN}a - sl\cite{10.1039/c6cp07997k} & P\textsuperscript{A}-ns-nfoc-L-532/1064 & 10    & 30    & 200, 270 & H$_2$O  & CH$_4$N$_2$O  & 1.5 \\ \hline
    \Cref{ExpSetups_NPLIN}a - sl\cite{10.1021/acs.cgd.6b00882} & P\textsuperscript{A}-ns-nfoc-L-1064 & 0.1    & 1-3 pulses & 12    & H$_2$O & NH$_4$Cl & 1.2 \\ \hline
    \Cref{ExpSetups_NPLIN}a - sl\cite{10.1063/1.4917022} & P\textsuperscript{A}-ns-nfoc-L-532 & 0.05    & 5 pulses & 2.4-14.5 & H$_2$O & CO$_2$ and sucrose & 4.3 \\ \hline
    \Cref{ExpSetups_NPLIN}a - sl\cite{10.1039/c7cp03146g} & P\textsuperscript{A}-ns-nfoc-L/C-1064 & 10    & 60    & 210   & H$_2$O & C$_2$H$_5$NO$_2$ & 1.4-1.7 \\ \hline
       - sl\cite{10.1063/1.3637946}   &  P\textsuperscript{A}-ns-nfoc-L/C-1064 &   -    & 4-20 pulses & 161-350 & NaClO$_3$ & NaClO$_3$ & - \\ \hline
    \Cref{ExpSetups_NPLIN}c - sl/al\cite{10.1107/S160057671401098X} & P\textsuperscript{A}-ns-nfoc-L/C-532 &   10    & 60    & 80-910 & H$_2$O & C$_2$H$_5$NO$_2$ & 1.35-1.6 \\ \hline
    \Cref{ExpSetups_NPLIN}c - sl/al\cite{10.1021/cg500163c} & P\textsuperscript{A}-ns-nfoc-L/C-532 & 10    & 60    & 300-450 & C$_2$H$_3$N + CH$_3$OH & C$_{15}$H$_{12}$N$_2$O & 1.1-1.3 \\ \hline
    \Cref{ExpSetups_NPLIN}c - sl/al\cite{10.1021/acs.cgd.5b01526} & P\textsuperscript{A}-ns-nfoc-L/C-532 & 10    & 1-60    & 140-300 & H$_2$O + C$_2$H$_5$OH & C$_9$H$_9$N$_3$O$_2$S$_2$ & 1.1-1.7 \\ \hline
    \Cref{ExpSetups_NPLIN}a - sl\cite{10.1063/1.3582897} & P\textsuperscript{A}-ns-nfoc-L-355/532/1064 & 1     & 10 pulses & -      & H$_2$O & CO$_2$ & 2.5 \\ \hline
    \Cref{ExpSetups_NPLIN}a - sl\cite{10.1021/acs.cgd.7b01277} & P\textsuperscript{A}-ns-nfoc-L-355/532/1064 & -    & single pulse & 0.5-55 & H$_2$O & KCl   & 1.027, 1.049 \\ \hline
    \Cref{ExpSetups_NPLIN}a - sl\cite{10.1021/acs.cgd.6b00046}  & P\textsuperscript{A}-ns-nfoc-L-1064 & 10    & 60    & 470   & H$_2$O & C$_2$H$_5$NO$_2$ & 1.4-1.6 \\ \hline
    \Cref{ExpSetups_NPLIN}a - sl\cite{10.1021/acsomega.0c04902}  & P\textsuperscript{A}-ns-nfoc-L-532 & 10    & single pulse    & 16   & H$_2$O + poly-(epoxysuccinic acid) & CsCl & 1.15 \\ \hline
    \Cref{ExpSetups_NPLIN}a - sl\cite{10.1021/acs.cgd.0c01415}  & P\textsuperscript{A}-ns-nfoc-L/C-532/1064 & 10    & 1, 600 pulses    & 450   & H$_2$O & C$_2$H$_5$NO$_2$ & 1.5-1.7 \\ \hline
              &       &       &       &       &       &       &  \\
              &       &       &       &       &       &       &  \\
    \end{tabular}}%
  \begin{tablenotes}\footnotesize
  \item \textbf{Setup classification:} The passage of the laser through an interface is indicated as follows: air-liquid (al), liquid-liquid(ll) and solid-liquid(sl). The liquid is usually the solution except in case of ll - in which the other liquid is a sealant. The solid represents the glass walls of the container. 
\item \textbf{Laser specification:} P = pulsed, ns = nanosecond, fs = femtosecond, foc = focused, nfoc = non-focused, L/C = linearly/circularly polarised light, 355/532/1064 = wavelengths in nm. Superscripts: \textsuperscript{A} = Q-switched Nd:YAG laser, \textsuperscript{B} = Quantel YG501, \textsuperscript{C} = diode-pumped, frequency doubled TEM\textsubscript{00} Nd:YAG laser.
\item \textbf{Solvent:} Buffer sol. = 6\% NaCl in \SI{0.1}{M} acetate aqueous buffer solution of pH 4.35.
\end{tablenotes}
\end{minipage}
}
\end{table}


\clearpage
\begin{table}[htbp!]
\rotatebox{90}{
\begin{minipage}{2.8\linewidth}
 \caption{Compilation of experimental conditions for HILIN}%
  \label{table_LIN}%
  \resizebox{\linewidth}{!}{%
  \centering
    \begin{tabular}{cccccccc}
    \hline
    \multicolumn{1}{m{6em}}{\centering \textbf{Setup \newline{}classification}} & \multicolumn{1}{m{10.43em}}{\centering \textbf{Laser specification}} & \multicolumn{1}{m{6.285em}}{\centering \textbf{Exposure frequency \newline{}(\SI{}{Hz})}} & \multicolumn{1}{m{8.785em}}{\centering \textbf{Exposure time\newline{}(\SI{}{s})}} & \multicolumn{1}{m{8.07em}}{\centering \textbf{Laser peak-intensity \newline{}(\SI{}{PW/cm^2})}} & \multicolumn{1}{m{14.015em}}{\centering \textbf{Solvent}} & \multicolumn{1}{m{10.355em}}{\centering \textbf{Solute}} & \multicolumn{1}{m{5.43em}}{\centering \textbf{Super-\newline{}saturation}} \\
    \hline \hline
    \Cref{fig8}f\cite{10.1143/jjap.42.l798} & P\textsuperscript{A}-fs-foc-780 & 1000    &   60    & 0.00263 & H$_2$O & Lysozome  & - \\ \hline
    \Cref{fig8}g\cite{10.1021/cg049709y} & P\textsuperscript{A}-fs-foc-800 & 20,1000    &    30000 pulses   & 318   & CH$_3$OH & DAST & 1.44-1.64 \\ \hline
    \Cref{fig8}d\cite{10.1021/cg060631q} & P\textsuperscript{A}-fs-foc-800 & 125    & 1 pulse    & 0-21.8   & Cyclohexane & Anthracene & 1.1-1.9 \\ \hline
    \Cref{fig8}f\cite{10.1016/j.apsusc.2007.01.046} & P\textsuperscript{A}-fs-foc-800 & 1000    & 1 pulse    & \makecell{5.11, 34.1 \\ 51.1, 85.3} & Buffer sol$^{1}$ & HEWL  & 1.23 \\ \hline
    \Cref{fig8}f\cite{10.1007/s00339-008-4790-x} & P\textsuperscript{B}-fs-foc-780 & 1000    & 67 pulses    & 0.191, 0.0566   & Buffer sol$^{2}$ & Lysozome & 5.5,5.9,6.3 \\ \hline
    \Cref{fig8}f\cite{10.1007/s00339-008-4790-x} & P\textsuperscript{B}-fs-foc-780 & 1000    & 128 pulses    &   11.6    &   Buffer sol$^{3}$ & Lysozome & -   \\ \hline
    \Cref{fig8}f\cite{10.1021/cg2000014} & P\textsuperscript{C}-ns-foc-532 & 10   & 1 pulse & 0.00083   & H$_2$O & \makecell{(NH$_{4}$)$_{2}$SO$_{4}$ \\ KMnO$_{4}$}   & \makecell{1.002,1.004 \\ 1.07,1.14,1.21} \\ \hline
    \Cref{fig8}c\cite{10.1016/j.jcrysgro.2010.10.068} & P\textsuperscript{B}-fs-foc-780 & 1000    &   5 pulses    & 6.3   & Buffer sol$^{4}$ & Glucose Isomerase & - \\ \hline
    \Cref{fig8}i\cite{10.1021/cg301024t} & P\textsuperscript{C}-fs-foc-780 & 10    &   1 pulse    & \makecell{0.000015\\0.000007}   & H$_2$O & KNO$_{3}$ & \makecell{1-1.3 \\ 1.5-2} \\ \hline
    \Cref{fig8}f\cite{10.1021/cg301618h} & P\textsuperscript{B}-fs-foc-780/800 & 1000 & 100 pulses & 13.4,26.8  & Buffer sol$^{5}$ & HEWL & - \\ \hline
    \Cref{fig8}f\cite{10.1021/cg301618h} & P\textsuperscript{B}-fs-foc-780/800 & 1000 & 100 pulses & 26.8 & H$_2$O & Paracetamol & - \\ \hline
    \Cref{fig8}e\cite{10.1016/j.jcrysgro.2012.11.018} & P\textsuperscript{A}-fs-foc-800 & 1-1000    & 10 min & 79.6 & H$_2$O & Glycine & 1.0-1.33 \\ \hline
    \Cref{fig8}g\cite{10.1021/acs.cgd.9b00951} & P\textsuperscript{D}ns-foc-532/1064 &   10    & 1 pulse & 0.087, 0.068 & H$_2$O & NaClO$_{3}$   & 1.21 \\ \hline
    \Cref{fig8}h\cite{10.7567/apex.8.045501} & P\textsuperscript{B}fs-foc-800 &   1000    & 60000 pulses & 0.000084 & Acetonitrile & Indomethacine   & 3.5 \\ \hline
    \Cref{fig8}i\cite{10.1021/acs.cgd.9b00123} & P\textsuperscript{E}fs-foc-800-as &   1000    & 120 & 0.00218-0.01383 & H$_2$O & Acetaminophen   & 1.5 \\ \hline
    \Cref{fig8}i\cite{10.1021/acs.cgd.0c01476} & P\textsuperscript{F}fs-foc-800-as &   1000    & 60000 pulses & 0.00145-0.01164 & H$_2$O + C$_2$H$_5$OH & Sulfathiazole   & 1.1-1.5 \\ \hline
              &       &       &       &       &       &       &  \\
              &       &       &       &       &       &       &  \\
    \end{tabular}}%
    \begin{tablenotes}\footnotesize
  \item 
\item \textbf{Laser specification:} P = pulsed, ns = nanosecond, fs = femtosecond, foc = focused, 532/780/800/1064 = wavelengths in nm. Superscripts: \textsuperscript{A} = Ti:sapphire femtosecond laser, \textsuperscript{B} = IFRIT femtosecond laser, \textsuperscript{C} = Q-switched Nd:YAG laser, \textsuperscript{D} = Q-switched Nd$^{+3}$:YAG laser, \textsuperscript{E} = Spitfire Pro femtosecond laser , \textsuperscript{F} = Coherent, Legend Eliter femtosecond laser.

\item \textbf{Solvent:} Buffer sol.$^{1}$ = 100 mM sodium acetate buffer at pH 4.5 including sodium chloride (NaCl, 3.5 wt$\%$) and PEG 6000 (20 wt$\%$). Buffer sol.$^{2}$ = 100 mM sodium acetate buffer at pH 4.5 including sodium chloride (NaCl, 2.5 wt$\%$). Buffer sol.$^{3}$ = 10 mg/ml lysozyme (F lysozyme 0.025 mg/ml), 1$\%$ agarose gel, 50 mM sodium acetate buffer (pH 4.5),and 6.0 wt$\%$ NaCl as a precipitant at a concentration of 1–5$\%$. Buffer sol.$^{4}$ = 0.2–2.6 mg/ml Glucose Isomerase in 50mM Tris–HCl buffer (pH 4.5) and 0.1 M  CaCl$_{2}$ were prepared. Polyethylene glycol (PEG) 6000 was used  as a  precipitant. Buffer sol.$^{5}$ = 20 mg/mL in 0.1 M sodium acetate buffer (pH 4.5) and precipitating solutions at 10$\%$ (w/v) NaCl in 0.1 M sodium acetate  buffer (pH 4.5).

\end{tablenotes}
\end{minipage}
}
\end{table}%

\clearpage
\begin{table}[htbp!]
 \centering
\rotatebox{90}{
\begin{minipage}{2.8\linewidth}
\caption{Compilation of experimental conditions for LTIC.}
    \label{table_laser_trap}%
    \resizebox{\linewidth}{!}{%
    \begin{tabular}{cccccccc}
    \hline
    \multicolumn{1}{m{11.07em}}{\centering \textbf{Setup classification}} & 
    \multicolumn{1}{m{10.43em}}{\centering \textbf{Laser \newline{}specification}} & 
    \multicolumn{1}{m{8.785em}}{\centering \textbf{Exposure time\newline{}(\SI{}{s})}} &
    \multicolumn{1}{m{8.07em}}{\centering \textbf{Laser intensity \newline{}(\SI{}{GW/cm^2})}} & 
    \multicolumn{1}{m{9.215em}}{\centering \textbf{Solvent}} & 
    \multicolumn{1}{m{10.355em}}{\centering \textbf{Solute}} & 
    \multicolumn{1}{m{5.43em}}{\centering \textbf{Super- \newline{}saturation}} &
    \multicolumn{1}{m{6.285em}}{\centering \textbf{Phenomena \newline{}observed}} \\\hline
    \hline
    Fig.\ref{fig8} b as \cite{10.1246/cl.2007.1480} & F-CW\textsuperscript{A}-1064 & \multicolumn{1}{c}{16} & \multicolumn{1}{c}{0.40} & D$_2$O & Glycine & 0.3g/1g & N, Ge \\ \hline
    \multicolumn{1}{c}{Not shown - as\cite{10.1143/JJAP.46.L1234}} & F-CW\textsuperscript{A}-1064 & 3600-7200 & \multicolumn{1}{c}{} & D$_2$O & HEWL  & \multicolumn{1}{c}{} & N \\ \hline
    Not shown - ns \cite{10.1246/cl.2009.482} & F-CW\textsuperscript{A}-1064 & $>$ 18 & \multicolumn{1}{c}{0.40} & D$_2$O & Glycine & 0.3g/1g & Ge \\ \hline
    Fig.\ref{fig8} a - as \cite{10.1021/jz900370x} & F-CW\textsuperscript{A}-L-1064 & $>$ 400 & 0.28 - 0.49\textsuperscript{*} & D$_2$O & Glycine & 0.5, 0.68 & N, P \\ \hline
    Fig.\ref{fig8} a - as \cite{10.1021/cg100830x} & F-CW\textsuperscript{A}-L-1064 & $>$ 400 & 0.28 - 0.49\textsuperscript{*} & D$_2$O & Glycine & 0.68, 1.36 & N, Ge, D \\ \hline
    Fig.\ref{fig8} a - as \cite{10.1021/jp9072334} & F-CW\textsuperscript{A}-1064 & \multicolumn{1}{c}{1800} & \multicolumn{1}{c}{} & D$_2$O & Multiple aminoacids & \multicolumn{1}{c}{} & N \\ \hline
    Fig.\ref{fig8} a - as \cite{10.1021/cg300065x} & F-CW\textsuperscript{A}-L/C-1064 & \multicolumn{1}{c}{1800} & 0.28 - 0.49 & D$_2$O & Glycine & 0.68, 1.36 & N, P \\ \hline
    Fig.\ref{fig8} a - as \cite{10.1021/jz401122v} & F-CW\textsuperscript{A}-1064 & \multicolumn{1}{c}{195} & \multicolumn{1}{c}{0.39\textsuperscript{*}} & H$_2$O & L-Phenylalanine & \multicolumn{1}{c}{0.83} & N \\ \hline
    Fig.\ref{fig8} a - as \cite{10.1039/c3pp50276g} & F-CW\textsuperscript{A}-1064 & $>$ 700 & \multicolumn{1}{c}{0.39\textsuperscript{*}} & D$_2$O \& H$_2$O & L-Phenylalanine & \multicolumn{1}{c}{1} & N \\ \hline
    Fig.\ref{fig8} a - as \cite{10.1021/acs.cgd.5b01505} & F-CW\textsuperscript{A}-1064 & \multicolumn{1}{c}{600} & 0.07, 0.21, 0.39\textsuperscript{*} & H$_2$O & L-Phenylalanine & 0.67 - 0.92, & N, Gc \\ \hline
    Fig.\ref{fig8} b - as \cite{10.1039/c7cp06990a} & F-CW\textsuperscript{B}-L-1064 & 1800-5400 & 0.18, 0.28, 0.39\textsuperscript{*} & D$_2$O & HEWL  & \multicolumn{1}{c}{3} & N, Gc \\ \hline
    Fig.\ref{fig8} a - as \cite{10.7567/1882-0786/ab4a9e}  & F-CW\textsuperscript{B}-1064 & $>$ 500 & \multicolumn{1}{c}{0.39\textsuperscript{*}} & H$_2$O & L-Phenylalanine & \multicolumn{1}{c}{0.58} & N, Gc \\ \hline
    Fig.\ref{fig8} b - gs \cite{10.1021/cg401065h} & F-CW\textsuperscript{A}-L-1064 & \multicolumn{1}{c}{1800} & 0.21, 0.39\textsuperscript{*} & D$_2$O & HEWL  & \multicolumn{1}{c}{} & Gc \\ \hline
    Fig.\ref{fig8} b - gs \cite{10.1021/cg501860k} & F-CW\textsuperscript{A}-L/C-1064 & \multicolumn{1}{c}{1800} & 0.21, 0.39\textsuperscript{*} & D$_2$O & HEWL  & \multicolumn{1}{c}{} & N \\ \hline
    Fig.\ref{fig8} a - as \cite{10.1021/acs.jpcc.9b11651} & F-CW\textsuperscript{A}-L/C-1064 & \multicolumn{1}{c}{1800} & 0.17 - 0.49\textsuperscript{*} & D$_2$O & KCl   & \multicolumn{1}{c}{1.03} & N, Mc \\ \hline
    Fig.\ref{fig8} a - as \cite{10.1021/acs.cgd.8b00796} & F-CW\textsuperscript{A}-L/C-1064 & \multicolumn{1}{c}{1800} & 0.25 - 0.53\textsuperscript{*} & D$_2$O & L-Phenylalanine &  & N, P \\ \hline
    Fig.\ref{fig8} b - as \cite{10.1021/acs.cgd.7b01116} & F-CW\textsuperscript{C}-C-532 & $>$ 6 & \multicolumn{1}{c}{0.35\textsuperscript{*}} & H$_2$O & NaClO$_3$ & $<$ 1 & N, P \\ \hline
    Fig.\ref{fig8} b - as \cite{10.1021/acs.cgd.9b00600} & F-CW\textsuperscript{C}-C-532 & $>$ 3 & \multicolumn{1}{c}{} & H$_2$O & NaClO$_3$ & saturated & Gc \\ \hline
    Fig.\ref{fig8} b - as \cite{10.1002/anie.201806079} & F-CW-800/1064 & $>$ 100 & \multicolumn{1}{c}{} & DMF   & MAPbX$_3$ & $<$ 1 & N, Gc \\ \hline
    Fig.\ref{fig8} a - as \cite{10.1117/12.929381} & F-CW\textsuperscript{A}-C-1064 & \multicolumn{1}{c}{1800} & 0.35, 0.49\textsuperscript{*} & D$_2$O & L-Alanine & \multicolumn{1}{c}{1.1} & N, R \\ \hline
    Fig.\ref{fig8} b - as, gs \cite{10.1117/12.860241} & F-CW\textsuperscript{A}-L-1064 & $>$ 470 & \multicolumn{1}{c}{} & D$_2$O & Urea  & 0.28 - 1.36 & LLPS, Gc \\ \hline
    Fig.\ref{fig8} b - as, gs \cite{10.1021/jz100266t} & F-CW\textsuperscript{A}-1064 & $>$ 200 & \multicolumn{1}{c}{0.40} & D$_2$O & Glycine & 23 wt\% & LLPS, N  \\ \hline
    Fig.\ref{fig8} a - as \cite{10.35848/1347-4065} & F-CW\textsuperscript{D}--800 & $>$ 720 & \multicolumn{1}{c}{} & H$_2$O & L-Phenylalanine & 0.8-1.2 & P \\ \hline
    Fig.\ref{fig8} a - as \cite{10.1351/PAC-CON-10-09-32} & F-CW\textsuperscript{A}-L-1064 & $>$ 60 & 0.28 - 0.49\textsuperscript{*} & H$_2$O & Glycine & 23 wt\% & N, G, D \\ \hline
    Fig.\ref{fig8} b - as, gs \cite{10.1021/jz100266t} & F-CW\textsuperscript{A}-1064 & $>$ 200 & \multicolumn{1}{c}{0.40} & D$_2$O & Glycine & 23 wt\% & LLPS, N  \\ \hline
    Fig.\ref{fig8} a - as \cite{10.35848/1347-4065} & F-CW\textsuperscript{D}--800 & $>$ 720 & \multicolumn{1}{c}{} & H$_2$O & L-Phenylalanine & 0.8-1.2 & P \\ \hline
    Fig.\ref{fig8} b - as \cite{10.1021/acs.cgd.1c00822} & F-CW\textsuperscript{A}-L/C-1064 & $>$ 2400 & 38 - 81\textsuperscript{*} & D$_2$O & $\beta$-Cyclodextrin & 0.14 - 0.84 & N, D, G, P \\ \hline
    \end{tabular}}%
  \label{tab:lasertrap}%
    \begin{tablenotes}\footnotesize
\item \textbf{Setup specification:} as = focus at the air-solution interface, gs = focus at the glass-solution interface, ns = focus near seed crystal.
\item \textbf{Laser specification:} F = focused laser beam, CW = continuous wave, \textsuperscript{A} = Cw Nd$^{3+}$-YVO$^4$, \textsuperscript{B} = Cw Near infrared (NIR) Laser; \textsuperscript{C} = CW green laser; \textsuperscript{D} = Ti:sapphire femtosecond laser, L = linearly polarized light, C = circularly polarized light, L/C = linearly and circularly polarised light, 355/532/1064 = wavelengths in nm.
\item \textbf{Phenomena observed:} D = dissolution, Gc = growth control, Ge = growth enhancement, LLPS = liquid-liquid phase separation, Mc = morphology control, N = nucleation, P = polymorph control, R = crystal rotation.
\item \textbf{\textsuperscript{*}} Calculated based on given laser powers and focusing objectives magnifications.
\end{tablenotes}
\end{minipage}
}
\end{table}%


\clearpage

\bibliography{refs}

\end{document}